\documentclass{svjour3-HAL}                     
\usepackage[a4paper]{geometry}
\usepackage[utf8]{inputenc}
\usepackage{extarrows}
\usepackage{verbatim}
\usepackage{stackengine}
\usepackage{amsfonts}
\usepackage{amsmath}
\usepackage{graphicx}
\usepackage{comment}
\usepackage[noend]{algorithmic}
\usepackage{orcidlink}
\usepackage{wasysym}
\usepackage{booktabs}

\usepackage[ruled]{algorithm}
\usepackage{algorithmic}

\usepackage{url}
\newcommand{\true}{{\sf True}}

\newcommand{\cons}{{\sf Cons}}
\newcommand{\inc}{{\sf Confl}}
\newcommand{\ans}{{\sf Ans}}
\newcommand{\consans}{{\sf C\_ans}}

\newcommand{\sconsans}{{\sf C\_ans}^-}
\newcommand{\wconsans}{{\sf C\_ans}^+}

\newcommand{\rep}{{\sf Rep}}

\newcommand{\tuples}{{\sf tuples}}

\usepackage{todonotes}

\usepackage{url}
\urldef{\mailsa}\path|dominique.laurent@cyu.fr|
\urldef{\mailsb}\path|nicolas.spyratos@lri.fr|

\begin{document}

\title{A Chase-based Approach to Consistent Answers of \\Analytic Queries in Star Schemas}
\titlerunning{Consistent Query Answering in Star Schemas}
\author{Dominique Laurent \and Nicolas Spyratos}

\institute{Dominique Laurent {\orcidlink{https://orcid.org/0000-0002-7264-9576}} \at ETIS Laboratory - ENSEA, CY Cergy Paris University, CNRS\\F-95000 Cergy-Pontoise, France\\ \mailsa \\
\and
Nicolas Spyratos {\orcidlink{https://orcid.org/0000-0002-3432-8608}}\at LISN Laboratory - University Paris-Saclay, CNRS\\
F-91405 Orsay, France\\
\mailsb\\~\\
{\bf Acknowledgment:} Work conducted while the second author was visiting at FORTH Institute of Computer Science, Crete, Greece (https://www.ics.forth.gr/)}

\maketitle


\begin{abstract}
We present an approach to computing consistent answers to queries possibly involving an aggregation operator in databases operating under a star schema and possibly containing missing values and inconsistent data. Our approach is based on earlier work concerning consistent query answering for standard queries (with no aggregate operator) in multi-table databases. In that work, we presented polynomial algorithms for computing either the exact consistent answer to a query or bounds of the exact answer, depending on whether the query involves a selection condition or not.

In the present work, we consider databases operating under a star schema. Calling data warehouses such databases, we extend our previous work to queries involving aggregate operators, called analytic queries. In this context, we propose specific algorithms for computing exact consistent answers to queries, whether analytic or not, provided that the selection condition in the query satisfies the property of independency (i.e., the condition can be expressed as a conjunction of conditions each involving a single attribute). We show that the overall time complexity of these specific algorithms is in $O(W\cdot \log(W))$, where $W$ is the size of the data warehouse. Moreover, the case of analytic queries involving a having clause associated with a group-by clause is discussed in the context of our approach.
\end{abstract}


\maketitle

\section{Introduction}\label{sec:intro}
The efficient processing of queries involving an aggregation operator, hereafter referred to as {\em analytic queries}, is an important issue in databases and has motivated considerable research work since the last three decades. The main purpose of analytic queries is to extract relevant `statistics' from huge volumes of data, resulting from the integration of heterogeneous databases and stored in what is called a {\em data warehouse} \cite{inmon96,Kimball}. For efficiency reasons, the data stored in a data warehouse is generally organized according to a non-normalized schema, called a {\em star schema}. A star schema consists of two types of relation schemas (also called {\ tables}): a {\em fact table} and a number of {\em dimension tables}. In this context, an analytic query can be seen as an SQL Group-by query involving some aggregate operator such as $min$, $max$, $count$ or $sum$ operating over attributes in the fact table called {\em measure attributes} (or simply measures). Let us see an example to illustrate the concepts of star schema and analytic query. 
\begin{example}\label{ex:intro}
{\rm
Consider a company structured into branches located all over the world and selling products of various types to customers. To analyze the efficiency of the company operations, one may for instance be interested in the quantities of products sold in each branch during the past year. In order to answer efficiently such a query, knowing that the data warehouse may contain billions of sales, the data are organized according to the following {\em star schema}:
\begin{itemize}
     \item {\em Fact table.} This table, denoted by $F$ is meant to store all sales by the company. In our example, $F$ is defined over attributes $K_B$, $K_P$, $K_C$, $K_D$ and $Qty$, with  $K_BK_PK_CK_D$ being its (primary) key, which means that $F$ must satisfy the functional dependency $K_BK_PK_CK_D \to Qty$. In other words, there can't be two distinct sales concerning the same branch, the same product, the same customer and the same date, associated with {\em more than one} quantity.
    \item {\em Dimension tables.} There are four dimension tables, one for each of the attributes $K_B$, $K_P$, $K_C$, $K_D$:
    \begin{itemize}
    \item {\sf Branch}, defined over the attributes $K_B$, $B\_Town$,  $B\_Ctry$ standing respectively for the branch identifier, the town in which the branch is located and the country which this town belongs to. The attribute $K_B$ is the (primary) key of {\sf Branch}, meaning that the table {\sf Branch} must satisfy the functional dependencies $K_B \to B\_Town$ and $K_B \to B\_Ctry$.
    \item {\sf Prod}, defined over the attributes $K_P$, $P\_Type$, $Price$, where $K_P$ is the product identifier, and $P\_Type$ and $Price$ are respectively the type and the price of a product. The attribute $K_P$ is the (primary) key of {\sf Prod}, meaning that the table {\sf Prod} must satisfy the functional dependencies $K_P \to P\_Type$ and $K_B \to Price$.
    \item {\sf Cust}, defined over the attributes $K_C$, $C\_Name$, $C\_Addr$ standing respectively for customer identifier, name and address of the customer. The attribute $K_C$ is the (primary) key of the table {\sf Cust}, meaning that {\sf Cust} must satisfy the functional dependencies $K_C \to C\_Name$ and $K_C \to C\_Addr$.
    \item {\sf Date}, defined over the attributes $K_D$, $Month$, $Year$ standing respectively for the date identifier or key, the month and the year of the date. The attribute $K_D$ is the (primary) key of {\sf Date}, meaning that {\sf Date} must satisfy the functional dependencies $K_D \to Month$ and $K_D \to Year$.
    \end{itemize}
\end{itemize}
Moreover, referential constraints are generally enforced in order to ensure that any key value occurring in $F$ also occurs in the corresponding dimension table. In the case of our example these constraints are expressed through the following inclusions: $\pi_{K_B}(F) \subseteq \pi_{K_B}({\sf Branch})$, $\pi_{K_P}(F) \subseteq \pi_{K_P}({\sf Prod})$, $\pi_{K_C}(F) \subseteq \pi_{K_C}({\sf Cust})$ and $\pi_{K_D}(F) \subseteq \pi_{K_D}({\sf Date})$.

In this setting a typical analytic query is to display  the total quantity of each product sold during the year $2024$. This query can be expressed in SQL as follows:

\smallskip
{\tt select} $K_P$, $sum(Qty)$ {\tt from} $J$ {\tt where} $Year=2024$ {\tt group by} $K_P$

\smallskip\noindent
Here $J$ denotes the (lossless) join of all dimension tables with the fact table $F$ (although the join can be simplified by involving only $F$, {\sf Prod} and {\sf Date}).
}
\hfill$\Box$
\end{example}
How to {\em efficiently} evaluate analytic queries against huge volumes of data has been widely investigated and lies outside the scope of the present paper; the reader is referred to \cite{Ullman} regarding standard SQL query optimization and to \cite{BellatrecheGL05} regarding more specific optimization techniques for analytic queries. 

Now, most approaches to optimize the evaluation of analytic queries assume that the functional dependencies and referential constraints are satisfied by the data warehouse. However, in practice, the situation is quite different as the data warehouse may contain inconsistencies and also missing data. For instance, in the above example, a customer may appear in the data warehouse with two distinct addresses (one in $Paris$ and one in $Athens$), thus violating the functional dependency $K_C \to C\_Addr$; or the price of a product may be missing in the table {\sf Prod}. We draw attention on that, in the case of the above query, these `inconsistencies' should {\em not} affect the computation of its answer, because the query does not refer to customer addresses, nor to product prices. Notice also that, if a product identifier occurs in the fact table $F$ but not in the dimension table {\sf Prod} - thus violating the referential constraint $\pi_{K_P}(F) \subseteq \pi_{K_P}({\sf Prod})$, all sales involving this product can be processed when computing the answer to the above query. This is so because, when computing the answer to this query, the only needed attribute value among all attributes of the table {\sf Cust} is the $K_C$-value of the tuple in $F$ being processed.

A more problematic situation is if the selection condition in the query is $Year=2024$ {\bf and} $C\_Addr= Paris$. This is so because among all transactions regarding customers whose address may be Paris, some concern customers whose address may violate the dependency $K_C \to C\_Addr$ in the table {\sf Cust}. Dealing with such inconsistencies, known as the problem of computing the {\em consistent answer} to an analytic query, is not trivial, and as argued in \cite{ArenasBCHRS03,FuxmanFM05}, techniques used for standard non analytic queries cannot be used for analytic queries.

\smallskip
To cope with inconsistencies and missing values in data warehouses,  our approach is based on our earlier work \cite{LauS25} dealing with consistent query answering for standard, non analytic queries in multi-table databases. In that work, we presented polynomial algorithms for computing either the exact consistent answer to a standard non analytic query or bounds of the exact answer, depending on whether the query involves a selection condition or not. 
In  the present paper, we show how to compute effectively the exact consistent answer in the case of a star schema, under the restriction that the selection condition satisfies the  property of independency  (i.e., the condition can be expressed as a conjunction of conditions each involving a single attribute).

\smallskip
The main contributions of this paper can be summarized as follows: We propose specific algorithms for computing exact consistent answers to queries, whether analytic or not, over a star schema, provided that the selection condition in the query satisfies the property of independency. 
We also show that the overall time complexity of our specific algorithms is in $O(W\cdot \log(W))$, where $W$ is the size of the data warehouse. We discuss the case of analytic queries involving a having clause associated with a group-by clause in the context of our approach. 

\smallskip
The paper is organized as follows: In Section~\ref{sec:model} we recall the main features of our previous work in \cite{LauS25}, on which the present approach is based. In Section~\ref{sec:star schema} we first recall the definition of a star schema and argue that the approach in \cite{LauS25} applies in this context. In Section~\ref{sec:repairs} we investigate the concept of repairs in the context of star schemas. Section~\ref{sec:cqa} deals with consistent answers to queries in the case of standard projection-selection queries as well as in the case of analytic queries. In Section~\ref{ans-computation}, we propose algorithms for efficiently computing the consistent answers to analytic queries, or in some few cases an approximation to the consistent answer. In Section~\ref{sec:rel-work} we compare our approach to other approaches from the literature and in Section~\ref{sec:conclusion} we summarize the contents of the paper and suggest research directions for future work. 
\section{The Approach of \cite{LauS25}}\label{sec:model}
\subsection{Basic Definitions and Notation}
As done in \cite{JIIS,LauS25}, we rely on basic notions from relational databases as introduced in the literature \cite{Ullman}. As usual in this context, we consider a finite set of attributes $U= \{A_1, \ldots , A_n\}$, called the universe, and we assume that each attribute $A_i$ is associated with a {\em domain} of values denoted $dom(A_i)$. A nonempty subset $R=\{A_{i_1},\ldots ,A_{i_p}\}$ of $U$ is called a (relation) schema and simply denoted by $A_{i_1}\ldots A_{i_p}$ (i.e., by the concatenation of its attributes). Given such a schema $R=A_{i_1}\ldots A_{i_p}$, a tuple $t$ over $R$ is an element of the cross product $dom(A_{i_1}) \times \ldots \times dom(A_{i_p})$, and $t$ is denoted by the concatenation of its values, that is $t=a_{i_1}\ldots a_{i_p}$ where $a_{i_j}$ is in $dom(A_{i_p})$, for every $j=1, \ldots ,p$. In this case, $R$ is called the schema of $t$ and is also denoted by $sch(t)$. Moreover, if $S$ is a subset of $R$, the restriction of $t$ over $S$, denoted by $t.S$ is the tuple over $S$ whose values over attributes in $S$ are those of $t$. More formally, given tuple $t$ over $R$, $S \subseteq R$, and tuple $q$ over $S$, the fact that $q=t.S$ means that $sch(q)=S$ and that for every $A \in S$, $q.A=t.A$.

In this setting a relation, also called a table, over schema $R$ is a finite set of tuples over $R$, and a multi-table database over universe $U$ is a set of such relations. Additionally, relations in a database are associated with {\em constraints}, whose goal is to restrict their content so as to model given properties. Functional dependencies, that we consider in our work, are recognized as the most popular constraints. 

Given a relation $r$ over schema  $R$, a functional dependency over $R$ (or simply `dependency over $R$') is an expression of the form $X \to Y$ where $X$ and $Y$ are subsets of $R$. A relation $r$ over $R$ is said to satisfy the dependency $X \to Y$ if for all tuples $t$ and $t'$ in $r$ the following holds: if $t.X=t'.X$ then $t.Y=t'.Y$; and $r$ is said to satisfy a set $FD$ of dependencies over $R$ if $r$ satisfies every dependency of $FD$. A basic result from functional dependency theory \cite{Armstrong} says that: $r$ satisfies a set $FD$ of dependencies over $R$ if and only if $r$ satisfies the set $FD'= \{X \to A~|~X \to Y \in FD, A \in Y\}$.  For example, if $r$ is a relation over $R= ABCDE$ then $r$ satisfies $FD= \{AB\to CD, D\to EC\}$ if and only if $r$ satisfies $FD'= \{AB\to C, AB\to D, D\to E, D\to C\}$. This explains why in our work we restrict our attention to dependencies having a right hand-side consisting of just one attribute. Next, we address the issue of checking whether a given multi-table database satisfies a given set of dependencies.
\subsection{From $Chase$ to $m\_Chase$}
Traditionally, to verify the consistency of a multi-table database with respect to a set $FD$ of functional dependencies one applies the well-known {\em Chase} algorithm \cite{Ullman}. The input of this algorithm is a table $T$ over the set $U$ of all attributes appearing in the database. $T$ has as many rows as there are tuples in the database and each tuple is placed on a separate row, eventually with missing values. The algorithm derives new tuples by applying the dependencies of $FD$ as long as no pair of tuples is in conflict with some dependency; and stops as soon as such a conflict is encountered. Let $Chase(T)$ denote the result upon termination. 

We recall here that a dependency application, also known as the {\em Lossless-Join} rule is defined as follows \cite{FaginMU82,Ullman}:

\smallskip
{\bf for all} $t$ and $t'$ in the current value of $Chase(T)$

\indent\hspace{.3cm}
{\bf if} there exists $X \to A$ in $FD$ such that $t$ and $t'$ are defined over $XA$ and $t.X=t'.X$ {\bf then}

\indent\hspace{.6cm}
{\bf if} $t.A$ and $t'.A$ are {\em distinct} domain values, {\bf then} {\em fail}

\indent\hspace{.6cm}
{\bf else if} $t.A=a$ and $t'.A$ is null {\bf then} assign $a$ to $t'.A$

\smallskip\noindent
When running the {\em Chase} algorithm, a tuple $t$ in the current value of $Chase(T)$ 
is said to be {\em conflicting} if the following holds: there is a tuple $t'$ in the {\em current} value of $Chase(T)$ and a dependency $X \to A$ in $FD$ such that $t$ and  $t'$ are both defined over $XA$, $t.X= t'.X$ and $t.A \ne t'.A$. A tuple $t$ is called {\em non-conflicting} if $t$ is not a conflicting tuple. 

\smallskip
Now, if {\em Chase} is successful (i.e., no conflicts are encountered and no more new tuples can be derived) then the database is declared consistent else conflicting. 
If the database is consistent then processing of queries (whether standard queries or analytic queries) proceeds as usual, else the following question arises: can we still extract useful (i.e., non conflicting) information from the conflicting database? The work in \cite{LauS25} gives a positive answer to this question based on an extended version of the {\em Chase} algorithm, called the $m\_Chase$ algorithm (to be presented formally in the following subsection). As will shall see, the $m\_Chase$ algorithm operates over the set of all tuples that can be built up from constants from the active domains of the database, rather than over the tuples in $T$. We recall that the active domain of an attribute $A$ in a multi-table database is the set of all values of the domain of $A$ that are present in the database. Note that every tuple appearing in the database is built up from constants in the active domains of database attributes but there are tuples that are built-up from constants in the active domains that are not present in the database. For example, if a database relation has only two tuples, $ab$ and $a'b'$, then the tuples  $ab'$ and $a'b$ are built up from constants in the active domains (here $\{a, a'\}$ and $\{b, b'\}$) but they don't belong to the relation.

In order to deal with this important difference between {\em Chase} and $m\_Chase$, we introduce the following notation: given a set of tuples $S$, we denote by $\mathcal{S}$ the set of all tuples that can be built up from constants occurring in $S$. In our previous example, where $S=\{ab', a'b\}$, we have $\mathcal S=\{ab, ab',a'b, a'b', a,a',b,b'\}$. It is important to note that when considering the table $T$, the set $\mathcal{T}$ is precisely the set  of all tuples that can be built up from constants in the active domains of attributes in $U$. Hereafter, we shall refer to $\mathcal T$ as the {\em universe of discourse of the $m\_Chase$}. To summarize, the $Chase$ algorithm operates over the set $T$ of all tuples in the database, whereas the $m\_Chase$ algorithm operates over the set $\mathcal T$. Moreover, in the $m\_Chase$ algorithm the notion of `conflicting tuple' remains the same as for the {\em Chase} algorithm, but now this notion concerns all tuples of $\mathcal T$, not just those of $T$. More precisely, it is shown in \cite{LauS25} that the tuples of $\mathcal{T}$ can be characterized in two orthogonal ways: a tuple of $\mathcal{T}$ can be {\em either true or false}, and it can be {\em either conflicting or non-conflicting}. This characterization can be intuitively described as follows:
\begin{itemize}
    \item If {\em Chase} algorithm terminates successfully and $Chase(T)$ denotes the output table, then a tuple $t$ of $\mathcal T$ is true if it appears in $Chase(T)$ and false otherwise (i.e., $t$ is false if it appears in $\mathcal T \setminus Chase(T)$). In this case no tuple in $\mathcal{T}$ is conflicting, entailing that all tuples in $\mathcal{T}$ are non-conflicting.
    \item However, if the {\em Chase} algorithm fails then we don't know which tuples are true and which are non-conflicting. The $m\_Chase$ algorithm remedies this deficiency by modifying the {\em Chase} algorithm as follows: instead of stopping the application of functional dependencies on table $T$ when a conflict is encountered, the application continues (and the true tuples are stored) until no more tuples are found. In doing so, all true tuples and all conflicting tuples are computed - and therefore each tuple of $\mathcal{T}$ can be characterized as true/false and as conflicting/non-conflicting. 
\end{itemize}
It follows from the above definition of conflicting tuple that if $t$ is conflicting then every true super-tuple of $t$ is also conflicting. Therefore the conflicting tuples can be retrieved as true super-tuples of true tuples of the form $xa$ over $XA$ such that: $(a)$ $X \to A$ is a dependency in $FD$ and $(b)$ $a$ and $a'$ are in $adom(A)$ such that $a \ne a'$ and $xa'$ is true (here $adom(A)$ stands for `active domain' of $A$). Then assuming that all true tuples and all conflicting tuples are known, we can define a tuple $t$ of $\mathcal{T}$ to be {\em consistent} if $t$ is true and non-conflicting. Note that every sub-tuple of a true tuple is true and that every sub-tuple of a consistent tuple is consistent. Then, a set $S$ of true tuples of $\mathcal{T}$ is said to be a {\em consistent subset} of $\mathcal{T}$ if the set of all tuples inferred from $S$ using the functional dependencies contains no conflicting tuples in $\mathcal{S}$.  Let us illustrate these concepts using the following example.
\begin{example}\label{ex:repair}
 \rm{
  Consider a universe of three attributes $A$, $B$ and $C$ over which the functional dependencies $FD=\{A \to C, B \to C\}$ are assumed to hold, along with  a database consisting of three relations, namely $r_1=\{ab\}$, $r_2=\{bc\}$ and $r_3=\{ac'\}$, respectively defined over schemas $AB$, $BC$ and $AC$. Since no dependency can be enforced on $r_1$, and since $r_2$ and $r_3$ respectively satisfy $B \to C$ and $A \to C$, the database  satisfies the dependencies in $FD$.
 
 However, when `chasing' this database, its tuples  are first collected in a single table $T$ defined over the full universe $ABC$ and containing the three tuples $\{ab, bc, ac'\}$, in which missing values occur (for example, $ab$ is not defined over attribute $C$). At this stage, all tuples in $T$ are true (since they are  in the database) and they are non conflicting (since the dependencies in $FD$ are satisfied). However, the application of these functional dependencies on $T$ allows to infer the tuples $abc$ and $abc'$, which are conflicting tuples of $\mathcal{T}$ inferred from $T$. In fact it can be seen that $(a)$ the true tuples of $\mathcal{T}$ are $abc$, $abc'$ and all their sub-tuples, implying that any other tuple of $\mathcal{T}$ is false in $\mathcal{T}$, and $(b)$ the conflicting tuples in $\mathcal{T}$ are $abc$, $abc'$, $ac$, $ac'$, $bc$ and $bc'$, implying that any other tuple in $\mathcal{T}$ is non conflicting in $\mathcal{T}$. In this example, the consistent tuples of $\mathcal{T}$ (i.e., the tuples that are true and non conflicting in $\mathcal{T}$) are $ab$, $a$, $b$, $c$ and $c'$.
 
 \smallskip
In this context, for $R=\{abc,c'\}$, we denote by  $\mathcal{R}$ the set of all tuples that can be built up using the constants occurring in $R$, namely $a$, $b$, $c$ and $c'$. Thus, we have $\mathcal{R}=\mathcal{T}$ and it can be seen that applying the $Chase$ algorithm to the tuples in $R$ does not change $R$, and does not fail. Therefore, no tuples of $\mathcal{R}$ are conflicting, and since $R$ is a set of tuples true in $\mathcal{R}$, $R$ is said to be a {\em consistent subset of $\mathcal{T}$}. We notice that  although $abc$ is conflicting in $\mathcal{T}$, $abc$ is {\em not} conflicting in $\mathcal{R}$ because $ac'$, $bc'$ and $abc'$ are not true in $\mathcal{R}$.
}
\hfill$\Box$
\end{example}
 Based on the concepts introduced so far, a {\em repair} of $T$ in \cite{LauS25} is defined to be a maximal and consistent subset of $\mathcal{T}$ containing a maximal and consistent set of tuples which are consistent in $\mathcal{T}$. In our previous example, the subset $R=\{abc,c'\}$ of $\mathcal{T}$ is a repair of $T$ because $(a)$ as we have just seen, $R$ is a consistent subset of $\mathcal{T}$, $(b)$ $R$ is maximal because adding to $R$ a tuple true of $\mathcal{T}$ either does not bring any new true tuple in $\mathcal{R}$ (e.g., adding the tuple $ac$) or generates a conflicting tuple in $\mathcal{R}$ (e.g., adding the tuple $ac'$), and $(c)$ all consistent tuples of $\mathcal{T}$ are true in $\mathcal{R}$. Note that similar arguments show that the set $S=\{bc,ac'\}$ is a maximal and consistent subset of $\mathcal{T}$, however, $S$ is not a repair of $T$, because $ab$ is a consistent tuple of $\mathcal{T}$ which is not true in $\mathcal{S}$.
 
 By the way, as we shall see in Section~\ref{subsec:def-rep}, our definition of repair is more restrictive than the usual definition \cite{ArenasBC99,Wijsen19} in which a repair is defined to be a maximal and consistent subset of tuples true in $\mathcal{T}$. In the context of the previous example, since the set $S=\{bc,ac'\}$ is a maximal and consistent subset of tuples true in $\mathcal{T}$, $S$ is seen as a repair of $T$ following  \cite{ArenasBC99,Wijsen19}, whereas $S$ is not a repair of $T$ following our approach, because $ab$ is consistent in $\mathcal{T}$ but not true in $\mathcal{S}$.
\subsection{The m-Chase Algorithm}
Clearly to apply the $m\_Chase$-based approach described above, one has to answer the following questions:
\begin{itemize}
\item 
Does the $m\_Chase$ algorithm terminate?
\item 
Is the result independent of the order in which the functional dependencies are applied?
\item 
Does the result contain all true tuples and all conflicting tuples that the dependencies can derive? In other words: which underlying semantics ensure that $m\_Chase$ algorithm is correct?
\end{itemize}
All these questions find positive answers in \cite{LauS25}, based on the set theoretic semantics introduced in  \cite{CKS86,Spyratos87}, under the assumption that the set $FD$ is {\em normalized}. Following \cite{LauS25}, given a set of attributes $U$ and a set of functional dependencies $FD$ over $U$, if $FD^+$ denotes the closure of $FD$ under the Armstrong's axioms \cite{Armstrong}, then $FD$ is said to be {\em normalized} if it contains all dependencies in $FD^+$  such that:

\begin{description}
    \item 
    [{\sf FD1}:] every dependency in $FD$ is of the form $X \to A$ where $A$ is an attribute in $U$ not in $X$
    \item 
    [{\sf FD2}:] for every $X \to A$ in $FD$, there is no $Y \subset X$ such that $Y \to A$ is implied by $FD$ (i.e., such
that $Y \to A$ is in $FD^+$)
\end{description}
As shown in \cite{LauS25}, every set $FD$ of functional dependencies can be put in an equivalent normalized form. Moreover, a set $FD$ of functional dependencies is said to be {\em cyclic} if there exist $X \to A$ and $Y \to B$ in $FD$ such that $A$ is in $Y$ and $B$ in $X$. It is shown in \cite{LauS25} that cyclic sets of functional dependencies raise important difficulties when it comes to computing consistent answers.    
It is easy to see that the sets $FD$ considered in Example~\ref{ex:intro} and in Example~\ref{ex:repair} are both normalized and acyclic.

\smallskip
In this section, we recall  from \cite{LauS25} the basic formalism on which the algorithm $m\_Chase$ relies, namely that of {\em multi-valued tuple}. A {\em multi-valued tuple}, or {\em m-tuple}, extends the notion of tuple in the sense that an m-tuple associates every attribute $A$ with a possibly empty {\em subset} of the active domain of $A$ as opposed to a {\em single} value from the active domain.
\begin{definition}\label{def:m-tuple}
A {\em multi-valued tuple} $\sigma$ over universe $U$, or m-tuple, is a function from $U$ to the cross product ${\Large{\sf X}}_{A \in U}\mathcal{P}(adom(A))$, where $\mathcal{P}(adom(A))$ is the power set of $adom(A)$. The set of all attributes $A$ such that $\sigma(A) \ne \emptyset$, is called the {\em schema  of $\sigma$}, denoted by $sch(\sigma)$. 
Given $\sigma$ and a subset  $X$  of $sch(\sigma)$, the restriction of $\sigma$ to $X$, denoted $\sigma(X)$, is the m-tuple defined by $(\sigma(X))(A)=\sigma(A)$ for every $A$ in $X$ and $(\sigma(X))(A)=\emptyset$ for any $A$ not in $X$.

Given an m-tuple $\sigma$, the set $\tuples(\sigma)$ denotes the set of all tuples $t$ such that $sch(t)=sch(\sigma)$ and for every $A$ in $sch(t)$, $t.A$ belongs to $\sigma(A)$.\hfill$\Box$
\end{definition}
Given an m-tuple $\sigma$ and an attribute $A$ in $sch(\sigma)$, the set $\sigma(A)$ is denoted by the concatenation of its elements between parentheses, and $\sigma$ is denoted by the concatenation of all $\sigma(A)$ such that $\sigma(A) \ne \emptyset$. Moreover,  $\sigma \sqsubseteq \sigma'$ denotes the `component-wise inclusion' of $\sigma$ in $\sigma'$, that is $\sigma \sqsubseteq \sigma'$ holds if for every $A \in sch(\sigma)$, $\sigma(A) \subseteq \sigma'(A)$.  Considering that a tuple $t$ can be seen as an m-tuple $\widetilde{t}$ whose components are either empty or singletons (i.e., $t.A=a$ if and only if $\widetilde{t}(A)=(a)$), we consider that $\sqsubseteq$ may  be applied indifferently to tuples and m-tuples.

Similarly, the set theoretic union operation is extended to m-tuples in a `component-wise' manner, that is: the union of m-tuples $\sigma$ and $\sigma'$ is the m-tuple denoted by $\sigma \sqcup \sigma'$ and for every $A$ in $U$, $(\sigma \sqcup \sigma')(A)= \sigma(A) \cup \sigma'(A)$. 

\smallskip
For example, if $A$, $B$ and $C$ are three attributes, $\sigma = (a a')(b)$ and $\sigma' = (a)$ are two m-tuples such that $sch(\sigma)=AB$ and $sch(\sigma')=A$. In this case, we have $\sigma' \sqsubseteq \sigma$ because $\sigma'(A) \subseteq \sigma(A)$ and $\sigma'(B) =\emptyset$ whereas $\sigma(B)=(b)$, and so, it holds that $\sigma \sqcup \sigma' = \sigma$. Moreover, if $\sigma''=(a)(b b')(c)$, then $sch(\sigma'')=ABC$ and we have $\sigma \sqcup \sigma''= (a a')(b b')(c)$.

\smallskip
We call {\em m-table} over $U$ any finite set of m-tuples over $U$. For all $\sigma$ and $\sigma'$ in an m-table $\Sigma$, and for every functional dependency $X \to A$ over $U$ such that $XA \subseteq sch(\sigma)$ and $XA \subseteq sch(\sigma')$, the following rule called {\sf m\_Chase rule} generalizes the chase rule.

\smallskip\noindent
\begin{tabular}{ll}
$\bullet$ {\sf m\_Chase rule:}
&
{\tt Let $\sigma_1=\sigma \sqcup \sigma'(A)$ and $\sigma'_1=\sigma' \sqcup \sigma(A)$}
\\
&{\tt Case of $\sigma_1 \sqsubseteq \sigma'_1$: replace $\sigma$  with $\sigma'_1$, and remove $\sigma_1$}
\\
&{\tt Case of $\sigma'_1 \sqsubseteq \sigma_1$: replace $\sigma'$  with $\sigma_1$, and remove $\sigma'_1$}
\\
&{\tt Otherwise: replace $\sigma$ and $\sigma'$  with $\sigma_1$ and $\sigma'_1$, respectively.}
\end{tabular}

\smallskip\noindent
As shown in Algorithm~\ref{algo:chase}, our approach consists in applying the above {\sf m\_Chase rule} whenever $\tuples(\sigma(X)) \cap \tuples(\sigma'(X)) \ne \emptyset$ until no further transformation is possible. According to this algorithm, at each step, if $\sigma$ and $\sigma'$ are two m-tuples in the current state of $\Sigma$ whose schemas contain $XA$ where $X \to A$ is a functional dependency in $FD$, if their $X$-value share at least one tuple then their $A$-values are set to the same value $\sigma(A) \cup \sigma'(A)$ (see the first and last line of the above  {\sf m\_Chase rule}), and in case these changes imply that one m-tuple subsumes the other one, the redundancy is removed (as stated in the two intermediate lines of the above {\sf m\_Chase rule}). Similarly to the $Chase$ algorithm, this process is applied in Algorithm~\ref{algo:chase}, until no further transformation of $\Sigma$ is possible. The output of Algorithm~\ref{algo:chase} is then an m-table denoted by $m\_Chase(T)$.

\smallskip
To illustrate this step in the context of Example~\ref{ex:repair}, consider first $\sigma=(a)(b)$ and $\sigma'=(b)(c)$ with the dependency $B \to C$. In this case $\sigma_1=(a)(b)(c)$ and $\sigma'_1=\sigma'$. As $\sigma'_1 \sqsubseteq \sigma_1$, only  $\sigma_1$ is kept for the next computation step. Thus, for $\sigma=(a)(b)(c)$ and $\sigma'=(a)(c')$ with the dependency $A \to C$, we have $\sigma_1=(a)(b)(cc')$ and $\sigma'_1=(b)(cc')$. In this case again, we have $\sigma'_1 \sqsubseteq \sigma_1$, and so, $\sigma'_1$ is discarded, and we are left with the only m-tuple $(a)(b)(cc')$. Since no further application of the  {\sf m\_Chase rule} is possible, we obtain that $m\_Chase(T)=\{(a)(b)(cc')\}$.

To see a situation where the two m-tuples $\sigma_1$ and $\sigma'_1$ are incomparable, consider $T = \{abc, ab'c'\}$ and $FD=\{A \to C\}$. Here, for $\sigma = (a)(b)(c)$ and $\sigma' = (a)(b')(c')$, we have $\sigma_1=  (a)(b)(cc')$ and $\sigma'_1=  (a)(b')(cc')$. Since these two m-tuples are not comparable with respect to $\sqsubseteq$, the two of them are kept and we obtain $m\_Chase(T)=\{(a)(b)(cc'),(a)(b')(cc')\}$.

\smallskip
It has been shown in \cite{LauS25} that Algorithm~\ref{algo:chase} always terminates and that the partition semantics of tuples in $\mathcal{T}$ (as introduced in \cite{Spyratos87} and extended in \cite{JIIS,LauS25}), can be defined based on $m\_Chase(T)$ according to the following proposition. To state these basic results, we use the following notation given a table $T$: the sets of all true tuples in $\mathcal{T}$, all conflicting tuples in $\mathcal{T}$, all consistent tuples in $\mathcal{T}$ are respectively denoted by $\true(\mathcal{T})$, $\inc(\mathcal{T})$ and $\cons(\mathcal{T})$.

\algsetup{indent=1em}
\begin{algorithm}[t]
\caption{The m-Chase Algorithm \label{algo:chase}}
\begin{algorithmic}[1]
{\footnotesize
\REQUIRE

A table $T$ over $U$ and a normalized set $FD$ of functional dependencies over $U$.

\ENSURE An m-table denoted by $m\_Chase(T)$.

\STATE{$\Sigma := \{\sigma_t~|~t \in T\}$ \COMMENT{$\sigma_t$ is the m-tuple such that $\sigma_t(A)=\{t.A\}$ for $A \in sch(t)$}}
\STATE{$change := true$}
\WHILE{$change = true$}\label{line:main-loop-chase}
    \STATE{$change := false$}
    \FORALL{$\sigma$ and $\sigma'$ in $\Sigma$}
        \FORALL{$X \to A$ in $FD$ such that $XA \subseteq sch(\sigma)$ and $XA \subseteq sch(\sigma)$}
               	 \IF{$\tuples(\sigma(X)) \cap \tuples(\sigma'(X)) \ne \emptyset$}
                        \STATE{apply the {\sf m\_Chase rule} to $\sigma$ and $\sigma'$}
                        \STATE{$change := true$}
               	\ENDIF
        \ENDFOR
     \ENDFOR
\ENDWHILE
\STATE{$m\_Chase(T):= \Sigma$}
\RETURN{$m\_Chase(T)$}
}
\end{algorithmic}
\end{algorithm}

\begin{proposition}\label{prop:truth-value}
Let $T$ be a table over universe $U$ with $FD$ as set of functional dependencies. The following holds:
\begin{itemize}
\item 
A tuple $t$ in $\mathcal{T}$ is in $\true(\mathcal{T})$ if and only if there exists $\sigma$ in $m\_Chase(T)$ such that:\\
$-$ $sch(t) \subseteq sch(\sigma)$ (i.e.,  $\sigma$ has nonempty components over attributes in $sch(t)$),\\
$-$ $t \sqsubseteq \sigma$.
\item 
A tuple $t$ in $\mathcal{T}$ is in $\inc(\mathcal{T})$ if and only if there exists $\sigma$ in $m\_Chase(T)$ such that:\\
$-$ $sch(t) \subseteq sch(\sigma)$ and $t \sqsubseteq \sigma$,\\
$-$ there exists $X \to A$ in $FD$ such that $XA \subseteq sch(t)$ and $|\tuples(\sigma(A)|>1$.
\item 
A tuple $t$ in $\mathcal{T}$ is in $\cons(\mathcal{T})$ if and only if there exists $\sigma$ in $m\_Chase(T)$ such that:\\
$-$ $sch(t) \subseteq sch(\sigma)$ and $t \sqsubseteq \sigma$,\\
$-$ for every $X \to A$ in $FD$ such that $XA \subseteq sch(t)$, $|\tuples(\sigma(A))|=1$.
\item 
For every $\sigma$ in $m\_Chase(T)$ and every $S \subseteq sch(\sigma)$, either $\tuples(\sigma(S))\subseteq \cons(\mathcal{T})$ or  $\tuples(\sigma(S))\subseteq \inc(\mathcal{T})$.\hfill$\Box$
\end{itemize}
\end{proposition}
We notice that when $\inc({T}) =\emptyset$, i.e., when no tuples in $\mathcal{T}$ are conflicting, then $T$ satisfies all functional dependencies in $FD$, which is denoted by $T \models FD$. In this case, as shown by Proposition~\ref{prop:truth-value}, all components of all m-tuples in $m\_Chase(T)$ contain one value, and so, $m\_Chase(T)$ can be seen as a table, which is in fact the table $Chase(T)$.

\smallskip
Regarding the time complexity of Algorithm~\ref{algo:chase}, it has been shown in \cite{LauS25} that the computation of $m\_Chase(T)$ is in $\mathcal{O}(|m\_Chase(T)|^3\cdot \delta^2)$, where $\delta$ is the maximal cardinality of the components of m-tuples in $m\_Chase(T)$, which is precisely the maximum number of $A$-values associated with $X$-values when $X \to A$ is a functional dependency in $FD$. As Algorithm~\ref{algo:chase} shows that $|m\_Chase(T)| \leq |T|$, we state that the computation of $m\_Chase(T)$ is in $\mathcal{O}(|T|^3\cdot \delta^2)$, i.e., {\em polynomial in the size of $T$}.
We notice that in this complexity result, the size of the universe $U$ is a constant, contrary to the size of m-tuples. In fact the size of every m-tuple in $m\_Chase(T)$ is bounded by  $\delta\cdot |U|$. As will be seen shortly, these bounds can be refined when considering star-schemas.

\smallskip
On the other hand, based on the fact that if a tuple $t$ is conflicting then all its true super-tuples are conflicting as well, the set $\inc(\mathcal{T})$ can be characterized by means of its minimal tuples with respect to $\sqsubseteq$. More precisely, denoting this set by $\inc_{\min}(\mathcal{T})$, we have $\inc(\mathcal{T})=\{t \in \true(\mathcal{T})~|~(\exists q \in \inc_{\min}(\mathcal{T}))(q \sqsubseteq t)\}$.

Thus, using Proposition~\ref{prop:truth-value}, the set  $\inc_{\min}(\mathcal{T})$ is characterized as follows: $t$ is in $\inc_{\min}(\mathcal{T})$ if and only if one of the following two statements holds:
\begin{itemize}
\item there exist $\sigma$ in $m\_Chase(T)$, $i_0$ in $\{1, \ldots ,n\}$ and $A$ in $sch^*(D_{i_0}) \cap sch(\sigma) \cap sch(t)$ such that $t=k_{i_0}a$ is in $\tuples(\sigma(K_{i_0}A))$ and  $|\tuples(\sigma(A))|>1$,
\item 
there exists $\sigma$ in $m\_Chase(T)$ such that $\mathbb{K} \subseteq sch(\sigma) \cap sch(t)$, there exists $M_i$ in $\mathbb{M} \cap sch(\sigma)\cap sch(t)$ such that $t=km_i$ is in $\tuples(\sigma(\mathbb{K}M_i))$ and  $|\tuples(\sigma(M_i))|>1$.
\end{itemize}
By complementation with respect to $\true(\mathcal{T})$, a tuple $t$ is in $\cons(\mathcal{T})$ if and only if it has no sub-tuple satisfying one of the above statements.

\smallskip
To illustrate Algorithm~\ref{algo:chase} and Proposition~\ref{prop:truth-value}, consider again the context of Example~\ref{ex:repair} where $U=\{A,B,C\}$, $FD= \{A \to C, B \to C\}$ and $T =\{ab, bc, ac'\}$. As seen earlier, running Algorithm~\ref{algo:chase} yields the following steps:

\smallskip
$-$ The algorithm starts with the m-table $\Sigma=\{(a)(b), (b)(c), (a)(c')\}$.

$-$ Applying $B \to C$ to the first two m-tuples, we obtain $\Sigma=\{(a)(b)(c), (a)(c')\}$.

$-$ Applying now $A \to C$ to these two m-tuples, we obtain $\Sigma=\{(a)(b)(cc')\}$.

\smallskip\noindent
Since no new m-tuple can be generated from $\Sigma$, $m\_Chase(T)=\{(a)(b)(cc')\}$ is returned by Algorithm~\ref{algo:chase}, and so, by Proposition~\ref{prop:truth-value}, it follows that

\begin{itemize}
    \item 
    $\true(\mathcal{T})$ is the set of all sub-tuples of tuples in $\tuples((a)(b)(c,c'))$, that is $\true(\mathcal{T})=\{abc, abc', ab,$ $ ac, ac', bc, bc',a, b, c,c'\}$. In other words, there are no false tuples in this example.
    \item 
    $\inc_{min}(\mathcal{T})=\{ac,ac',bc,bc'\}$ and so, we have $\inc(\mathcal{T})=\{abc, abc', ac, ac', bc, bc'\}$.
    \item 
    $\cons(\mathcal{T})= \true(\mathcal{T}) \setminus \inc(\mathcal{T})$, that is $\cons(\mathcal{T})= \{ab, a,  b, c, c'\}$.
\end{itemize}
In the following section, we first recall the notion of star schema and then we show that the results from \cite{LauS25} that have just been recalled apply in this context as well.
\section{Star Schemas}\label{sec:star schema}
\subsection{The Context of our Approach}
We first recall from the literature \cite{inmon96,Ullman} that a star schema, as considered in our approach, consists of the following tables and constraints:
\begin{itemize}
\item $n$ dimension tables $D_1, \ldots , D_n$. For $i=1, \ldots , n$, $D_i$ is defined over attributes $K_i$, $A_i^1, \ldots ,$ $A_i^{d_i}$. For $i=1, \ldots ,n$, the schema of $D_i$ is denoted by $sch(D_i)$, and the set $sch(D_i) \setminus \{K_i\}$ is denoted by $sch^*(D_i)$.
\item 
a fact table $F$ defined over $K_1, \ldots , K_n, M_1, \ldots , M_p$. The attributes $M_1, \ldots , M_p$ are called measures, and we denote by $\mathbb{M}$ the set of all measures, that is $\mathbb{M}=\{M_1, \ldots , M_p\}$.
\item 
$FD= \bigcup_{i=1}^{i=n}\{K_i \to A_i^j~|~j=1, \ldots ,d_i\} \cup \{K_1 \ldots  K_n \to M_k~|~k=1, \ldots ,p\}$.
\\
In other words, for $i=1, \ldots ,n$, $K_i$ is the key of $D_i$ and $K_1\ldots K_n$ is the key of $F$. We denote by $\mathbb{K}$ the set of all dimension keys that is $\mathbb{K} = \{K_1, \ldots , K_n\}$.
\end{itemize}
It is easy to see that if $FD$ is defined as above, then for every non trivial functional dependency $X \to A$ in $FD^+$, it holds that $X \cap \mathbb{K} \ne \emptyset$ and $A \not\in \mathbb{K}$.  More precisely, given an attribute $A$, only one of the following three cases is possible: $(i)$ $A$ is in $\mathbb{K}$, or $(ii)$ $A$ is in $sch^*(D_i)$ for some $i=1 ,\ldots ,n$, or $(iii)$ $A$ is in $\mathbb{M}$. For each of these cases, we have the following: $(i)$ if $A$ is in $\mathbb{K}$ then $A$ must occur in $X$, in which case $X \to A$ is trivial (because $FD$ contains no dependency whose right hand-side is in $\mathbb{K}$); $(ii)$ if $A$ is in $sch^*(D_i)$ then $X$ must contain $K_i$; $(iii)$ if $A$ is in $\mathbb{M}$ then $X$ must contain $\mathbb{K}$. Thus, for every non trivial functional dependency $X \to A$ in $FD^+$, there exists $X_0 \to A$ in $FD$ such that $X_0 \subseteq X$. 

Since the left hand-sides of the dependencies in $FD$ can not be reduced further, this means that $FD$ is normalized. On the other hand as the left hand-sides of functional dependencies in $FD$ are attributes in $\mathbb{K}$, not occurring in the right hand-sides of these dependencies, $FD$ is acyclic. As a consequence, all results in \cite{LauS25} apply in the context of star schemas. 

\smallskip
In what follows, we call {\em data warehouse} a database whose schema is a star schema. Moreover, we use the terms `data warehouse' and `table' instead of `multi-relation database' and `relation', to better fit the usual terminology when dealing with data warehouses. In our approach, it is thus possible to deal with data warehouses in which some of the tables $D_i$ or $F$ have missing values for some of their attributes. However, in order to consider cases that make sense in practice, we restrict missing values in the warehouse tables as follows: 
\begin{enumerate}
\item 
For every $i=1,\ldots, n$, every $t$ in $D_i$ is defined over the key attribute $K_i$ and over at least one non-key attribute in $sch^*(D_i)$. We consider that storing  a key value with no associated non-key value makes no sense.
\item 
For every $t$ in $F$, $t$ is defined over $\mathbb{K}$ and over at least one measure attribute in $\mathbb{M}$. We consider that storing a fact in which some key values are missing makes no sense, since we consider that measure values depend on {\em all} dimensions. Similarly, considering a fact with no associated measure value makes no sense.
\end{enumerate}
We notice that the above restrictions follow common sense considerations (storing values without corresponding key values or storing key values alone with no other attribute values make no sense). However, we also notice that relaxing these restrictions can be supported by our approach, at the cost of technical complications.

\begin{figure}[t]
\begin{center}
{\footnotesize
\begin{tabular}{c|ccc}
$D_1$&$K_1$&$A_1^1$&$A_1^2$\\
\hline
&$k_1$&$a_1$&$a_2$\\
&$k_1$&$a'_1$&$a_2$\\
&$k'_1$&$a_1$&\\
&$k''_1$&$a'_1$&$a'_2$
\end{tabular}
\qquad\qquad
\begin{tabular}{c|ccc}
$D_2$&$K_2$&$A_2^1$&$A_2^2$\\
\hline
&$k_2$&$b_1$&$b_2$\\
&$k'_2$&$b_1$&\\
\end{tabular}
\qquad\qquad
\begin{tabular}{c|ccc}
$F$&$K_1$&$K_2$&$M_1$\\
\hline
&$k_1$&$k_2$&$m_1$\\
&$k_1$&$k'_2$&$m'_1$\\
&$k_1$&$k'_2$&$m''_1$\\
&$k'_1$&$k''_2$&$m_1$
\end{tabular}
}
\end{center}
\caption{The tables of the data warehouse in our running example\label{fig:tables}}
\end{figure}

\begin{example}\label{ex:star}
{\rm
We illustrate the above concepts using a toy example that will serve as a running example in the remainder of this paper. We consider two dimensions $D_1$ and $D_2$ such that $sch(D_1)=K_1A_1^1A_1^2$ and $sch(D_2)=K_2A_2^1A_2^2$. Moreover the fact table $F$ is such that $sch(F)=K_1K_2M_1$ meaning that we consider one measure attribute $M_1$. As specified above, we have $FD=\{K_1 \to A_1^1, K_1 \to A_1^2, K_2 \to A_2^1, K_2 \to A_2^2, K_1K_2 \to M_1\}$. The content of the tables $D_1$, $D_2$ and $F$ is shown in Figure~\ref{fig:tables}. 

We observe that these tables are indeed those of a star schema and that they comply with the two restrictions above regarding missing values. Moreover, it should be emphasized that $D_1$ and $F$ do {\em not} satisfy $FD$. Indeed, the first two tuples in $D_1$ violate $K_1 \to A_1^1$ and the second and third tuples in $F$ violate $K_1K_2 \to M_1$. On the other hand $D_2$ satisfies its two associated functional dependencies $K_2 \to A_2^1$ and $K_2 \to A_2^2$.

We also stress that the above two restrictions on missing values are satisfied and that key value $k''_1$ occurs in $D_1$ but not in $F$, whereas the key value $k''_2$ over $K_2$ occurs in $F$ but not in $D_2$. These two cases respectively illustrate that key values in a dimension table may not occur in the fact table and that the foreign key constraint between a dimension table and the fact table may not be satisfied (contrary to what is generally assumed in the literature).

\begin{figure}[t]
{\footnotesize
\begin{center}
\qquad \qquad\qquad 
\begin{tabular}{c|ccccccc}
$T$&\,$K_1$\,&\,$K_2$\,&~~~$A_1^1$\quad&\quad$A_1^2$\quad&~~\quad$A_2^1$\quad&\quad$A_2^2$\quad&~$M_1$~\quad\\
\hline
&$k_1$&&~~$a_1$&\quad$a_2$&&&\\
&$k_1$&&~~$a'_1$&\quad$a_2$&&&\\
&$k'_1$&&~~$a_1$&&&&\\
&$k''_1$&&~~$a'_1$&\quad$a'_2$&&&\\
&&$k_2$&&&\quad$b_1$&~$b_2$&\\
&&$k'_2$&&&\quad$b_1$&&\\
&$k_1$&$k_2$&&&&&$m_1$\\
&$k_1$&$k'_2$&&&&&$m'_1$\\
&$k_1$&$k'_2$&&&&&$m''_1$\\
&$k'_1$&$k''_2$&&&&&$m_1$
\end{tabular}
\\~\\~\\~\\
\quad~
\begin{tabular}{c|ccccccc}
$m\_Chase(T)$&$K_1$&$K_2$&$A_1^1$&$A_1^2$&$A_2^1$&$A_2^2$&$M_1$\\
\hline
&$(k_1)$&$(k_2)$&$(a_1\, a'_1)$&$(a_2)$&$(b_1)$&$(b_2)$&$(m_1)$\\
&$(k_1)$&$(k'_2)$&$(a_1\, a'_1)$&$(a_2)$&$(b_1)$&&$(m'_1 \, m''_1)$\\
&$(k'_1)$&$(k''_2)$&$(a_1)$&&&&$(m_1)$\\
&$(k''_1)$&&$(a'_1)$&$(a'_2)$&&&
\end{tabular}
\end{center}
}
\caption{The star-table $T$ and the m-table $m\_Chase(T)$ of Example~\ref{ex:star}
\label{fig:m-chase}}
\end{figure}

In this context, the associated table $T$ over $U$ and its corresponding m-table $m\_Chase(T)$ are shown in Figure~\ref{fig:m-chase}.
Applying  Proposition~\ref{prop:truth-value}, the sets $\true(\mathcal{T})$, $\inc(\mathcal{T})$ and $\cons(\mathcal{T})$ are as follows:
\begin{itemize}
\item $\true(\mathcal{T})$ is the set of all sub-tuples of the tuples in $\tuples(\sigma)$ for every $\sigma$ in $m\_Chase(T)$. Thus, $\true(\mathcal{T})$ is the set of all sub-tuples of:\\
$-$ $k_1k_2a_1a_2b_1b_2m_1$, $k_1k_2a'_1a_2b_1b_2m_1$, \\
$-$ $k_1k'_2a_1a_2b_1m'_1$, $k_1k'_2a'_1a_2b_1m'_1$, $k_1k'_2a_1a_2b_1m''_1$, $k_1k'_2a'_1a_2b_1m''_1$,\\
$-$ $k'_1k''_2a_1m_1$,\\
$-$ $k''_1a'_1a'_2$.
\item $\inc(\mathcal{T})$ is the set of all true super-tuples of tuples in $\inc_{\min}(\mathcal{T})= \{k_1a_1$, $k_1a'_1$, $k_1k'_2m'_1$,  $k_1k'_2m''_1\}$. 
The maximal tuples in $\inc(\mathcal{T})$ are:\\
$-$ $k_1k_2a_1a_2b_1b_2m_1$, $k_1k_2a'_1a_2b_1b_2m_1$,
\\
$-$ $k_1k'_2a_1a_2b_1m'_1$, $k_1k'_2a'_1a_2b_1m'_1$, $k_1k'_2a_1a_2b_1m''_1$, $k_1k'_2a'_1a_2b_1m''_1$.
\item $\cons(\mathcal{T})= \true(\mathcal{T}) \setminus \inc(\mathcal{T})$. Thus $\cons(\mathcal{T})$ is the set of all sub-tuples of:\\
$-$ $k_1k_2a_2b_1b_2m_1$, $k_2a_1 a_2b_1b_2m_1$, $k_2a'_1 a_2b_1b_2m_1$,\\
$-$ $k_1k'_2a_2b_1$, $k'_2a_1 a_2b_1m'_1$, $k'_2a'_1 a_2b_1m'_1$, $k'_2a_1 a_2b_1m''_1$, $k'_2a'_1 a_2b_1m''_1$, $k_1 a_2b_1m'_1$, $k_1 a_2b_1m''_1$,\\
$-$ $k'_1k''_2a_1m_1$,\\
$-$ $k''_1a'_1a'_2$.\hfill$\Box$
\end{itemize}
}
\end{example}
\subsection{$m\_Chase$ Computation in a Star Schema}
In this section, we show that when considering a star schema, the time complexity of computing $m\_Chase(T)$ for a star-table $T$ can be drastically lowered, as compared with the general case shown in \cite{LauS25} and recalled above. To do so, we first state important properties regarding m-tuples of $m\_Chase(T)$ where $T$ is the table collecting all tuples in the data warehouse. In the remainder of this paper, we refer to such table as a {\em star-table}. The following proposition states important properties of the m-tuples in $m\_Chase(T)$.
\begin{proposition}\label{prop:star-m-chase}
Let $T$ be a star-table over universe $U$. The following hold:
\begin{enumerate}
\item 
For every $\sigma$ in $m\_Chase(T)$ and every $i=1, \ldots ,n$, if $K_i \in sch(\sigma)$ then $|\tuples(\sigma(K_i))|=1$. Consequently,  if $\mathbb{K} \subseteq sch(\sigma)$ then $|\tuples(\sigma(\mathbb{K}))|=1$.
\item 
For every tuple $k$ over $\mathbb{K}$ in $\mathcal{T}$, there exists at most one $\sigma$ in $m\_Chase(T)$ such that $\mathbb{K} \subseteq sch(\sigma)$ and $\tuples(\sigma(\mathbb{K})) = (k)$.
\item
Moreover, $m\_Chase(T)$ contains the following two kinds of m-tuples:
\begin{enumerate}
\item $\sigma$ for which there exists $i_0 \in \{1, \ldots ,n\}$ such that:
\begin{itemize}
\item $sch(\sigma) \subseteq sch(D_{i_0})$, $\tuples(\sigma(K_{i_0}))= (k_{i_0})$ and for every $t \in F$, $t.K_{i_0} \ne k_{i_0}$,
\item for every $A \in sch^*(D_{i_0})$, $\sigma(A) = \{a~|~(\exists q \in D_{i_0})(q.K_{i_0} = k_{i_0} \wedge q.A = a)\}$.
\end{itemize}
\item $\sigma$ such that $\mathbb{K} \subseteq sch(\sigma)$ and $\tuples(\sigma(\mathbb{K}))=(k)$, and
\begin{itemize}
\item for every $M_j \in \mathbb{M}$, $\sigma(M_j)=\{m_j~|~(\exists t \in F)(t.\mathbb{K}=\sigma(\mathbb{K}) \wedge t.M_j = m_j)\}$,
\item for every $i =1, \ldots , n$, for every $A \in sch^*(D_i)$,\\
\rightline{$\sigma(A) = \{a~|~(\exists t \in D_i)(t.K_i = k.K_i \wedge t.A = a)\}$.\qquad\qquad\qquad\quad}
\end{itemize}
\end{enumerate}
\end{enumerate}
\end{proposition}
{\sc Proof.} See Appendix~\ref{append:prop-star-m-chase}.\hfill$\Box$ 

\medskip\noindent
We point out that item 3 in Proposition~\ref{prop:star-m-chase} holds thanks to the restrictions on missing values earlier stated. Moreover, an important consequence of Proposition~\ref{prop:star-m-chase} is that $m\_Chase(T)$ can be computed by an algorithm whose time complexity is significantly lower than that of Algorithm~\ref{algo:chase}. This new algorithm, displayed  in Algorithm~\ref{algo:chase-star}, is shown to be correct in the following proposition.

\algsetup{indent=1em}
\begin{algorithm}[t]
\caption{The m-Chase Algorithm for a star-table\label{algo:chase-star}}
\begin{algorithmic}[1]
{\footnotesize
\REQUIRE
A data warehouse composed of $n$ dimensional tables $D_1, \ldots , D_n$ and a fact table $F$.

\ENSURE The m-table $m\_Chase(T)$.

\FORALL{$i=1, \ldots ,n$}\label{main-dim-loop-start}
	\STATE{sort the m-tuples in $D_i$ according to their $K_i$-value}\label{line:sort-dim}
	\STATE{$D_i^* := \emptyset$}
	\FORALL{$K_i$-value $k_i$ occurring in $D_i$}\label{line:dim-loop-start}
		\STATE{$\sigma := (k_i)s_i^1\ldots s_i^{d_i}$ where  
		$s_i^j = \{a_i^j\in adom(A_i^j)~|~(\exists t_i \in D_i)(t_i.{K_i}=k_i \wedge A_i^j \in sch(t_i) \wedge t_i.A_i^j=a_i)\}$ for every $j=1, \ldots , d_i$}
		\STATE{$D_i^* := D_i^* \cup\{\sigma\}$}
	\ENDFOR\label{line:dim-loop-end}
\ENDFOR\label{main-dim-loop-end}
\STATE{sort the m-tuples in $F$ according to their $\mathbb{K}$-value}\label{line:sort-fact}
\STATE{$F^* := \emptyset$}
\FORALL{$\mathbb{K}$-value $k = k_1\ldots k_n$ occurring in $F$}\label{line:K-loop-start}
	\STATE{$\sigma := (k)\mu_1\ldots \mu_p$ where 
		$\mu_q = \{m_q \in adom(M_q)~|~(\exists t \in F)(t.{\mathbb{K}}=k \wedge M_q \in sch(t) \wedge t.M_q=m_q)\}$ for every $q=1, \ldots , p$, and $\sigma(A)=\emptyset$ for every $A \in U \setminus (\mathbb{K} \cup \mathbb{M})$}\label{line:srep1}
	\FORALL{$i=1, \ldots ,n$}
		\IF{$k_i$ occurs in $D_i^*$}\label{line:test-ki}
			\STATE{Let $\sigma_i$ be the unique m-tuple in $D_i^*$ such that $\sigma(K_i)=(k_i)$}
			\STATE{Mark $\sigma_i$ in $D_i^*$}
			\FORALL{$j=1, \ldots , d_i$}
				\STATE{$\sigma(A_i^j):= \sigma_i(A_i^j)$}
			\ENDFOR
		\ENDIF
	\ENDFOR
	\STATE{$F^* := F^* \cup\{\sigma\}$}
\ENDFOR\label{line:K-loop-end}
\FORALL{$i=1, \ldots ,n$}\label{line:second-loop-start}
	\FORALL{$\sigma_i$ in $D_i^*$}
		\IF{$\sigma_i$ is not marked}
			\STATE{$F^* := F^* \cup\{\sigma_i\}$}
		\ENDIF
	\ENDFOR
\ENDFOR\label{line:second-loop-end}
\STATE{$m\_Chase(T):= F^*$}
\RETURN{$m\_Chase(T)$}
}
\end{algorithmic}
\end{algorithm}

\begin{proposition}\label{prop:m-chase-star}
Given a data warehouse composed of $n$ dimensional tables $D_1, \ldots , D_n$ and a fact table $F$, and its associated star-table $T$, Algorithm~\ref{algo:chase-star} correctly computes $m\_Chase(T)$.

The time complexity of Algorithm~\ref{algo:chase-star} is in
\begin{center}
$\mathcal{O}\left( \Sigma_{i=1}^{i=n} (|D_i|\cdot \log(|D_i|)) + |F|\cdot (\log(|F|))  \right).$
\end{center}
\end{proposition}
{\sc Proof.} It can be seen that the loop lines~\ref{line:second-loop-start}-\ref{line:second-loop-end} in Algorithm~\ref{algo:chase-star} computes all m-tuples of kind (a) in Proposition~\ref{prop:star-m-chase}(3), whereas the loop lines~\ref{line:K-loop-start}-\ref{line:K-loop-end} in Algorithm~\ref{algo:chase-star} computes all m-tuples of kind (b). The first part of the proof is therefore complete.

Regarding the time complexity of Algorithm~\ref{algo:chase-star}, we first notice that the statements line~\ref{line:sort-dim} and \ref{line:sort-fact} are respectively in $\mathcal{O}(|D_i]\cdot \log(|D_i|))$ and $\mathcal{O}(|F]\cdot \log(|F|))$. Moreover, the loop lines~\ref{line:dim-loop-start}-\ref{line:dim-loop-end} is linear in the size of $D_i$ as the table is sorted according to its key values. Thus the loop lines~\ref{main-dim-loop-start}-\ref{main-dim-loop-end}  is in $\mathcal{O}( \Sigma_{i=1}^{i=n} (|D_i|\cdot \log(|D_i|)))$. Similarly, the loop lines~\ref{line:K-loop-start}-\ref{line:K-loop-end} is in $\mathcal{O}(|F]\cdot \log(|F|))$ and the last loop lines~\ref{line:second-loop-start}-\ref{line:second-loop-end} is in $\mathcal{O}(|D_i^*|) $. Since $\mathcal{O}(|D_i^*|) \approx \mathcal{O}(|D_i|)$, we obtain that Algorithm~\ref{algo:chase-star} is in $\mathcal{O}\left(\Sigma_{i=1}^{i=n} (|D_i|\cdot \log(|D_i|))\right) + \mathcal{O}(|F]\cdot \log(|F|)) + \mathcal{O}\left( \Sigma_{i=1}^{i=n} (|D_i|)\right)$, that is in $\mathcal{O}\left( \Sigma_{i=1}^{i=n} (|D_i|\cdot \log(|D_i|))\right) + \mathcal{O}(|F]\cdot \log(|F|))$. The proof is therefore complete.
\hfill$\Box$

\medskip\noindent
We emphasize that, denoting by $W$ the overall size of the underlying data warehouse (i.e., the sum of the sizes of all its tables), the complexity result in Proposition~\ref{prop:m-chase-star} can be written as $\mathcal{O}(W\cdot \log(W))$, which is significantly lower than the generic time complexity stated in \cite{LauS25}, which can be approximated to $\mathcal{O}(W^3\cdot \delta^2)$. Moreover, considering Algorithm~\ref{algo:chase-star} instead of Algorithm~\ref{algo:chase} to assess the size of the m-table $m\_Chase(T)$ shows that the number of m-tuples in $m\_Chase(T)$ is {\em exactly} the number of distinct $\mathbb{K}$-values present in the fact table $F$, plus for every $i=1, \ldots ,n$, the number of all $K_i$-values occurring in $D_i$ but not in $F$.  On the other hand it is easy to see that, if $T$ is a star-table, the size of any m-tuple in $m\_Chase(T)$ is bounded by $|\mathbb{K}|+ \delta\cdot |U \setminus \mathbb{K}|$, which is less than $\delta\cdot |U|$ as earlier stated.

\smallskip
We now illustrate Proposition~\ref{prop:m-chase-star} in the context of Example~\ref{ex:star}.

\begin{figure}[t]
\begin{center}
{\footnotesize
\begin{tabular}{c|ccc}
$D_1^*$&$K_1$&$A_1^1$&$A_1^2$\\
\hline
&$(k_1)$&$(a_1\, a'_1)$&$(a_2)$\\
&$(k'_1)$&$(a_1)$&\\
&$(k''_1)$&$(a'_1)$&$(a'_2)$
\end{tabular}
\qquad\qquad
\begin{tabular}{c|ccc}
$D_2^*$&$K_2$&$A_2^1$&$A_2^2$\\
\hline
&$(k_2)$&$(b_1)$&$(b_2)$\\
&$(k'_2)$&$(b_1)$&\\
\end{tabular}
\qquad\qquad
\begin{tabular}{c|ccc}
$F^*_0$&$K_1$&$K_2$&$M_1$\\
\hline
&$(k_1)$&$(k_2)$&$(m_1)$\\
&$(k_1)$&$(k'_2)$&$(m'_1\, m''_1)$\\
&$(k'_1)$&$(k''_2)$&$(m_1)$
\end{tabular}
\\~\\~\\~\\
\quad~
\begin{tabular}{c|ccccccc}
$F^*$&$K_1$&$K_2$&$A_1^1$&$A_1^2$&$A_2^1$&$A_2^2$&$M_1$\\
\hline
&$(k_1)$&$(k_2)$&$(a_1\, a'_1)$&$(a_2)$&$(b_1)$&$(b_2)$&$(m_1)$\\
&$(k_1)$&$(k'_2)$&$(a_1\, a'_1)$&$(a_2)$&$(b_1)$&&$(m'_1 \, m''_1)$\\
&$(k'_1)$&$(k''_2)$&$(a_1)$&&&&$(m_1)$\\
&$(k''_1)$&&$(a'_1)$&$(a'_2)$&&&
\end{tabular}
}
\end{center}
\caption{The m-tables  when running Algorithm~\ref{algo:chase-star} in the context of Example~\ref{ex:star}\label{fig:tables-algo-2}}
\end{figure}

\begin{example}\label{ex:star-1}
{\rm
Applying Algorithm~\ref{algo:chase-star} to the tables shown in Figure~\ref{fig:tables}, yields the following m-tables shown in Figure~\ref{fig:tables-algo-2}:
\begin{itemize}
\item $D_1^*$ and $D_2^*$, when running the loop lines~\ref{main-dim-loop-start}-\ref{main-dim-loop-end},
\item $F_0^*$ when considering the m-tuples $\sigma$ on line~\ref{line:srep1},
\item the first three rows in $F^*$ when  completing the loop lines~\ref{line:first-loop-start}-\ref{line:first-loop-end},
\item the whole m-table $F^*$, after computing the loop lines~\ref{line:second-loop-start}-\ref{line:second-loop-end}. Indeed, as $k''_1$ does not occur in $F$, the last m-tuple in $D_1^*$ is not marked because the test line~\ref{line:test-ki} fails.
\end{itemize}
It is easy to see that the obtained m-table $F^*$ is equal to $m\_Chase(T)$ as shown in Figure~\ref{fig:m-chase}. Moreover, the number of m-tuples in $m\_Chase(T)$ is $5$, corresponding respectively  to the four $\mathbb{K}$-values occurring in $F$ and to the $K_1$-value $k_1''$ occurring in $D_1$ but not in $F$.
\hfill$\Box$
}
\end{example}
\section{Repairs and their Basic Properties}\label{sec:repairs}
In this section we adapt the definition of repair given in \cite{LauS25} to the case of a star-table and then we further investigate repairs of star-tables.
\subsection{Definition of Repairs}\label{subsec:def-rep}
As explained in \cite{LauS25}, contrary to most approaches to consistent query answering in the literature, it is not appropriate to define a repair of $T$ as a {\em maximal subset  of $\true(\mathcal{T})$ satisfying $FD$}. This is so because, although it is intuitively justified to impose that all tuples of $\cons(\mathcal{T})$ be true in every repair, it is shown in \cite{LauS25} that in case of cyclic sets of functional dependencies (i.e., if there exist $X \to A$ and $Y \to B$ in $FD$ such that $A \in Y$ and $B \in X$), the set $\cons(\mathcal{T})$ may {\em not} satisfy $FD$.
To cope with this issue, given universe $U$, a set $FD$ of functional dependencies over $U$ and a table $T$ over $U$, a repair $R$ of $T$ is defined in \cite{LauS25} as a table $R$ over $U$ such that:
\begin{enumerate}
\item
$\true(\mathcal{R}) \subseteq \true(\mathcal{T})$.
\item $R \models FD$.
\item $\true(\mathcal{R}) \cap \cons(\mathcal{T})$ is a maximal consistent subset of $\cons(\mathcal{T})$.
\item For every table $R'$ satisfying 1, 2 and 3 above, and such that $\true(\mathcal{R})\subseteq \true(\mathcal{R}')$, we have $\true(\mathcal{R}) = \true(\mathcal{T})$.
\end{enumerate}
Moreover, it is also shown in \cite{LauS25} that when $FD$ is acyclic,  the set $\cons(\mathcal{T})$ satisfies $FD$. Hence, it turns out from the above definition that when $FD$ is acyclic, every tuple in $\cons(\mathcal{T})$ must be true in every repair. It is important to notice that, as shown in Example~\ref{ex:repair}, even when $FD$ is acyclic, there exist maximal and consistent subsets of $\true(\mathcal{T})$ that do {\em not} contain all tuples in $\cons(\mathcal{T})$. This shows that, even when considering acyclic sets of functional dependencies, our notion of repair does {\em not} exactly coincide with that of traditional approaches, in which repairs would be defined as maximal subsets  of $\true(\mathcal{T})$ satisfying $FD$. However, we show in the next sub-section (see Proposition~\ref{prop:standard-repair}) that these two notions of repairs coincide, when considering star-tables as we do in the present approach.

It has also been shown in \cite{LauS25} that when the set $FD$ is acyclic, then the following holds: 
\begin{itemize}
\item 
For every table $T$, $\rep(T) \ne \emptyset$.
\item 
If $T \models FD$ then for every $R$ in $\rep(T)$,  $\true(\mathcal{R}) = \true(\mathcal{T})$. In this case $R$ and $T$ carry the same information, but  the tables $R$ and $T$ might not be equal. For example for $FD=\emptyset$ and $T=\{abc, ac\}$, $R = \{abc\}$ is a repair of $T$.
\end{itemize}
Elaborating briefly on the remark in the second item above, we notice that two distinct tables defined over the same universe and the same set of functional  dependencies can have the {\em same sets of true tuples.} Two tables $T$ and $R$ over the same universe $U$ are said to be equivalent if they have the same set of true tuples, that is $\true(\mathcal{T})=\true(\mathcal R)$; and when we refer to a table $T$ we in fact refer to any table $\widetilde{T}$ such that $\true(\mathcal{T})=\true(\mathcal{\widetilde{T}})$.

Moreover, the following basic theorem has been shown in \cite{LauS25} to hold whenever the set $FD$ is acyclic.
\begin{theorem}\label{theo:repair}
Let $T$ be a table over universe $U$ and $FD$ an acyclic set of functional dependencies over $U$. Then:
$$\cons(\mathcal{T})= \bigcap_{R \in \rep(T)}\true(\mathcal{R}).$$\hfill$\Box$
\end{theorem}
Coming back to star-tables,  the set $\cons(\mathcal{T})$ satisfies $FD$, because as seen earlier,  the set $FD$ of functional dependencies is acyclic. We thus define repairs of star-tables as follows.
\begin{definition}\label{def:repair}
Let $T$ be a star-table over universe $U$. A repair $R$ of $T$ is a table over $U$ such that:
\begin{enumerate}
\item $\cons(\mathcal{T}) \subseteq \true(\mathcal{R}) \subseteq \true(\mathcal{T})$.
\item $R \models FD$.
\item For every table $R'$ satisfying 1, and 2 above, and such that $\true(\mathcal{R}) \subseteq  \true(\mathcal{R}')$, we have $\true(\mathcal{R}) = \true(\mathcal{R}')$.
\end{enumerate}
The set of all repairs of $T$ is denoted by $\rep(T)$.
\hfill$\Box$
\end{definition}
It should be clear that all results recalled just above concerning acyclic sets of functional dependencies hold in the case of star-tables.
\subsection{Repairs of Star-Tables}\label{subsec:rep-star}
In this section, we show that, in the case of star-tables, repairs satisfy important properties that don't hold in general.
 
 An important specificity of star-tables is that our notion of repair coincides with that in the literature, applied to $\true(\mathcal{T})$. More precisely, we show that if $T$ is a star-table and $S$ is a maximal consistent subset  of $\true(\mathcal{T})$, then $S$ is a repair of $T$ in the sense of Definition~\ref{def:repair}, implying in particular that $\true(\mathcal{S})$ contains $\cons(\mathcal{T})$. As a consequence, in Definition~\ref{def:repair}, the first item can be more simply stated as $\true(\mathcal{R}) \subseteq \true(\mathcal{T})$. This result relies on the following preliminary two lemmas.
\begin{lemma}\label{lemma:repair1}
Let $T$ be a star-table over universe $U$, $S$ a subset  of $\true(\mathcal{T})$ and $t$ a tuple in $\cons(\mathcal{T})$. If $S \models FD$ then $S \cup \{t\} \models FD$.
\end{lemma}
{\sc Proof.} See Appendix~\ref{append:lemma-repair1}.\hfill$\Box$
\begin{lemma}\label{lemma:repair2}
Let $T$ be a star-table over universe $U$, $S$ a subset  of $\true(\mathcal{T})$. If $S \models FD$ then $S \cup \cons(\mathcal{T}) \models FD$.
\end{lemma}
{\sc Proof.} If $N$ denotes the number of tuples in $\cons(\mathcal{T})$, applying Lemma~\ref{lemma:repair1} successively to each tuple of $\cons(\mathcal{T})$ generates a sequence of sets $S_i$  ($i = 1, \ldots ,N$)  such that $S_i \models FD$ and $\cons(\mathcal{T}) \subseteq \true(\mathcal{S}_N)$. Since $\true(\mathcal{S}_N)= \true(S \cup \cons(\mathcal{T}))$, the proof is complete.\hfill$\Box$

\medskip\noindent
We now state the expected result that every maximal and consistent subset of $\true(\mathcal{T})$ defines a repair of $T$.
\begin{proposition}\label{prop:standard-repair}
Let $T$ be a star-table over universe $U$, and $S$ a subset  of $\true(\mathcal{T})$ such that $S \models FD$ and there is no subset  $S'$ of $\true(\mathcal{T})$ such that $S' \models FD$ and $\true(\mathcal{S})$ is a strict subset  of $\true(\mathcal{S}')$. Then $S$ is in $\rep(T)$.
\end{proposition}
{\sc Proof.} By Definition~\ref{def:repair}, in order to show that $S \in \rep(T)$, we only have to prove that $\cons(\mathcal{T}) \subseteq \true(\mathcal{S})$. If such is not the case, by Lemma~\ref{lemma:repair2}, $\true(\mathcal{S}) \cup \cons(\mathcal{T}) \models FD$. We thus obtain a consistent strict super-set of $\true(\mathcal{S})$, which is a contradiction with the last statement in the proposition. The proof is therefore complete.
\hfill$\Box$

\medskip\noindent
We emphasize that, Proposition~\ref{prop:standard-repair} shows that, if  $\true(\mathcal{T})$ is seen as a table over $U$, then a repair $R$ in $\rep(T)$ can be seen as a repair of $\true(\mathcal{T})$, in the sense of traditional approaches \cite{ArenasBC99}.

\smallskip
On the other hand, we show below that  Proposition~\ref{prop:standard-repair} implies that,  in the case of a star-table, given a true tuple $t$, there exists a repair in which $t$ is true, and that moreover, if $t$ is conflicting, there exists a repair in which $t$ is {\em not} true. It should be noticed here that, the second result is shown in \cite{LauS25}, whereas the first one is not.
\begin{proposition}\label{prop:repairs}
Let $T$ be a star-table over universe $U$. Then the following statements hold:

\smallskip
$\bullet$ For every $t$ in $\true(\mathcal{T})$ there exists $R$ in $\rep(T)$ such that $t$ is in $\true(\mathcal{R})$.

$\bullet$ For every $t$ in $\inc(\mathcal{T})$ there exists $R$ in $\rep(T)$ such that $t$ is not in $\true(\mathcal{R})$.
\end{proposition}
{\sc Proof.} To show the first statement, based on the fact that for every tuple $t$ in $\mathcal{T}$, $\{t\} \models FD$ trivially holds, as $\true(\mathcal{T})$ is finite, there exists a maximal subset $S$ of $\true(\mathcal{T})$ such that $t \in S$ and $S \models FD$. By Proposition~\ref{prop:standard-repair}, $S$ is in $\rep(T)$. Thus the proof of the first statement of the proposition is complete. The second statement of this proposition has been proved in \cite{LauS25} as Proposition~4. Therefore the proof of this proposition is complete.\hfill$\Box$

\medskip\noindent
As a last property of repairs of star-tables, we show below that if tuples are conflicting, then for every repair $R$, one of the tuples involved in the conflict is true in $R$.
\begin{proposition}\label{new-prop-repairs}
Let $T$ be a star-table over universe $U$. For a given $X \to A$ in $FD$ and a given tuple $x$ over $X$, let $\{xa_1, \ldots , xa_q\}$ be the set of all tuples over $XA$ that belong to $\true(\mathcal{T})$. Then, for every $R$ in $\rep(T)$, there exists $q_0$ in $\{1, \ldots ,q\}$ such that $xa_{q_0}$ is in $\true(\mathcal{R})$.
\end{proposition}
{\sc Proof.} See Appendix~\ref{append:new-prop-repairs}.\hfill$\Box$ 

\medskip\noindent
We now show how to characterize all repairs of a star-table $T$, based on $m\_Chase(T)$. To this end, we define the following three-step process {\sf (P)}:

\begin{description}
\item {\sf Step~P1.} For every $i$ in $\{1, \ldots ,n\}$ and every $A_i^j$ in $sch^*(D_i)$, if $k_i$ is a $K_i$-value occurring in $m\_Chase(T)$, we choose one $A_i^j$-value  in  the set $\sigma(A_i^j)$ where $\sigma$ is any m-tuple in $m\_Chase(T)$ such that $K_iA_i^j \subseteq sch(\sigma)$ and $\sigma(K_i)=(k_i)$. Denoting by $\varphi_i^j(k_i)$ this value, we notice that $\varphi_i^j(k_i)$ is well defined because, thanks to $K_i \to A_i^j$ in $FD$, all $\sigma$ in $m\_Chase(T)$ such that $K_iA_i^j \subseteq sch(\sigma)$ and $\sigma(K_i)=(k_i)$ have the same value over $A_i^j$.
\item 
{\sf Step P2.} For every $l$ in $\{1, \ldots , p\}$, if $k$ is a $\mathbb{K}$-value occurring in $m\_Chase(T)$, we choose one $M_l$-value in the set $\sigma(M_l)$ where $\sigma$ is any m-tuple in $m\_Chase(T)$ such that $\mathbb{K}M_l\subseteq sch(\sigma)$ and $\sigma(\mathbb{K})=(k)$. Denoting by $\varphi_l(k)$ this value, we notice that, as above, $\varphi_l(k)$ is well defined because, thanks to $\mathbb{K} \to M_l$ in $FD$, all $\sigma$ in $m\_Chase(T)$ such that $\mathbb{K}M_l \subseteq sch(\sigma)$ and $\sigma(\mathbb{K})=(k)$ have the same value over $M_l$.
\item 
{\sf Step P3.} Denoting by $\varphi$ the set of all $\varphi_i^j(k_i)$ and all $\varphi_l(k)$ defined above, for every $\sigma$ in $m\_Chase(T)$, let $t_\varphi(\sigma)$ be the tuple such that $sch(t_\varphi(\sigma)) = sch(\sigma)$, defined by:
\begin{itemize}
    \item For every $i=1, \ldots ,n$, if $K_i \in sch(\sigma)$ and $\sigma(K_i)=(k_i)$, then $t_\varphi(\sigma).K_i=k_i$ and for every $j$ such that $A_i^j \in sch(\sigma)$, $t_\varphi(\sigma).A_i^j=\varphi_i^j(k_i)$.
    \item If $\mathbb{K} \subseteq sch(\sigma)$ and $\sigma(\mathbb{K})=(k)$, for every $l =1 ,\ldots ,p$ such that $M_l \in sch(\sigma)$, let $t_\varphi(\sigma).M_l =\varphi_l(k)$.
\end{itemize}
\end{description}
Denoting by $R_\varphi$ the table $R_\varphi =\{t_\varphi(\sigma)~|~\sigma \in m\_Chase(T)\} \cup \cons(\mathcal{T})$, the following proposition states that $R_\varphi$ is a repair and that all repairs are obtained through such process.
\begin{proposition}\label{prop:all-repairs}
Let $T$ be a star-table over $U$. $R$ is a repair of $T$ if and only if there is $\varphi$ as defined above such that $\true(\mathcal{R})=\true(\mathcal{R}_\varphi)$.
\end{proposition}
{\sc Proof.} See Appendix~\ref{append:all-repairs}.\hfill$\Box$

\medskip\noindent 
We illustrate repairs of star-tables in the context of our running Example~\ref{ex:star} as follows.

\begin{figure}[t]
{\footnotesize
\begin{center}
\begin{tabular}{c|ccccccc}
$R_1$&$K_1$&$K_2$&$A_1^1$&$A_1^2$&$A_2^1$&$A_2^2$&$M_1$\\
\hline
&$k_1$&$k_2$&$a_1$&$a_2$&$b_1$&$b_2$&$m_1$\\
&&$k_2$&$a'_1$&$a_2$&$b_1$&$b_2$&$m_1$\\
&$k_1$&$k'_2$&$a_1$&$a_2$&$b_1$&&$m'_1$\\
&$k_1$&&$a_1$&$a_2$&$b_1$&&$m''_1$\\
&&$k'_2$&$a_1$&$a_2$&$b_1$&&$m''_1$\\
&&$k'_2$&$a'_1$&$a_2$&$b_1$&&$m'_1$\\
&&$k'_2$&$a'_1$&$a_2$&$b_1$&&$m''_1$\\
&$k'_1$&$k''_2$&$a_1$&&&&$m_1$\\
&$k''_1$&&$a'_1$&$a'_2$&&&\\
\end{tabular}
\qquad
\begin{tabular}{c|ccccccc}
$R_2$&$K_1$&$K_2$&$A_1^1$&$A_1^2$&$A_2^1$&$A_2^2$&$M_1$\\
\hline
&$k_1$&$k_2$&$a'_1$&$a_2$&$b_1$&$b_2$&$m_1$\\
&&$k_2$&$a_1$&$a_2$&$b_1$&$b_2$&$m_1$\\
&$k_1$&$k'_2$&$a'_1$&$a_2$&$b_1$&&$m'_1$\\
&$k_1$&&$a'_1$&$a_2$&$b_1$&&$m''_1$\\
&&$k'_2$&$a'_1$&$a_2$&$b_1$&&$m''_1$\\
&&$k'_2$&$a_1$&$a_2$&$b_1$&&$m'_1$\\
&&$k'_2$&$a_1$&$a_2$&$b_1$&&$m''_1$\\
&$k'_1$&$k''_2$&$a_1$&&&&$m_1$\\
&$k''_1$&&$a'_1$&$a'_2$&&&\\
\end{tabular}
\\~\\~\\~\\
\begin{tabular}{c|ccccccc}
$R_3$&$K_1$&$K_2$&$A_1^1$&$A_1^2$&$A_2^1$&$A_2^2$&$M_1$\\
\hline
&$k_1$&$k_2$&$a_1$&$a_2$&$b_1$&$b_2$&$m_1$\\
&&$k_2$&$a'_1$&$a_2$&$b_1$&$b_2$&$m_1$\\
&$k_1$&$k'_2$&$a_1$&$a_2$&$b_1$&&$m''_1$\\
&$k_1$&&$a_1$&$a_2$&$b_1$&&$m'_1$\\
&&$k'_2$&$a_1$&$a_2$&$b_1$&&$m'_1$\\
&&$k'_2$&$a'_1$&$a_2$&$b_1$&&$m''_1$\\
&&$k'_2$&$a'_1$&$a_2$&$b_1$&&$m'_1$\\
&$k'_1$&$k''_2$&$a_1$&&&&$m_1$\\
&$k''_1$&&$a'_1$&$a'_2$&&&\\
\end{tabular}
\qquad
\begin{tabular}{c|ccccccc}
$R_4$&$K_1$&$K_2$&$A_1^1$&$A_1^2$&$A_2^1$&$A_2^2$&$M_1$\\
\hline
&$k_1$&$k_2$&$a'_1$&$a_2$&$b_1$&$b_2$&$m_1$\\
&&$k_2$&$a_1$&$a_2$&$b_1$&$b_2$&$m_1$\\
&$k_1$&$k'_2$&$a'_1$&$a_2$&$b_1$&&$m''_1$\\
&$k_1$&&$a'_1$&$a_2$&$b_1$&&$m'_1$\\
&&$k'_2$&$a'_1$&$a_2$&$b_1$&&$m'_1$\\
&&$k'_2$&$a_1$&$a_2$&$b_1$&&$m''_1$\\
&&$k'_2$&$a_1$&$a_2$&$b_1$&&$m'_1$\\
&$k'_1$&$k''_2$&$a_1$&&&&$m_1$\\
&$k''_1$&&$a'_1$&$a'_2$&&&\\
\end{tabular}
\end{center}
}
\caption{The repairs of the star-table $T$ of Example~\ref{ex:star}
\label{fig:repairs}}
\end{figure}

\begin{example}\label{ex:star2}
{\rm
In the context of Example~\ref{ex:star},  $T$ has four repairs as shown in Figure~\ref{fig:repairs}.
It can be seen from the m-table $m\_Chase(T)$ in Figure~\ref{fig:m-chase} that, in each of these four tables:
\begin{itemize}
\item the first two rows are generated using the first m-tuple in $m\_Chase(T)$, namely\\
$(k_1)(k_2)(a_1\,a'_1)(a_2)(b_1)(b_2)(m_1)$,
\item the next five rows are generated using the second m-tuple in $m\_Chase(T)$, namely\\$(k_1)(k'_2)(a_1\,a'_1)(a_2)(b_1)(m'_1\,m''_1)$,
\item the last two rows are generated using respectively the last two m-tuples in $m\_Chase(T)$, namely $(k'_1)(k''_2)(a_1)(m_1)$ and $(k''_1)(a'_1)(a'_2)$.
\end{itemize}
To illustrate how process {\sf (P)} works, we apply it to generate the repair $R_1$ shown in Figure~\ref{fig:repairs}.
\begin{itemize}
\item {\sf Step P1.} Regarding attributes in $sch^*(D_1)$, for $A_1^1$, the three $K_1$-values $k_1$, $k'_1$ and $k''_1$ are respectively associated with $a_1$, $a_1$ and $a'_1$, while for $A_1^2$, the two $K_1$-values $k_1$ and $k''_1$ are associated with $a_2$ and $a'_2$, respectively (no $A_1^2$-value is associated with $k'_1$ in $T$).\\
Regarding attributes in $sch^*(D_2)$, for $A_2^1$, the two $K_2$-values $k_2$ and $k'_2$ are both associated with $b_1$ (no $A_2^1$-value is associated with $k''_2$ in $T$), and for $A_2^2$, the only $K_2$-values $k'_2$ is associated with $b_2$ (no $A_2^2$-value is associated with $k'_2$ or with $k''_2$ in $T$).
\item {\sf Step P2.} The only measure attribute is here $M_1$ and the $\mathbb{K}$-values to be considered are $k_1k_2$, $k_1k'_2$ and $k'_1k''_2$ which are respectively associated with $M_1$-values $m_1$, $m''_1$ and $m_1$.
\item {\sf Step P3.} Considering every m-tuple $\sigma$ of the table $m\_Chase(T)$ shown in Figure~\ref{fig:m-chase}, we obtain the following four tuples: $k_1k_2a_1a_2b_1b_2m_1$, $k_1k'_2a_1a_2b_1m'_1$, $k'_1k''_2a_1m_1$ and $k''_1a'_1a'_2$.
\end{itemize}
Hence, $R_\varphi$ contains the four tuples above along with the tuples in $\cons(\mathcal{T})$ that have been characterized in Example~\ref{ex:star}. Moreover, inspecting thoroughly the table $R_1$ and the set $R_\varphi$ shows that the two sets yield the same set of true tuples, corresponding to the same repair of $T$.\hfill$\Box$
}
\end{example}
\section {Consistent Query Answering}\label{sec:cqa}
Hereafter, given a star-table $T$, we consider different kinds of queries on $T$, all based on the SQL syntax:
\begin{enumerate}
    \item {\em Standard queries} (or simply {\em queries}), to be defined in the next section. These queries are simple projection-selection queries of the form {\tt select} $X$  {\tt from} $T$ {\tt where $\Gamma$}, where $X$ is a list of attributes and $\Gamma$ an optional selection condition.
    \item {\em Analytic queries with no {\tt group-by} clause}, defined in Section~\ref{subsec:analytic}. These queries are of the form {\tt select} $X, aggr(M_i)$  {\tt from} $T$ {\tt where $\Gamma$}, where $X$ and $\Gamma$ are as above, and where $aggr$ is an aggregation operator and $M_i$ is a measure attribute.
    \item {\em Analytic queries with a {\tt group-by} clause}, also defined in Section~\ref{subsec:analytic}. These queries are of the form {\tt select} $X, aggr(M_i)$  {\tt from} $T$ {\tt where $\Gamma$} {\tt group by} $X$, where $X$, $\Gamma$, $aggr$ and $M_i$ are as above.
    \item {\em Analytic queries with a {\tt group-by-having} clause}, dealt with in Section~\ref{sec:groupbyhaving}. These queries are of the form {\tt select} $X, aggr(M_i)$  {\tt from} $T$ {\tt where $\Gamma$} {\tt group by} $X$ {\tt having} $\Upsilon$, where $X$, $\Gamma$, $aggr$ and $M_i$ are as above, and where $\Upsilon$ is a boolean expression involving aggregates.
\end{enumerate}
\subsection{Standard Queries}                                                        
As mentioned just above, we use SQL as the query language and, as we query a single table $T$, the (standard) queries $Q$ that we consider have one of the following two forms:

\smallskip\centerline{
$Q:$ {\tt select} $X$ {\tt from} $T$\qquad or \qquad $Q:$ {\tt select} $X$  {\tt from} $T$ {\tt where $\Gamma$}}

\smallskip\noindent
In either of these forms, $X$ is an attribute list seen as a relation schema, and in the second form the {\tt where} clause specifies a selection condition $\Gamma$. As in SQL the where clause in a query is optional, the generic form of a query $Q$ is denoted by 

\smallskip\centerline{$Q:$ {\tt select} $X$  {\tt from} $T$ {\tt [where $\Gamma$]}} 

\smallskip\noindent
The set of all attributes occurring in $\Gamma$ is called the {\em schema of $\Gamma$}, denoted by $sch(\Gamma)$; and the attribute set $X \cup sch(\Gamma)$ is called the {\em schema of $Q$}, denoted by $sch(Q)$. Moreover, in what follows, given an attribute set $X$, $\mathcal{T}(X)$ denotes the set of all tuples in $\mathcal{T}$ whose schema is $X$.

\smallskip
A selection condition $\Gamma$ is a well-formed formula involving the usual connectors $\neg$, $\vee$ and $\wedge$ and built up from atomic Boolean comparisons of one of the following forms: $A \,\theta\, a$ or $A \,\theta\, A'$, where $\theta$ is a comparison predicate, $A$ and $A'$ are attributes in $U$ whose domain elements are comparable through $\theta$, and $a$ is in $dom(A)$. 

Given a selection condition $\Gamma$,  the set of tuples satisfying $\Gamma$, denoted by $Sat(\Gamma)$, is inductively defined as follows:

\smallskip
$\bullet$~if $\Gamma$ is of the the form $A \,\theta\, a$, $Sat(\Gamma)=\{t \in \mathcal{T}(sch(\Gamma)) ~|~t.A \,\theta\, a\}$,

$\bullet$~if $\Gamma$ is of the the form $A \,\theta\, B$, $Sat(\Gamma)=\{t\in \mathcal{T}(sch(\Gamma)) ~|~t.A \,\theta\, t.B\}$,

$\bullet$~if $\Gamma$ is of the form $\Gamma_1 \vee \Gamma_2$, $Sat(\Gamma)= Sat(\Gamma_1) \cup Sat(\Gamma_2)$,

$\bullet$~if $\Gamma$ is of the form $\Gamma_1 \wedge \Gamma_2$, $Sat(\Gamma)= Sat(\Gamma_1) \cap Sat(\Gamma_2)$,

$\bullet$~if $\Gamma$ is of the form $\neg \Gamma_1$, $Sat(\Gamma)= \mathcal{T}(sch(\Gamma)) \setminus Sat(\Gamma_1)$.

\smallskip\noindent
As usual in the literature \cite{ArenasBC99}, we define the {\em consistent answer} of such a query $Q$ as the intersection of the answers to the query in every repair. The formal definition follows.
\begin{definition}\label{def:cons-answer}
Let $T$ be a table over universe $U$ and $FD$ an acyclic set of functional dependencies over $U$.  Given the query $Q:$ {\tt select}~$X$~{\tt from}~$T$~{\tt [where~}$\Gamma${\tt ]}, the {\em consistent answer to $Q$ in $T$}, denoted by $\consans(Q)$, is defined by:
$$\consans(Q)= \bigcap_{R \in \rep(T)}\ans\left( Q^{[R]}\right) $$
where $Q^{[R]}$ is the query  {\tt select}~$X$~{\tt from}~$R$~{\tt [where~}$\Gamma${\tt ]}, and $\ans(Q^{[R]})$ is the answer to $Q^{[R]}$ in $R$. Formally, $\ans(Q^{[R]})$ is the set of all tuples $x$ over $X$ such that there exits $\gamma$ in $Sat(\Gamma)$ such that $x\gamma$ is a tuple over $sch(Q)$ that belongs to $\true(\mathcal{R})$.
\hfill$\Box$
\end{definition}
It is important to notice that, as a consequence of Theorem~\ref{theo:repair}, given a query $Q$, we have $\consans(Q) \subseteq \cons(\mathcal{T})$. This is so because every $x$ in $\consans(Q)$ is true in every repair of $T$. We also point out that if $Q$ involves no selection condition then Theorem~\ref{theo:repair} allows for characterizing $\consans(Q)$ as the set of all tuples in $\cons(\mathcal{T})$ whose schema is $X$.

We however recall that the issue of easily characterizing $\consans(Q)$ in the case of a query $Q$ involving a selection condition remains open. We recall in this respect from \cite{LauS25}, that  $\consans(Q)$ can be {\em bounded} as follows.
\begin{proposition}\label{prop:approx}
Let $T$ be a table over universe $U$ and its associated acyclic set of functional dependencies $FD$. Given a query $Q: {\tt select}$~$X$~{\tt where}~$\Gamma$, let:
\begin{itemize}
\item $\sconsans(Q)=\{x \in \mathcal{T}(X)~|~(\exists t \in \mathcal{T}(sch(Q))(t \in \cons(\mathcal{T}) \wedge t.X = x \wedge t.sch(\Gamma) \in Sat(\Gamma)\}$
\item $\wconsans(Q)=\{x \in \mathcal{T}(X)~|~(\exists t \in \mathcal{T}(sch(Q))(t \in \true(\mathcal{T}) \wedge t.X = x \, \wedge $\\
\qquad \rightline{$x \in \cons(\mathcal{T}) \wedge t.sch(\Gamma) \in Sat(\Gamma)\}$\quad~\qquad}
\end{itemize}
Then, $\sconsans(Q)$ and $\wconsans(Q)$ are respectively a lower bound and an upper bound of $\consans(Q)$, i.e., the following inclusions hold: $\sconsans(Q) \subseteq \consans(Q) \subseteq \wconsans(Q)$.
\hfill$\Box$
\end{proposition}
Moreover, based on Proposition~\ref{prop:truth-value}, it is shown in \cite{LauS25} that the two bounds $\sconsans(Q)$ and $\wconsans(Q)$ can be easily computed by means of one scan of the m-table $m\_Chase(T)$. It turns out that approximating the consistent answer to any query in our approach is polynomial.
For example, in the context of Example~\ref{ex:star},
\begin{itemize}
\item 
For $Q_1:$ {\tt select} $K_1$, $K_2$, $M_1$  {\tt from} $T$ {\tt where ($A_1^1=a_1$)}, we have $\sconsans(Q_1)=\{k'_1k''_2m_1\}$ and $\wconsans(Q_1)=\{k_1k_2m_1, k'_1k''_2m_1\}$. On the other hand, it can be seen from Figure~\ref{fig:repairs} that $\consans(Q_1)=\sconsans(Q_1)=\{k'_1k''_2m_1\}$.
\item 
For $Q_2:$ {\tt select} $K_1$, $K_2$, $A_2^1$, $M_1$,  {\tt from} $T$ {\tt where ($A_1^1=a_1$ or $A_1^1=a'_1$)}, then we have $\sconsans(Q_2)=\emptyset$ and $\wconsans(Q_2)=\{k_1k_2b_1m_1\}$. On the other hand, it can be seen from Figure~\ref{fig:repairs} that $\consans(Q_2)=\wconsans(Q_2)=\{k_1k_2b_1m_1\}$.
\end{itemize}
\subsection{Consistent Query Answering and Star Schemas}
As mentioned in our introductory section, we can compute efficiently exact consistent answers (instead of bounds of exact consistent answers as in \cite{LauS25}) if the selection condition is {\em independent}, as stated in the following definition.
\begin{definition}\label{def:indpdt}
A selection condition $\Gamma$ is said to be {\em independent} if $\Gamma$ is equivalent to the conjunction  $\Gamma(A_1) \wedge \ldots \wedge \Gamma(A_k)$ where for every $i=1, \ldots ,k$, $\Gamma(A_i)$ is a selection condition involving only attribute $A_i$.
\hfill$\Box$
\end{definition}
By Definition~\ref{def:indpdt}, if $\Gamma = \Gamma(A_1) \wedge \ldots \wedge \Gamma(A_k)$ is independent, then $sch(\Gamma)=A_1 \ldots A_k$ and a tuple $\gamma$ over  $sch(\Gamma)$ is in $Sat(\Gamma)$ if and only if for every $i=1, \ldots ,k$, $\gamma.A_i$ is in $Sat(\Gamma(A_i))$. 

\smallskip
For example, let $\Gamma=(A_1 \leq 0) \vee ((A_1 \geq 4) \wedge (A_2 =a))$. Clearly, the conjunctive normal form of $\Gamma$ is $((A_1 \leq 0) \vee (A_1 \geq 4))\wedge ((A_1 \leq 0) \vee (A_2 =a))$. Due to the second conjunct, it turns out that $\Gamma$ cannot be written as the conjunction of disjunctions involving only one attribute. Therefore, $\Gamma$ is not independent.

On the other hand, $\Gamma'=((A_1 \leq 4) \vee (A_1 \geq 0)) \wedge (A_2 =a)$ is obviously independent with $\Gamma'(A_1)=(A_1 \geq 4) \vee (A_1 \leq 0)$ and $\Gamma'(A_2)=(A_2 =a)$, and we have $Sat(\Gamma')=\{(a_1,a_2)\in adom(A_1) \times adom(A_2)~|~(a_1 \leq 0 \vee a_1 \geq 4) \wedge (a_2=a)\}$.

We emphasize that, as shown by the example just above, independent selection conditions may involve disjunctions. Therefore, and contrary to most existing approaches to consistent query answering \cite{ArenasBC99,Wijsen09}, our approach does {\em not} restrict selection conditions to be conjunctive.
\begin{itemize}
\item []
{\em From now on, it is assumed, even when not explicitly mentioned, that selection conditions are independent.}
\end{itemize}
The following proposition states one of the main contributions of the paper, namely that, under the restriction above,  the consistent answer to any query having an independent selection condition can be easily computed.
\begin{proposition}\label{prop:cons-ans-ana}
Let $T$ be a star-table over universe $U$, and let $Q: {\tt select}$~$X$~{\tt from~$T$~where~}$\Gamma$ be a query such that $\Gamma$ is an independent selection condition. Then $\consans(Q)$ is the set of all tuples $x$ over $X$ for which there exists $\sigma$ in $m\_Chase(T)$ such that 

\begin{enumerate}
\item $sch(Q) \subseteq sch(\sigma)$, $x \in \sigma(X)$ and $\sigma(sch(\Gamma)) \cap Sat(\Gamma) \ne \emptyset$,
\item for every $Y \to B$ in $FD$ such that $YB \subseteq sch(Q)$:
\begin{enumerate}
\item if $B \in X$, then $|\tuples(\sigma(B))|=1$,
\item if $B \in sch(\Gamma)$, then $\tuples(\sigma(B)) \subseteq Sat(\Gamma(B))$.
\end{enumerate}
\end{enumerate}
\end{proposition}
{\sc Proof.} See Appendix~\ref{append:cons-ans-ana}.\hfill$\Box$

\medskip\noindent
An important consequence of Proposition~\ref{prop:cons-ans-ana} is that the time complexity of consistent query answering to projection-selection-join queries on star schemes is {\em quasi-linear} (i.e., in $\mathcal{O}(W\cdot \log(W))$, where $W$ is the size of the data warehouse), under the restriction that the selection condition is independent.
The following corollary shows that the  items in Proposition~\ref{prop:cons-ans-ana} can be more simply stated when $\mathbb{K}$ is a subset  of $X$.
\begin{corollary}\label{coro:cons-ans-ana}
Let $T$ be a star-table over universe $U$, and let $Q: {\tt select}$~$X$~{\tt from~$T$~where~}$\Gamma$ be a query such that $\Gamma$ is an independent selection condition. If $\mathbb{K} \subseteq X$, $\consans(Q)$ is the set of all tuples $x$ over $X$ for which there exists $\sigma$ in $m\_Chase(T)$ such that 
\begin{enumerate}
\item $sch(Q) \subseteq sch(\sigma)$,
\item $|\tuples(\sigma(X))|=1$, i.e., $\tuples(\sigma(X))=\{x\}$,
\item $\tuples(\sigma(sch(\Gamma))) \subseteq Sat(\Gamma)$.
\end{enumerate}
\end{corollary}
{\sc Proof.} The inclusion of $\mathbb{K}$ in $X$ implies that for every $B$ in $sch(Q) \setminus \mathbb{K}$, there exists $Y \to B$ in $FD$ such that $YB \subseteq sch(Q)$. Considering moreover that for every $Y \subseteq \mathbb{K}$, $|\tuples(\sigma(Y))|=1$ always holds, items 1 and 2(a) in Proposition~\ref{prop:cons-ans-ana} are equivalent to item 2 in the corollary.
Similarly, for every attribute $B$ in $sch(\Gamma)$, either $B$ is in $\mathbb{K}$ or there exists $Y \to B$ in $FD$  such that $Y \subseteq \mathbb{K}$. Thus, considering the latter case, item 2(b) in Proposition~\ref{prop:cons-ans-ana} is equivalent to: for every $B \in sch(\Gamma) \setminus \mathbb{K}$, $\tuples(\sigma(B)) \subseteq Sat(\Gamma(B))$. On the other hand if $B$ is in $\mathbb{K}$, we have $\tuples(\sigma(B))=(b)$ where $b$ is in $Sat(\Gamma(B))$, by item 1 in  Proposition~\ref{prop:cons-ans-ana}. Hence item 1 and item 2(b) imply that for every $B \in sch(\Gamma)$, $\tuples(\sigma(B) \subseteq Sat(\Gamma(B))$. Since $\Gamma$ is independent, as mentioned right after Definition~\ref{def:indpdt}, this is equivalent $\tuples(\sigma(sch(\Gamma))) \subseteq Sat(\Gamma)$. The proof is therefore complete.
\hfill$\Box$

\medskip\noindent
In the following example, we illustrate Proposition~\ref{prop:cons-ans-ana} and Corollary~\ref{coro:cons-ans-ana}, along with the importance of our restriction on the selection conditions.
\begin{example}
{\rm
In the context of Example~\ref{ex:star}, we consider the star-table $T$, and its  associated m-table $m\_Chase(T)$ as shown in Figure~\ref{fig:m-chase}, let $Q_2$ be the query  defined by:

\smallskip
$Q_2:$ {\tt select} $K_1$, $K_2$, $A_2^1$, $M_1$,  {\tt from} $T$ {\tt where ($A_1^1=a_1$ or $A_1^1=a'_1$)}.

\smallskip\noindent
As already  noticed after stating Proposition~\ref{prop:approx}, we have $\consans(Q_2)=  \{k_1k_2b_1m_1\}$. As the selection condition in this query is obviously independent, applying Proposition~\ref{prop:cons-ans-ana} or Corollary~\ref{coro:cons-ans-ana} (because $K_1K_2 \subseteq sch(Q_2)$) yields the following:
\begin{itemize}
\item
The m-tuples $\sigma$ in $m\_Chase(T)$ such that $sch(Q_i) \subseteq sch(\sigma)$, $x \in \sigma(X)$ and $\sigma(sch(\Gamma)) \cap Sat(\Gamma) \ne \emptyset$, are
$(k_1)(k_2)(a_1\,a'_1)(a_2)(b_1)(b_2)(m_1)$ and $(k_1)(k'_2)(a_1\,a'_1)(a_2)(b_1)(m'_1\,m''_1)$.
\item
Among these m-tuples, only $(k_1)(k_2)(a_1\,a'_1)(a_2)(b_1)(b_2)(m_1)$ is such that: for every $Y \to B$ in $FD$ such that $YB \subseteq X_i$, $|\tuples(\sigma(B))|=1$.
\item The unique m-tuple $\sigma$ of the previous item is such that $\tuples(\sigma(sch(\Gamma))) \subseteq Sat(\Gamma)$.
\end{itemize}
We therefore obtain the expected consistent answer, namely $\consans(Q_2)=  \{k_1k_2b_1m_1\}$.
Now, let $Q_3$ be defined by:

\smallskip
$Q_3:$ {\tt select} $K_1,K_2,A^2_1$  {\tt from} $T$ {\tt where}\\
\rightline{{\tt ($A_1^1=a_1$ and $M_1 = m_1$) or ($A_1^1=a'_1$ and $M_1 = m''_1$)}.\qquad\qquad}

\smallskip\noindent
In this case, the selection condition $\Gamma_3$ is clearly {\em not} independent, and  it can be seen from Figure~\ref{fig:repairs} that $\consans(Q_3)$ is in fact empty, because:

\noindent
~$-$ in $R_2$, $k_1k_2a_2$ is only associated with $a'_1m_1$ which does not belong to $Sat(\Gamma_3)$,

\noindent
~$-$ in $R_1$, $k_1k'_2a_2$ is only associated with $a_1m'_1$ which does not belong to $Sat(\Gamma_3)$.

\smallskip\noindent
On the other hand, considering the third item in Proposition~\ref{prop:cons-ans-ana}, we emphasize the following: $k_1k'_2a_2$ is discarded from the consistent answer because $a_1m'_1$ is in $\tuples(\sigma_2(A_1^1M_1))$ but not in $Sat(\Gamma_3)$. However, $\sigma_1$ can be seen as satisfying the third item in the proposition for the following reasons:  the dependencies to consider are $K_1 \to A_1^1$ and $K_1K_2 \to M_1$, and  $\tuples(\sigma_1(A^1_1))$, respectively $\tuples(\sigma_1(M_1))$, is a subset of the set of all $A_1^1$-values, respectively all $M_1$-values, occurring in $Sat(\Gamma_3)$.  A characterization of the consistent answer to a query involving a {\em non independent} selection condition is currently unknown to the authors.
\hfill$\Box$
}
\end{example}
In the next two sections, we consider analytic queries and see how their consistent answer can be defined and effectively computed, relying on Corollary~\ref{coro:cons-ans-ana}.
\subsection{Analytic Queries and their Consistent Answers}\label{subsec:analytic}
In the literature \cite{Ullman}, an analytic query is a query involving an aggregate operator among $count$, $min$, $max$, $sum$, and possibly a {\tt group-by} clause. Formally, given a data warehouse,  an analytic query $\mathcal{AQ}$ has one of the following two forms:
\begin{itemize}
\item 
$\mathcal{AQ}:  {\tt select}$~$aggr(M_i)$~{\tt from $\varphi$ where} $\Gamma$, or
\item 
$\mathcal{AQ}:  {\tt select}$~$X$, $aggr(M_i)$~{\tt from $\varphi$ where} $\Gamma$ {\tt group by $X$}
\end{itemize}
where $\varphi$ is the join of the fact table $F$ with all dimension tables $D_1, \ldots , D_n$, $X$ is a relation schema, and for $j=1, \ldots , p,$ $aggr$ is an aggregate operator such as $count$, $min$, $max$, $sum$ and $M_i$ is a measure attribute.

In all traditional approaches to data warehouses, it is generally assumed that the tables in the data warehouse have no missing values and that the key-foreign-key constraints between $F$ and all dimension tables $D_i$ are satisfied. In this context, the answer to $\mathcal{AQ}$ is as follows:
\begin{itemize}
\item 
If $\mathcal{AQ}$ involves no  {\tt group-by} clause, $\ans(\mathcal{AQ})$ is the value of the aggregation evaluated over the set of all tuples in $\varphi$ satisfying the condition $\Gamma$. We notice that the aggregate may involve no attribute, when  expressed as $count(*)$. 
\item 
If $\mathcal{AQ}$ involves a statement {\tt group-by} $X$, $\ans(\mathcal{AQ})$ is the set of all pairs $\langle x, v_x\rangle$ where $x$ is such that there exists a tuple in $\varphi$ satisfying $\Gamma$ and whose $X$-value is $x$, and where $v_x$ is the value of the aggregation evaluated over all tuples in $\varphi$ satisfying $\Gamma$ and whose $X$-value is $x$.
\end{itemize}
Now, if the underlying data warehouse does {\em not} satisfy all functional dependencies, this traditional semantics of answer to analytic queries has to be revisited. When no dependency of the form $\mathbb{K} \to M_i$ is considered, this has been done in \cite{ArenasBCHRS03}, in the case of queries with no {\tt group-by} clause and in \cite{FuxmanFM05} when a {\tt group-by} clause occurs in the query. As shown in \cite{DixitK22}, in either case, the `repair semantics' of the consistent answer to an analytic query consists intuitively in producing an interval in which aggregate values fall, when answering the query in any repair. The reader is referred to the forthcoming Section~\ref{sec:rel-work} for a more precise relationship between these approaches and our work.

In our approach, we follow the same line regarding analytic queries and their consistent answers, and we emphasize that, as in the previous section, all selection conditions are assumed to be independent.
\begin{definition}\label{def:anal-query}
Let $T$ be a star-table over universe $U$. We call {\em analytic query} with or without {\tt group-by} clause, a query of the generic form:

\smallskip\centerline{
$\mathcal{AQ}: ~${\tt select [$X$]}, $aggr(M_i)$~{\tt from $T$ where} $\Gamma$ {\tt [group by $X$]}}

\smallskip\noindent
where the {\tt group-by} clause may be omitted, in which case $X$ is not  in the {\tt select} clause.

\smallskip
The {\em consistent answer} to analytic query $\mathcal{AQ}$, denoted by $\consans(\mathcal{AQ})$, is defined as follows:

\smallskip\noindent
{\sf (A)} If $\mathcal{AQ}$ involves no {\tt group-by} clause, then:
 \begin{itemize}
 \item If there exists $R$ in $\rep(T)$ such that $\true(\mathcal{R})$ contains no tuple over $sch(Q)$ satisfying $\Gamma$, then $\consans(\mathcal{AQ})={\tt NULL}$ if $aggr \ne count$, and $\consans(\mathcal{AQ})=0$ if $aggr$ is $count$.
\item If for every $R$ in $\rep(T)$, $\true(\mathcal{R})$ contains at least one tuple over $sch(Q)$ satisfying $\Gamma$, then $\consans(\mathcal{AQ}) = [glb, lub]$ such that:
\begin{itemize}
\item for every $R$ in $\rep(T)$ there exists $d$ in $[glb, lub]$ such that $\ans(\mathcal{AQ}^{[R]})=d$,
\item there exist $R_1$ and $R_2$ in  $\rep(T)$ such that $\ans(\mathcal{AQ}^{[R_1]})=glb$ and $\ans(\mathcal{AQ}^{[R_2]})=lub$.
 \end{itemize}
  \end{itemize}
{\sf (B)} If $\mathcal{AQ}$ involves a {\tt group-by} clause, $\consans(\mathcal{AQ})$ is a set of pairs of the form $\langle x,[glb, lub]\rangle$ such that:
\begin{itemize}
\item for every $R$ in $\rep(T)$ there exists $d$ in $[glb, lub]$ such that $\ans(\mathcal{AQ}^{[R]})$ contains $(x\,d)$,
\item there exist $R_1$ and $R_2$ in  $\rep(T)$ such that $\ans(\mathcal{AQ}^{[R_1]})$ and $\ans(\mathcal{AQ}^{[R_2]})$ respectively contain $(x, glb)$ and $(x, lub)$.
\hfill$\Box$
\end{itemize}
\end{definition}
The following remarks are in order regarding Definition~\ref{def:anal-query}:
\begin{itemize}
\item In case {\sf (A)}, a {\tt NULL} answer means that the consistent answer to the query $Q:~{\tt select}$ $\mathbb{K}$ {\tt from} $T~{\tt where}~\Gamma$ is empty. In this case the aggregate value cannot be computed if the aggregate operator is different than $count$, and if the aggregate operator is $count$, then the expected answer is $0$. This explains why we return {\tt NULL} in the former case.
\item In case {\sf (B)}, all $X$-values $x$ occurring in $\consans(\mathcal{AQ})$ must be consistent, because the first condition requires that $x$ be true in every repair of $T$. On the other hand, if $x$ is consistent and if there is a repair $R$ such that $\true(\mathcal{R})$ contains no tuple of the form $x\,\gamma$ where $\gamma$ is in $Sat(\Gamma)$, there is no need to use a {\tt NULL} because is this case, no pair involving $x$ will appear in $\consans(\mathcal{AQ})$. Consequently, if no $X$-value fits the conditions in the definition, $\consans(\mathcal{AQ})$  is simply set to $\emptyset$.
\end{itemize}
An alternative way to deal with consistent answers to analytic queries consists of computing first  the consistent answer to an appropriate non analytic query and then, of applying the aggregation operators to this answer. More formally,  a given analytic query $\mathcal{AQ}: ~${\tt select [$X$]}, $aggr(M_i)$~{\tt from $T$ where} $\Gamma$ {\tt [group by $X$]} is associated with the non analytic query defined by

\smallskip\centerline{$Q_{\mathcal{AQ}}:$~{\tt select~$\mathbb{K},$[$X$]$,M_i$~from $T$ where $\Gamma$}}

\smallskip\noindent
and this alternative consistent answer  to $\mathcal{AQ}$ is defined as follows.
\begin{definition}\label{def:strong-answer}
Let $T$ be a star-table over universe $U$ and let $\mathcal{AQ}$ be an analytic query defined as follows: $\mathcal{AQ}: ~${\tt select [$X$]}, $aggr(M_i)$~{\tt from $T$ where} $\Gamma$ {\tt [group by $X$]}.  The {\em strongly consistent answer to} $\mathcal{AQ}$, denoted by $\consans^*(\mathcal{AQ})$, is the answer to the analytic query:

\smallskip\centerline{
{\tt select [$X$]$,aggr(M_i)$}~{\tt from $\consans(Q_{\mathcal{AQ}})$ [group by $X$]}}

\smallskip\noindent
where $\consans(Q_{\mathcal{AQ}})$ is the consistent answer to the (non-analytic) query $Q_{\mathcal{AQ}}$ as defined above.\hfill$\Box$
\end{definition}
We emphasize that, since the {\tt select} clause of $Q_{\mathcal{AQ}}$ contains $\mathbb{K}$, $\consans(Q_{\mathcal{AQ}})$ can be effectively computed based on Corollary~\ref{coro:cons-ans-ana}, assuming that $m\_Chase(T)$ is available. Moreover, the semantics of such answers are radically different than those defined in Definition~\ref{def:anal-query}, simply because strongly consistent answers do not involve intervals, whereas consistent answers do. Moreover, as shown in the following example, aggregate values in the strongly consistent answer may not belong to the corresponding interval of the consistent answer.
\begin{example}\label{ex:star3.1}
{\rm
In the context of Example~\ref{ex:star}, $\mathcal{AQ}_1:  {\tt select}$~$sum(M_1)$~{\tt from $T$ where~($A_1^1=a_1$)} is an example of analytic query with no {\tt group-by} clause and involving the aggregate $sum$. The expected result is the interval containing all possible sums of $M_1$-values among all tuples  defined over $K_1K_2$ and associated with $A_1^1$-value $a_1$ in all repairs. Referring to Figure~\ref{fig:repairs}, we have:

\smallskip
$\bullet$ $\ans(\mathcal{AQ}_1^{[R_1]})=m_1+m'_1+m_1$, and $\ans(\mathcal{AQ}_1^{[R_2]})=m_1$,

$\bullet$ $\ans(\mathcal{AQ}_1^{[R_3]})=m_1+m''_1+m_1$, and $\ans(\mathcal{AQ}_1^{[R_4]})=m_1$.

\smallskip\noindent
Assuming that $m_1$, $m'_1$ and $m''_1$ are positive numbers and that $m'_1 \leq m''_1$, we obtain that $\consans(\mathcal{AQ}_1)=[glb,lub]$  where $glb=m_1$ and $lub=2 \cdot m_1+m''_1$.

On the other hand, we have $Q_{\mathcal{AQ}_1}:~${\tt select $\mathbb{K},M_1$ from $T$~where ($A_1^1=a_1$)}, and it can be seen from Figure~\ref{fig:repairs},  that $\consans(Q_{\mathcal{AQ}_1})=\{k'_1k''_2m_1\}$. Therefore, $\consans^*(\mathcal{AQ}_1)=m_1$. Notice that in this case, the value in $\consans^*(\mathcal{AQ}_1)$ belongs to the interval in $\consans(\mathcal{AQ}_1)$.

\smallskip
As another example of analytic query with no {\tt group-by} clause and slightly different from $\mathcal{AQ}_1$, consider $\mathcal{AQ}'_1:  {\tt select}$~$sum(M_1)$~{\tt from $T$ where~($A_1^1=a'_1$)}. Referring to Figure~\ref{fig:repairs}, we have in this case:

\smallskip
$\bullet$ $\ans(\mathcal{AQ}_1'^{[R_1]})={\tt NULL}$, and $\ans(\mathcal{AQ}_1'^{[R_2]})=m_1$,

$\bullet$ $\ans(\mathcal{AQ}_1'^{[R_3]})={\tt NULL}$, and $\ans(\mathcal{AQ}_1'^{[R_4]})=m_1$.

\smallskip\noindent
The presence of {\tt NULL} for $R_1$ and $R_3$ above is due to the fact that these repairs contain no tuples satisfying the selection condition in $\mathcal{AQ}'_1$. In this case, by Definition~\ref{def:anal-query}{\sf (A)}, the expected consistent answer is {\tt NULL}.

On the other hand,  we have $Q_{\mathcal{AQ}'_1}:~${\tt select $\mathbb{K},M_1$ from $T$~where ($A_1^1=a'_1$)}, and it can be seen from Figure~\ref{fig:repairs},  that $\consans(Q_{\mathcal{AQ}'_1})=\emptyset$. Therefore, $\consans^*(\mathcal{AQ}'_1)={\tt NULL} $.

\smallskip
The query $\mathcal{AQ}_2:  {\tt select}$~$A_2^1$, $sum(M_1)$ {\tt from} $T$ {\tt where ($A_1^1=a_1$ or $A_1^1=a'_1$) group-by $A_2^1$} is an analytic query involving a {\tt group-by} clause. Since in this example $\true(\mathcal{T})$ contains only one $A_2^1$-value, namely $b_1$, at most one pair $\langle b_1,[glb,lub]\rangle$ is expected in $\consans(\mathcal{AQ}_2)$. Referring again to Figure~\ref{fig:repairs}, we have:

\smallskip
$\bullet$ $\ans(\mathcal{AQ}_2^{[R_1]})=\ans(\mathcal{AQ}_2^{[R_2]})=\{(b_1,m_1+m'_1)\}$, and

$\bullet$ $\ans(\mathcal{AQ}_2^{[R_3]})=\ans(\mathcal{AQ}_2^{[R_4]})=\{(b_1, m_1+m''_1)\}$.

\smallskip\noindent
Assuming as above that $m_1$, $m'_1$ and $m''_1$ are positive and that $m'_1 \leq m''_1$, we obtain that $\consans(\mathcal{AQ}_2)= \{\langle b_1,[glb, lub]\rangle\}$ where $glb=m_1+m'_1$ and $lub=m_1+m''_1$. 

On the other hand, we have $Q_{\mathcal{AQ}_2}:~${\tt select $\mathbb{K},A_1^2,M_1$ from $T$~where ($A_1^1=a_1$ or $A_1^1=a'_1$)}. It can be seen from Figure~\ref{fig:repairs},  that $\consans(Q_{\mathcal{AQ}_2})=\{k_1k_2b_1m_1\}$. Therefore, $\consans^*(\mathcal{AQ}_2)=\{(b_1, m_1)\}$.

It is important to notice that this case illustrates that if $\langle x, [glb, lub]\rangle$ is in $\consans(\mathcal{AQ})$ and $x\,m$ is in $\consans^*(\mathcal{AQ})$, then it does not always hold that $glb \leq m \leq lub$. Indeed, for $m_1=1$, $m'_1=2$ and $m''_1=3$, we have $glb=3$, $lub=4$ showing that for $x=b_1$, $m \not\in [glb,lub]$.
\hfill$\Box$
}
\end{example}
\section{Computing Consistent Answers to Analytic Queries}\label{ans-computation}
In this section, we give algorithms for computing consistent answers to analytic queries, first when no {\tt group-by} clause is present, then when a {\tt group-by} clause is present. We also consider in this section the case of analytic queries involving a {\tt group-by-having} clause, whereas the introduction of the clause {\tt distinct} is the subject of the last sub-section. We emphasize again that the selection conditions in queries are assumed to be independent.
\subsection{Basic Result}
Before presenting our algorithms, we state important properties of repairs and analytic queries. To this end,  we introduce the following additional notation. Given a star-table $T$ and an analytic query $\mathcal{AQ}:~${\tt select [$X$]},$\,aggr(M_i)$~{\tt from $T$ where} $\Gamma$ {\tt [group by~$X$]}, let:

\smallskip
$\bullet$ $\Sigma(\mathcal{AQ})=\{\sigma \in m\_Chase(T)~|~(\mathbb{K} \cup M_i \cup sch(\Gamma) \subseteq sch(\sigma)) \wedge (\sigma(sch(\Gamma)) \cap Sat(\Gamma) \ne \emptyset)\}$

\smallskip
$\bullet$ $\Sigma^+(\mathcal{AQ})=\{\sigma \in \Sigma(\mathcal{AQ})~|~\sigma(sch(\Gamma)) \subseteq Sat(\Gamma)\}$.

\begin{proposition}\label{prop:indpdt}
Let $T$ be a star-table over universe $U$ and $\mathcal{AQ}:~${\tt select [$X$]},$\,aggr(M_i)$~{\tt from $T$ where} $\Gamma$ {\tt [group by~$X$]} an analytic query where $\Gamma$ is independent, i.e., $\Gamma=\Gamma(A_1) \wedge\ldots \wedge  \Gamma(A_k)$ where $sch(\Gamma)=A_1 \ldots A_k$. If $\Sigma(\mathcal{AQ}) \setminus \Sigma^+(\mathcal{AQ}) \ne \emptyset$, then there exist $R_1$ and $R_2$ in $\rep(T)$ such that:
\begin{enumerate}
\item  $(\forall \sigma \in \Sigma(\mathcal{AQ}) \setminus \Sigma^+(\mathcal{AQ}))(\exists t \in \tuples(\sigma) \cap \true(\mathcal{R}_1))(t.sch(\Gamma)\not\in Sat(\Gamma))$,
\item  $(\forall \sigma \in \Sigma(\mathcal{AQ}) \setminus \Sigma^+(\mathcal{AQ}))(\exists t \in \tuples(\sigma) \cap \true(\mathcal{R}_2))(t.sch(\Gamma)\in Sat(\Gamma))$.
\end{enumerate}
\end{proposition}
{\sc Proof.} See Appendix~\ref{append:prop-indpt}.\hfill$\Box$

\medskip\noindent
The following example shows that Proposition~\ref{prop:indpdt} does not always hold for non independent  selection conditions.
\begin{example}\label{ex:indpdt}
\rm{
Let $U=\{K_1,K_2,A_1,A_2,M_1\}$ and $FD=\{K_1 \to A_1, K_2 \to A_2, K_1K_2 \to M_1\}$, and the following tables

\smallskip\centerline{
\footnotesize{
\hfill
\begin{tabular}{c|cc}
$D_1$&$K_1$&$A_1$\\
\hline
&$k_1$&$10$\\
&$k'_1$&$-5$
\end{tabular}}
\hfill
\footnotesize{
\begin{tabular}{c|cc}
$D_2$&$K_2$&$A_2$\\
\hline
&$k_2$&$20$\\
&$k'_2$&$-1$\\
&$k'_2$&$30$\\
\end{tabular}}
\hfill
\footnotesize{
\begin{tabular}{c|ccc}
$F$&$K_1$&$K_2$&$M_1$\\
\hline
&$k_1$&$k_2$&$-10$\\
&$k_1$&$k'_2$&$2$\\
&$k'1$&$k'_2$&$-100$\\
\end{tabular}
\hfill
}}

\smallskip\noindent
Thus $m\_Chase(T)$ contains three m-tuples $\sigma_1$, $\sigma_2$, $\sigma_3$ defined over $U$ as follows:

\smallskip\centerline{
$\sigma_1=(k_1)(k_2)(10)(20)(-10)$, $\sigma_2=(k_1)(k'_2)(10)(-1\,30)(2)$, $\sigma_3=(k'_1)(k'_2)(-5)(-1\,30)(-100)$}

\smallskip\noindent
and $T$ has two repairs $R_1$ and $R_2$. Moreover, the tuples in $\true(\mathcal{R}_i)$ ($i=1,2$) whose schemas contain $\mathbb{K}=K_1K_2$, respectively denoted $R^{\mathbb{K}}_1$ and $R^{\mathbb{K}}_2$, are:

\smallskip
$R^{\mathbb{K}}_1=\{(k_1,k_2,10,20,-10), (k_1,k'_2,10,-1,2), (k'_1,k'_2,-5,-1,-100)\}$

\smallskip
$R^{\mathbb{K}}_2=\{(k_1,k_2,10,20,-10), (k_1,k'_2,10,30,2), (k'_1,k'_2,-5,30,-100)\}$.

\smallskip\noindent
For $\Gamma= (M_1 \leq A_1) \wedge (M_1 \geq 0 \Rightarrow A_2 < 0) \wedge (M_1 < -50 \Rightarrow A_2 >20)$ and $\mathcal{AQ}$ an analytic query involving $\Gamma$, we have $\Sigma(\mathcal{AQ})=\{\sigma_1, \sigma_2, \sigma_3\}$ and $\Sigma^+(\mathcal{AQ})=\{\sigma_1\}$. More precisely:
\begin{enumerate}
\item $\tuples(\sigma_1)=\{(k_1,k_2,10,20,-10)\}$ where $(10,20,-10) \in Sat(\Gamma)$. Thus,  $\tuples(\sigma_1) \subseteq Sat(\Gamma)$ holds.
\item $\tuples(\sigma_2)=\{(k_1,k'_2,10,-1,2),(k_1,k'_2,10,30,2)\}$, where $(10,-1,2) \in Sat (\Gamma)$ and $(10,30,2)\not\in Sat(\Gamma)$. Thus $\tuples(\sigma_2) \cap Sat(\Gamma)\ne \emptyset$ holds but not $\tuples(\sigma_2) \subseteq Sat(\Gamma)$.
\item $\tuples(\sigma_3)=\{(k'_1,k'_2,-5,-1,-100),(k'_1,k'_2,-5,30,-100)\}$,  where $(-5,30,-100)\in Sat(\Gamma)$ and $(-5,-1,-100)\not\in Sat(\Gamma)$. Thus $\tuples(\sigma_3) \cap Sat(\Gamma)\ne \emptyset$ holds but not $\tuples(\sigma_3) \subseteq Sat(\Gamma)$.
\end{enumerate}
On the other hand, it should be clear that $\Gamma$ is not independent, and that neither $\true(\mathcal{R}_1)$ nor $\true(\mathcal{R}_2)$ contains the two tuples $(k_1,k'_2,10,30,2)$ and $(k'_1,k'_2,-5,-1,-100)$, simply because they form a set that does not satisfy $K_2 \to A_2$. Thus Proposition~\ref{prop:indpdt}(1) does not hold in this example. Similarly, Proposition~\ref{prop:indpdt}(2) is not satisfied either because neither $\true(\mathcal{R}_1)$ nor $\true(\mathcal{R}_2)$ contains the two tuples $(k_1,k'_2,10,-1,2)$ and $(k'_1,k'_2,-5,30,-100)$, again because of $K_2 \to A_2$ in $FD$.
}
\hfill$\Box$
\end{example}

\algsetup{indent=1.5em}
\begin{algorithm}[th]
\caption{Procedure {\sf Compute\_Aggregate}\label{algo:proc-aggr}}
{\footnotesize
\begin{algorithmic}[1]
\REQUIRE
An m-table $S \subseteq m\_Chase(T)$ and  $\mathcal{AQ}:~${\tt select [}$X${\tt ],} $aggr(M_i)$~{\tt from $T$ where} $\Gamma$  {\tt [group by $X$]}

\ENSURE~\\
\begin{tabular}{ll}
$proc\_change$&\COMMENT a boolean variable\\
$proc\_min$&\COMMENT{meant to be equal to the lower bound of $\consans(\mathcal{AQ})$ if $proc\_change$ has value {\tt true}}\\
$proc\_max$&\COMMENT{meant to be equal to the upper bound of $\consans(\mathcal{AQ})$ if $proc\_change$ has value {\tt true}}\\
$proc\_ans^*$&\COMMENT{meant to be equal to $\consans^*(\mathcal{AQ})$ if $proc\_change$ has value {\tt true}}
\end{tabular}

\IF{$aggr \ne count$}
	\STATE{$proc\_min:=$\,{\tt -dummy} ; $proc\_max:=$\,{\tt +dummy} ; $proc\_ans^*:=$\,{\tt -dummy}}\label{line:init}\\
	\COMMENT{{\tt -dummy} and {\tt +dummy} are values not in $adom(M_i)$ such that for every $m$ in $adom(M_i)$,}\\
	\COMMENT{$aggr(${\tt -dummy}$, m)=aggr(${\tt +dummy}$,m)=m$ and $\min(${\tt -dummy}$,m)=\max(${\tt +dummy}$,m)=m$}
\ENDIF
\IF{$aggr= count$}
    \STATE{$proc\_min:=0$ ; $proc\_max:=0$ ; $proc\_ans^*:=0$}
\ENDIF
\STATE{$proc\_change := {\tt false}$}
\FORALL{$\sigma$ in $S$}\label{line:proc-loop-start}
    \IF{$\mathbb{K} \cup sch(\mathcal{AQ}) \subseteq sch(\sigma)$ and $|\tuples(\sigma(X))|=1$ and $\tuples(\sigma(sch(\Gamma))) \cap Sat(\Gamma) \ne \emptyset$}\label{line:test-schema}
        \IF{$aggr \ne count$}\label{min-max-sigma-start}
		\IF{$M_i \in sch(\Gamma)$}
			\STATE{$min_\sigma:=\min(\tuples(\sigma(M_i)) \cap Sat(\Gamma(M_i)))$ ; $max_\sigma:=\max(\tuples(\sigma(M_i)) \cap Sat(\Gamma(M_i)))$}
		\ELSE
			\STATE{$min_\sigma:=\min(\tuples(\sigma(M_i)))$ ; $max_\sigma:=\max(\tuples(\sigma(M_i)))$}
		\ENDIF
	\ENDIF\label{min-max-sigma-end}
		\IF{$\tuples(\sigma(sch(\Gamma))) \subseteq Sat(\Gamma)$}
			\STATE{$proc\_change := {\tt true}$}\label{line:proc-check-change-1}
			\IF{$aggr = count$}
				\STATE{$proc\_min := proc\_min +1$}\label{Zcount-min}
			\ELSE
				\STATE{$proc\_min:=aggr(min_\sigma, proc\_min)$ ; $proc\_max :=aggr(max_\sigma, proc\_max)$}\label{line:set-max1}\label{Ymin-max}
			\ENDIF
			\IF{$\tuples(\sigma(M_i))=\{m_0\}$}
				\STATE{$proc\_ans^*:= aggr(proc\_ans^*, m_0)$}\label{line:setans*}
			\ENDIF
		\ELSE
			\IF{$aggr=min$}
				\STATE{$proc\_min:= \min(min_\sigma, proc\_min)$}\label{Xmin}
			\ENDIF
			\IF{$aggr=max$}
				\STATE{$proc\_max:= \max(max_\sigma, proc\_max)$}\label{Zmax}
			\ENDIF
			\IF{$aggr=count$}
				\STATE{$proc\_max := proc\_max +1$}\label{Zcount}
			\ENDIF
			\IF{$aggr=sum$}
				\IF{$min_\sigma <0$}\label{test-minsum}
					   \STATE{$proc\_min:= proc\_min+min_\sigma$}\label{Xsum}
			\ENDIF
			\IF{$max_\sigma > 0$}\label{test-maxsum}
				\STATE{$proc\_max:= proc\_max+max_\sigma$}\label{Zsum}
			\ENDIF
		\ENDIF
	\ENDIF
    \ENDIF
\ENDFOR\label{line:proc-loop-end}
\RETURN{$proc\_change$, $proc\_min$, $proc\_max$, $proc\_ans^*$}
\end{algorithmic}
}
\end{algorithm}

\noindent
In the following two sections~\ref{subsec:no-group-by} and \ref{subsec:with-group-by}, we successively introduce our algorithms for computing consistent answers to analytic queries, depending on whether they involve or not a {\tt group-by} clause. In each case, the main algorithm relies on a procedure, called {\sf Compute\_aggregate} and shown in Algorithm~\ref{algo:proc-aggr}, whose role is to scan (either entirely or partially) the m-table $m\_Chase(T)$ and return values that will appear in the answers. 

\subsection{Analytic Queries with no {\tt{Group-by}} Clause}\label{subsec:no-group-by}
If the query involves no {\tt group-by} clause, as shown in Algorithm~\ref{algo:cons-answer-no-groupby},  the procedure {\sf Compute\_aggregate} of Algorithm~\ref{algo:proc-aggr} is called to scan the whole m-table $m\_Chase(T)$. When running this call, line~\ref{line:proc-call-no-groupby} in Algorithm~\ref{algo:cons-answer-no-groupby}, the main loop lines~\ref{line:proc-loop-start}-\ref{line:proc-loop-end} of Algorithm~\ref{algo:proc-aggr} scans $m\_Chase(T)$ and computes values appearing in the answers. The following proposition shows that Algorithm~\ref{algo:cons-answer-no-groupby} is correct, except when $aggr=sum$. In other words, if $aggr\ne sum$ Algorithm~\ref{algo:cons-answer-no-groupby} returns $\consans(\mathcal{AQ})$. Moreover, it is also shown that $\consans^*(\mathcal{AQ})$ is correctly computed by Algorithm~\ref{algo:cons-answer-no-groupby} in any case.

\algsetup{indent=1.5em}
\begin{algorithm}[h]
\caption{Consistent answer to analytic queries with no {\tt group-by} clause\label{algo:cons-answer-no-groupby}}
{\footnotesize
\begin{algorithmic}[1]
\REQUIRE
The m-table $m\_Chase(T)$ and  $\mathcal{AQ}: ~${\tt select} $aggr(M_i)$~{\tt from $T$ where} $\Gamma$
\ENSURE~\\
\begin{tabular}{ll}
$[min\_ans, max\_ans]$&\COMMENT{meant to be equal to $\consans(\mathcal{AQ})$}\\
$ans^*$&\COMMENT{meant to be equal to $\consans^*(\mathcal{AQ})$}
\end{tabular}

\STATE{Call procedure {\sf Compute\_aggregate} with input parameters $m\_Chase(T)$, $\mathcal{AQ}$ and\\ with output parameters $change\_min\_max$, $min\_ans$, $max\_ans$, $ans^*$}\label{line:proc-call-no-groupby}
\IF{$change\_min\_max = {\tt true}$}\label{line:final-test}
	\RETURN{$([min\_ans, max\_ans]$, $ans^*$)}
\ELSE
	\IF{$aggr \ne count$}
		\RETURN{$({\tt NULL}, {\tt NULL})$}
	\ENDIF
	\IF{$aggr = count$}
		\RETURN{($[0, 0]$, $0$)}
	\ENDIF
\ENDIF

\end{algorithmic}
}
\end{algorithm}

\begin{proposition}\label{prop:compute-cons-ans-no-group-by}
Let $T$ be a star-table over universe $U$ and let $\mathcal{AQ} :~${\tt select} $aggr(M_i)$~{\tt from $T$ where} $\Gamma$ be an  analytic query with no {\tt group-by} clause. Then we have:
\begin{itemize}
\item 
If $aggr$ is $min$, $max$ or $count$, then $\consans(\mathcal{AQ})=[min\_ans,max\_ans]$ where $min\_ans$ and $max\_ans$ are returned by  Algorithm~\ref{algo:cons-answer-no-groupby}.
\item 
If $aggr=sum$ and if $\consans(\mathcal{AQ})=[glb, lub]$, then  $min\_ans$ and $max\_ans$ as returned by  Algorithm~\ref{algo:cons-answer-no-groupby} satisfy that $min\_ans \leq glb$ and $max\_ans \geq lub$.
\\
Moreover, if {\em for every $m \in adom(M_i)$, $m \geq 0$}, then $\consans(\mathcal{AQ})=[min\_ans,max\_ans]$.
\item 
For every aggregate operator and every selection condition $\Gamma$,  $ans^*$ as returned by  Algorithm~\ref{algo:cons-answer-no-groupby} is equal to $\consans^*(\mathcal{AQ})$.
\end{itemize}
\end{proposition}
{\sc Proof.} See Appendix~\ref{append:compute-cons-ans-no-group-by}.
\hfill$\Box$

\medskip\noindent
In the following example we show that Algorithm~\ref{algo:cons-answer-no-groupby} may fail to return exact values for $glb$ and $lub$ when the aggregate operator is $sum$, operating on positive values and negative values.
\begin{example}\label{ex:cons-answer-no-groupby}
\rm{
As in Example~\ref{ex:indpdt}, we consider $U=\{K_1,K_2,A_1,A_2,M_1\}$ and $FD=\{K_1 \to A_1, K_2 \to A_2, K_1K_2 \to M_1\}$, but here, with the following tables

\begin{center}
\footnotesize{
\begin{tabular}{c|cc}
$D_1$&$K_1$&$A_1$\\
\hline
&$k_1$&$10$\\
&$k'_1$&$-15$\\
&$k'_1$&$20$\\
\end{tabular}}
\qquad\qquad
\footnotesize{
\begin{tabular}{c|cc}
$D_2$&$K_2$&$A_2$\\
\hline
&$k_2$&$2$\\
&$k'_2$&$0$\\
&$k'_2$&$3$\\
\end{tabular}}
\qquad\qquad
\footnotesize{
\begin{tabular}{c|ccc}
$F$&$K_1$&$K_2$&$M_1$\\
\hline
&$k_1$&$k_2$&$30$\\
&$k'_1$&$k'_2$&$-10$\\
&$k_1$&$k'_2$&$100$\\
\end{tabular}
}
\end{center}
Thus $m\_Chase(T)$ contains three m-tuples $\sigma_1$, $\sigma_2$, $\sigma_3$ defined over $U$ as follows:

\smallskip\centerline{
$\sigma_1=(k_1)(k_2)(10)(2)(30)$, $\sigma_2=(k'_1)(k'_2)(-15\,20)(0\,3)(-10)$, $\sigma_3=(k_1)(k'_2)(10)(0\,3)(100)$}

\smallskip\noindent
and $T$ has four repairs denoted by $R_i$ for $i=1, \ldots ,4$, whose sets $R^\mathbb{K}_i$ of tuples over $U$ are defined by:

\smallskip
$R^\mathbb{K}_1=\{(k_1,k_2,10,2,30), (k'_1,k'_2,-15,0,-10), (k_1,k'_2,10,0,100)\}$

\smallskip
$R^\mathbb{K}_2=\{(k_1,k_2,10,2,30), (k'_1,k'_2,-15,3,-10), (k_1,k'_2,10,3,100)\}$

\smallskip
$R^\mathbb{K}_3=\{(k_1,k_2,10,2,30), (k'_1,k'_2,20,0,-10), (k_1,k'_2,10,0,100)\}$

\smallskip
$R^\mathbb{K}_4=\{(k_1,k_2,10,2,30), (k'_1,k'_2,20,3,-10), (k_1,k'_2,10,3,100)\}$.

\smallskip\noindent
For $\Gamma= (A_1 > 0) \wedge (A_2 > 0)$ and $\mathcal{AQ}:$ {\tt select $aggr(M_1)$~from~}$T$~{\tt where}~$\Gamma$, we have  $\Sigma(\mathcal{AQ})=\{\sigma_1, \sigma_2, \sigma_3\}$ and $\Sigma^+(\mathcal{AQ})=\{\sigma_1\}$, because $\tuples(\sigma_1(A_1A_2)) \subseteq Sat(\Gamma)$ holds,  $\tuples(\sigma_2(A_1A_2)) \cap Sat(\Gamma)\ne \emptyset$ holds but not $\tuples(\sigma_2(A_1A_2)) \subseteq Sat(\Gamma)$, and $\tuples(\sigma_3(A_1A_2)) \cap Sat(\Gamma)\ne \emptyset$ holds but not $\tuples(\sigma_3(A_1A_2)) \subseteq Sat(\Gamma)$.
On the other hand, $\Gamma$ is clearly independent.
\begin{itemize}
\item If $aggr=min$ then, for every $i=1,\ldots , 3$, $\ans(\mathcal{AQ}^{[R_i]})= 30$ and $\ans(\mathcal{AQ}^{[R_4]})= -10$. Thus, $\consans(\mathcal{AQ})=[-10, 30]$.\\
When running the call of the procedure {\sf Compute\_aggregate} in Algorithm~\ref{algo:cons-answer-no-groupby}, $min\_ans$ is first set to $30$ when processing $\sigma_1$, then to $\min(\{30, \min\{-10,30\}\})=-10$ when processing $\sigma_2$, and then to $\min(\{-10, \min\{100,-10\}\})=-10$ when processing $\sigma_3$. Similarly, $max\_ans$ is first set to $30$ when processing $\sigma_1$, then to $\max(\{30, \min\{-10,30\}\})=30$ when processing $\sigma_2$, and then to $\max(\{30, \min\{100,30\}\})=30$ when processing $\sigma_3$. Hence, Algorithm~\ref{algo:cons-answer-no-groupby}  returns $[-10, 30]$ as expected. 
\item
Similarly, if $aggr=max$, we have $\consans(\mathcal{AQ})=[30, 100]$, which is also returned by Algorithm~\ref{algo:cons-answer-no-groupby}.
\item If $aggr=count$ then we have $\ans(\mathcal{AQ}^{[R_1]})= 1$, $\ans(\mathcal{AQ}^{[R_2]})= 2$, $\ans(\mathcal{AQ}^{[R_3]})= 1$ and $\ans(\mathcal{AQ}^{[R_4]})= 3$. Thus, $\consans(\mathcal{AQ})=[1,3]$.\\
When running the call of the procedure {\sf Compute\_aggregate} in Algorithm~\ref{algo:cons-answer-no-groupby}, $min\_ans$ is first set to $1$ when processing $\sigma_1$ and then unchanged when processing $\sigma_2$ and $\sigma_3$. Moreover, $max\_ans$ is increased by $1$ for each m-tuple $\sigma_1$, $\sigma_2$ and $\sigma_3$. Thus Algorithm~\ref{algo:cons-answer-no-groupby} returns $[1, 3]$ as expected.
\item If $aggr=sum$ then we have $\ans(\mathcal{AQ}^{[R_1]})= 30$, $\ans(\mathcal{AQ}^{[R_2]})= 130$, $\ans(\mathcal{AQ}^{[R_3]})= 30$ and $\ans(\mathcal{AQ}^{[R_4]})= 120$. Thus, $\consans(\mathcal{AQ})=[30, 130]$.\\
When running the call of the procedure {\sf Compute\_aggregate} in Algorithm~\ref{algo:cons-answer-no-groupby}, $min\_ans$ is successively set to $30$, $(30-10)=20$ when processing $\sigma_1$ and then $\sigma_2$, and unchanged when processing $\sigma_3$. On the other hand, $max\_ans$ is successively set to $30$ when processing $\sigma_1$, unchanged when processing $\sigma_2$ and set to $(30+100)=130$ when processing $\sigma_3$. Thus the call of the procedure {\sf Compute\_aggregate} in Algorithm~\ref{algo:cons-answer-no-groupby} returns $[20, 130]$.
\item
On the other hand, assuming that the second tuple in $F$ is $(k'_1k'_2,10)$ (instead of $(k'_1k'_2,-10)$), then $\consans(\mathcal{AQ})=[30, 140]$, because  $\ans(\mathcal{AQ}^{[R_i]})$ is $30$ if $i=1$ or $i= 3$, $130$ if $i=2$ and $140$ if $i=4$. It can be seen that when running the call of the procedure {\sf Compute\_aggregate} in Algorithm~\ref{algo:cons-answer-no-groupby}, we obtain $min\_ans=30$ and $max\_ans=140$.
\hfill$\Box$
\end{itemize}
}
\end{example}

\algsetup{indent=1.5em}
\begin{algorithm}[t]
\caption{Consistent answer to analytic queries with a {\tt group-by} clause\label{algo:cons-answer-groupby}}
{\footnotesize
\begin{algorithmic}[1]
\REQUIRE
The m-table $m\_Chase(T)$ and  $\mathcal{AQ}: ~${\tt select} $X,aggr(M_i)$~{\tt from $T$ where} $\Gamma$ {\tt group by $X$}

\ENSURE~\\
\begin{tabular}{ll}
$Cons\_Ans$: a set of pairs $\langle x,[min\_ans, max\_ans]\rangle$&\COMMENT{meant to be equal to $\consans(\mathcal{AQ})$}\\
$Ans^*$: a set of pairs $(x,ans^*)$&\COMMENT{meant to be equal to $\consans^*(\mathcal{AQ})$}
\end{tabular}

\STATE{$Cons\_Ans := \emptyset$ ; $Ans^*:=\emptyset$ ; $Temp:= \emptyset$}

\FORALL{$\sigma$ in $m\_Chase(T)$}\label{line:first-loop-start}
	\IF{$\mathbb{K} \cup sch(Q) \subseteq sch(\sigma)$}
                 \IF{$(Y \to B$ in $FD$ and $YB \subseteq X \Rightarrow |\tuples(\sigma(B))|=1)$ and $(\tuples(\sigma(sch(\Gamma))) \cap Sat(\Gamma) \ne \emptyset$)}
                 	\STATE{$Temp := Temp \cup \{\sigma\}$}
		\ENDIF
	\ENDIF
\ENDFOR\label{line:first-loop-end}
\FORALL{$X$-value $x$ occurring  in $Temp$}\label{line:inner-loop-start}
	\STATE{$Temp(x):=\{\sigma \in Temp~|~x \in \tuples(\sigma(X))\}$}
	\STATE{Call procedure {\sf Compute\_aggregate} with input parameters $Temp(x)$, $\mathcal{AQ}$ and\\
            with output parameters $change\_min\_max$, $min\_ans$, $max\_ans$, $ans^*$}\label{line:call-proc-with-groupby}
	\IF{$change\_min\_max = {\tt true}$}
		\STATE{$Cons\_Ans := Cons\_Ans \cup \{\langle x, [min\_ans, max\_ans]\rangle\}$}
		\STATE{$Ans^*:= Ans^*\cup \{(x,ans^*)\}$}
	\ENDIF
\ENDFOR\label{line:inner-loop-end}
\RETURN{$(Cons\_Ans, Ans^*$)}
\end{algorithmic}
}
\end{algorithm}

\subsection{Analytic Queries with a {\tt Group-by} Clause}\label{subsec:with-group-by}
As in the case of analytic queries with no {\tt group-by} clause, when the query involves a {\tt group-by} clause, the computation of the consistent answers also involves a call of the procedure {\sf Compute\_aggregate}, as shown line~\ref{line:call-proc-with-groupby} of Algorithm~\ref{algo:cons-answer-groupby}. This algorithm works as follows: a first scan of $m\_Chase(T)$, lines~\ref{line:first-loop-start}-\ref{line:first-loop-end} retrieves in the set $Temp$ all relevant m-tuples $\sigma$, namely, such that $sch(\sigma)$ contains all attributes occurring in $\mathcal{AQ}$ and such that the $X$-values are consistent and may be associated with a tuple over $sch(\Gamma)$ satisfying the condition $\Gamma$. Then, a subsequent loop lines~\ref{line:inner-loop-start}-\ref{line:inner-loop-end} operates a call of the procedure {\sf Compute\_aggregate} for each $X$-value $x$  occurring in the m-tuples of the set $Temp$. For each such call, Algorithm~\ref{algo:proc-aggr} scans the subset  $Temp(x)$ of $Temp$ and computes the corresponding aggregate values appearing in the answers associated with $x$.

The following proposition shows that Algorithm~\ref{algo:cons-answer-groupby} is correct, except when $aggr=sum$. In other words, if $aggr\ne sum$ Algorithm~\ref{algo:cons-answer-groupby} returns $\consans(\mathcal{AQ})$. Moreover,  it is also shown that $\consans^*(\mathcal{AQ})$ is correctly computed by Algorithm~\ref{algo:cons-answer-groupby} in any case.
\begin{proposition}\label{prop:compute-cons-ans-group-by}
Let $T$ be a star-table over universe $U$ and $\mathcal{AQ}$ an  analytic query with a {\tt group-by} clause.
\begin{itemize}
\item 
If $aggr=min$, $aggr=max$ or $aggr=count$, then $Cons\_Ans$ as returned by Algorithm~\ref{algo:cons-answer-groupby} is equal to $\consans(\mathcal{AQ})$.
\item 
If $aggr=sum$, then  $\langle x, [min\_ans, max\_ans]\rangle$ is in $Cons\_Ans$ as returned by  Algorithm~\ref{algo:cons-answer-groupby} if and only if $\consans(\mathcal{AQ})$ contains $\langle x, [lub, glb]\rangle$ such that $min\_ans \leq glb$ and $max\_ans \geq lub$.
\\
Moreover, if {\em for every $m \in adom(M_i)$, $m \geq 0$}, then  $Cons\_Ans = \consans(\mathcal{AQ})$.
\item 
For every aggregate operator and every selection condition $\Gamma$,  $Ans^*$ as returned by  Algorithm~\ref{algo:cons-answer-groupby} is equal to $\consans^*(\mathcal{AQ})$.
\end{itemize}
\end{proposition}
{\sc Proof.} First, the loop lines~\ref{line:first-loop-start}-\ref{line:first-loop-end} scans $m\_Chase(T)$ and collects in the set $Temp$ the only m-tuples necessary to compute the consistent answer. Indeed, the collected m-tuples are all m-tuples $\sigma$ such that: they are defined over a super-set of $\mathbb{K} \cup sch(\sigma)$, their $X$-component contains consistent tuples, and their m-tuples over $sch(\Gamma)$ contain at least one tuple of $Sat(\Gamma)$. This is so because any m-tuple in $m\_Chase(T)$ not satisfying the above conditions can not contribute to the consistent answer. Then, in the next steps for every collected $X$-value, its associated aggregate value is evaluated as done in Algorithm~\ref{algo:cons-answer-no-groupby}. As a consequence of Proposition~\ref{prop:compute-cons-ans-no-group-by}, for $x$ in $Temp$, the loop lines~\ref{line:inner-loop-start}-\ref{line:inner-loop-end} in Algorithm~\ref{algo:cons-answer-groupby} generates correct answers. The proof is therefore complete.\hfill$\Box$

\medskip\noindent
We emphasize that Algorithm~\ref{algo:cons-answer-no-groupby} and Algorithm~\ref{algo:cons-answer-groupby} show that the consistent answers to a given analytic query $\mathcal{AQ}$, with or without a {\tt group-by} clause, are computed in linear time in the size of $m\_Chase(T)$. Recalling that the computation of $m\_Chase(T)$ has been shown earlier to be quasi-linear  with respect to the size of $T$ (or in $\mathcal{O}(W\cdot \log(W))$ where $W$ is the size of the data warehouse), it turns out that $\consans(\mathcal{AQ})$ and $\consans^*(\mathcal{AQ})$ can be computed in quasi-linear time with respect to the size of the data warehouse.
\subsection{Analytic Queries with a {\tt group-by-having} Clause}\label{sec:groupbyhaving}
Another important feature of our approach is that analytic queries with a {\tt group-by-having} clause can be handled in our framework. This is so because, intuitively a {\tt having} clause specifies through a boolean expression, which groupings should be in the answer. For example, in the context of Example~\ref{ex:star}, let $\mathcal{AQ}_3$ be the following query:

\smallskip\centerline{
\begin{tabular}{ll}
$\mathcal{AQ}_3:  {\tt select}$~$A_2^1$, $sum(M_1)$ {\tt from} $T$&{\tt where ($A_1^1=a_1$ or $A_1^1=a'_1$)}\\
&{\tt group by $A_2^1$ having ($max(M_1) < 10$)}
\end{tabular}} 

\smallskip\noindent
In this case, $\mathcal{AQ}_3$ can be associated with the analytic query $\mathcal{AQ}'_3$ defined by:

\smallskip\centerline{
$\mathcal{AQ}'_3: {\tt select}$~$A_2^1$, $sum(M_1)$, $max(M_1)$ {\tt from} $T$ {\tt where ($A_1^1=a_1$ or $A_1^1=a'_1$) {\tt group by $A_2^1$}}}

\smallskip\noindent
whose consistent answer $\consans(\mathcal{AQ}'_3)$ is computed as described above. Then, considering the triples of the form $\langle a_2,[glb_1, lub_1],[glb_2, lub_2]\rangle$ in $\consans(\mathcal{AQ}'_3)$, all pairs $\langle a_2,[glb_1, lub_1]\rangle$ such that $lub_2 < 10$ are inserted in $\consans(\mathcal{AQ}_3)$. In other words, we have:

\smallskip\centerline{
$\consans(\mathcal{AQ}_3)=\{\langle a_2,[glb_1, lub_1]\rangle~|~\langle a_2,[glb_1, lub_1],[glb_2, lub_2]\rangle\in \consans(\mathcal{AQ}'_3) \wedge lub_2 < 10\}$.}

\smallskip\noindent
More precisely,  applying Proposition~\ref{prop:compute-cons-ans-group-by} to the star-table $T$ in Figure~\ref{fig:m-chase}, we obtain the following:
\begin{itemize}
    \item If we assume that $m_1=5$, $m'_1=5$ and $m''_1=15$, then $\consans(\mathcal{AQ}'_3)=\{\langle a_2,[10,20], [5,15]\rangle\}$ and so, $\consans(\mathcal{AQ}_3)$ is empty. Notice incidentally that it is easily seen from Figure~\ref{fig:repairs} that there exist repairs $R$ in $\rep(T)$ for which $\ans(\mathcal{AQ}_3'^{[R]}) = \{\langle a_2,20, 15\rangle\}$, implying that $\ans(\mathcal{AQ}_3^{[R]}) =\emptyset$.
    \item If we assume that $m_1=5$, $m'_1=5$ and $m''_1=8$, then $\consans(\mathcal{AQ}'_3)=\{\langle a_2,[10,13], [5,8]\rangle\}$ and so, $\langle a_2,[10,13] \rangle$ is in $\consans(\mathcal{AQ}_3)$. In this case, it is easily seen from Figure~\ref{fig:repairs} that for every repair $R$ in $\rep(T)$, $\ans(\mathcal{AQ}_3'^{[R]})$ contains a pair $\langle a_2, \mu, \nu \rangle$ where $10 \leq \mu \leq 13$ and $\nu < 10$, implying that $\langle a_2, [10,13] \rangle$ is in $\consans(\mathcal{AQ}_3)$.
\end{itemize}
In the light of this example, we argue that our approach can deal with analytic queries involving a {\tt having} clause with condition $\Upsilon$, under the following restrictions:
\begin{enumerate}
    \item $\Upsilon$ is a {\em conjunctive} boolean expression built up from atoms of the form $aggr(M_i)\, \theta\, \alpha$ where $M_i$ is a measure attribute in $\mathbb{M}$, $\theta$ is a comparison predicate in $\{<,\, \leq,\, >,\, \geq\}$ and $\alpha$ is a number.
    \item For every aggregate term in $\mathcal{AQ}$ of the form  $sum(M_i)$, then all $M_i$-values are positive.
    \end{enumerate}
Calling $aggr(M_i)$ an {\em aggregate term}, the first item above implies that $\Upsilon$ can be written as $\Upsilon_{1} \wedge \ldots \wedge \Upsilon_{h}$ where for every $p=1, \ldots , h$, $\Upsilon_{p}$ is the conjunction of all atoms in $\Upsilon$ involving the same aggregate term.
Moreover, for every aggregate term $\lambda_p$ occurring in $\Upsilon$, given a number $\alpha$, let $\Upsilon_p^{[\alpha]}$ be the expression obtained by substituting in $\Upsilon_{p}$ every occurrence of $\lambda_p$ with $\alpha$. Then, it turns out that the set $Sat(\Upsilon_{p})=\{\alpha~|~\Upsilon_p^{[\alpha]}$ evaluates to {\tt true}$\}$ is an interval.

Based on this important remark, the following proposition characterizes the consistent answers of analytic queries involving a {\tt group-by-having} clause.
\begin{proposition}\label{prop:having}
Let $T$ be a star-table over universe $U$ and $\mathcal{AQ}$ an  analytic query with a {\tt group-by-having} clause defined by

\smallskip
$\mathcal{AQ}:  {\tt select}~X$, $aggr(M_i)$ {\tt from} $T$ {\tt where $\Gamma$} {\tt group by $X$ having $\Upsilon$}

\smallskip\noindent
such that the above two restrictions regarding $\Upsilon$ are met. If $\Upsilon=\Upsilon_{1} \wedge \ldots \wedge \Upsilon_{h}$, let $\mathcal{AQ}'$ be the following analytic query with no {\tt having} clause:

\smallskip
$\mathcal{AQ}':  {\tt select}~X$, $aggr(M_i)$, $aggr_1(M_{i_1})$, $\ldots$ , $aggr_h(M_{i_h})$ {\tt from} $T$ {\tt where $\Gamma$} {\tt group by $X$}

\smallskip\noindent
The tuple $\langle x, [glb, lub] \rangle$ belongs to $\consans(\mathcal{AQ})$ if and only if there exists a tuple of the form $\langle x, [glb,lub],$ $[glb_1,lub_1], \ldots ,[glb_h,lub_h] \rangle$ in $\consans(\mathcal{AQ}')$ such that $[glb_p,lub_p] \subseteq Sat(\Upsilon_{p})$ holds for every $p=1, \ldots ,h$.
\end{proposition}
{\sc Proof.} Let $\langle x, [glb,lub],$ $[glb_1,lub_1], \ldots ,[glb_h,lub_h] \rangle$ in $\consans(\mathcal{AQ}')$ such that for every $p=1, \ldots ,h$, $[glb_p,lub_p] \subseteq Sat(\Upsilon_{p})$. Given $R$ in $\rep(T)$, $\ans(\mathcal{AQ}'^{[R]})$ contains a tuple $\langle x,\mu, \mu_1, \ldots ,\mu_p \rangle$ such that $\mu \in [glb, lub]$ and $\mu_p \in [glb_p, lub_p]$ for $p=1,\ldots ,h$ hold. Hence, for $p=1,\ldots ,h$, $\mu_p$ is in $Sat(\Upsilon_p)$, implying that $\langle x, [glb,lub]\rangle$ is in $\consans(\mathcal{AQ})$.

Conversely, let $\langle x, [glb, lub] \rangle$ be in $\consans(\mathcal{AQ})$ and let $p$ such that $[glb_p,lub_p] \subseteq Sat(\Upsilon_{p})$ does not hold. Since $Sat(\Upsilon_{p})$ is an interval, this implies that $glb_p \not\in Sat(\Upsilon_p)$ or that $lub_p \not\in Sat(\Upsilon_p)$. Since there exists a repair $R_{min}$, respectively a repair $R_{max}$, such that $aggr_p(M_{i_p})=glb_p$, respectively $aggr_p(M_{i_p})=lub_p$, it turns out that there exists at least one repair $R$ such that $ans(\mathcal{AQ}'^{[R]})$ contains a tuple $\langle x, \mu, \mu_1, \ldots, \mu_h\rangle$ such that $\mu_p \not\in Sat(\Upsilon_p)$. Thus $\langle x, \mu \rangle$ is not in $ans(\mathcal{AQ}^{[R]})$, which implies that $\langle x, [glb, lub] \rangle$ is {\em not} in $\consans(\mathcal{AQ})$; a contradiction which completes the proof.\hfill$\Box$

\medskip\noindent
It should be clear from Proposition~\ref{prop:compute-cons-ans-group-by} and Proposition~\ref{prop:having} that, under the restrictions regarding $\Upsilon$ earlier stated, the consistent answer of an analytic query involving a {\tt group-by-having} clause can be computed in quasi-linear time with respect to the size of the data warehouse.

\smallskip
To end this section, we illustrate why in our approach, the sets $Sat(\Upsilon_p)$ ($p=1, \ldots ,h$) must be intervals. In the context of our earlier example, consider the query $\mathcal{AQ}_3$ with a {\tt having} clause defined by $\Upsilon ={\tt (}max(M_1) < 50~{\tt or}~max(M_1) > 100{\tt )}$, instead of $max(M_1) < 10$. Notice that in this case, the above restrictions are not satisfied, because $\Upsilon$ is not a conjunction, and $Sat(\Upsilon)$ is clearly not an interval. 

Let $m_1=20$, $m'_1=10$ and $m''_1=150$ and let $\langle a_2, [30, 170],[20,150] \rangle$ in $\consans(\mathcal{AQ}'_3)$, implying that the values of $max(M_1)$ range between $20$ and $150$, when varying the repair. Although there are repairs in which the maximum is $20$ or $150$, thus satisfying the {\tt having} condition, it is not possible to conclude that for {\em every} repair, the maximum $M_1$-value satisfies this condition: it could for example be that for a repair, $max(M_1)$ is equal to $80$. It is thus impossible to certify that $\langle a_2, [glb, lub]\rangle$ is in the consistent answer of $\mathcal{AQ}_3$.
\subsection{Handling {\tt distinct} Clauses}
The clause  {\tt distinct} is used in SQL queries to avoid the processing of duplicates. The most frequent cases in which this key word is used are (1)  associated with a {\tt select} clause, in which case repeated values are not displayed to the user, and (2) associated with aggregation operator $count$, in which case every value is counted only once.

In our approach, it turns out that the first case does not raise specific difficulties, since as usual, this case simply necessitates to check whether a given value has been displayed earlier or not. However, case (2) is much more involved because in order to compute the  interval in the consistent answer to a clause of the form {\tt select $count$(distinct $M_i$)...}, it must be known {\em for each repair} which $M_i$-values have been counted, and this is clearly  not  tractable.
In order to show how to handle  the {\tt distinct} clause associated with the operator $count$, we focus on analytic queries of the form $\mathcal{AQ}:$~{\tt select} $count({\tt distinct}~M_i)$ {\tt from} $T$ {\tt where} $\Gamma$,  because the presence of a {\tt group-by} clause does not raise any particular difficulties.

\smallskip
Similarly to the general case of the aggregate $sum$, an effective way of computing the interval $[glb, lub]$ of $\consans(\mathcal{AQ})$ is currently unknown to the authors. This is so because, as mentioned above, repairs must be considered separately in order to compute $lub$ and $glb$, which cannot be done when running  Algorithm~\ref{algo:proc-aggr}. Consequently, instead of computing the exact values of $lub$ and $glb$, we propose a  {\em tractable approximation} of $\consans(\mathcal{AQ})$, easily implemented based on  Algorithm~\ref{algo:proc-aggr} and  Algorithm~\ref{algo:cons-answer-no-groupby}.

To this end, we first notice that  the values to be counted for determining $lub$ are obtained by picking exactly one $M_i$-value in $\tuples(\sigma(M_i))$ for every $\sigma$ in $\Sigma(\mathcal{AQ})$. Therefore, the set resulting from such a choice can not contain $(a)$ more distinct values than the number of m-tuples in $\Sigma(\mathcal{AQ})$ and $(b)$ more distinct values than the number of distinct values in the union of all sets $\tuples(\sigma(M_i))$ where $\sigma$ is in $\Sigma(\mathcal{AQ})$. This can be formally stated as $(a)$ $lub \leq |\Sigma(\mathcal{AQ})|$ and $(b)$ $lub \leq |\bigcup_{\sigma \in \Sigma(\mathcal{AQ})}\tuples(\sigma(M_i))|$, which we write as

\smallskip\centerline{
 $lub \leq \inf \left( |\Sigma(\mathcal{AQ})|, |\bigcup_{\sigma \in \Sigma(\mathcal{AQ})}\tuples(\sigma(M_i))|\right)$.}

\smallskip\noindent
If $count\_max$ denotes the infimum defined above, we emphasize that Algorithm~\ref{algo:proc-aggr} can be modified so as to compute $count\_max$. Indeed,  it is easy to see that $|\Sigma(\mathcal{AQ})|$ is equal to the value of $proc\_max$ returned by Algorithm~\ref{algo:proc-aggr} in the case of  $count$. Moreover, the computation of $|\bigcup_{\sigma \in \Sigma(\mathcal{AQ})}\tuples(\sigma(M_i))|$ can be implemented in Algorithm~\ref{algo:proc-aggr} in a similar way, at the cost of an extra storage to ensure that every value in the union is counted only once. Then, once the two cardinalities are known, $count\_max$ can easily be computed and returned by modifying Algorithm~\ref{algo:cons-answer-no-groupby} accordingly.

\smallskip
Regarding the approximation of $glb$, we recall from Algorithm~\ref{algo:proc-aggr} that the values to be counted for determining $glb$ are distinct values obtained by picking exactly one $M_i$-value in $\tuples(\sigma(M_i))$ for every $\sigma$ in $\Sigma^+(\mathcal{AQ})$. Thus, if $\Sigma^+(\mathcal{AQ})=\emptyset$, $glb=0$, and otherwise, a trivial lower bound for $glb$ is $1$, because all sets $\tuples(\sigma(M_i))$ for $\sigma$ in $\Sigma^+(\mathcal{AQ})$, are nonempty.

To find a more accurate lower bound,  we notice that any choice of one value in $\tuples(\sigma(M_i))$ for every $\sigma$ in $\Sigma^+(\mathcal{AQ})$, concerns {\em at least} as many distinct $M_i$-values as there are in the union of all $\tuples(\sigma(M_i))$ of cardinality $1$, for $\sigma$ in $\Sigma^+(\mathcal{AQ})$. Denoting by  $\Sigma^+_1(\mathcal{AQ})$ the set of these m-tuples, that is $\Sigma^+_1(\mathcal{AQ})=\{\sigma \in \Sigma^+(\mathcal{AQ})~|~|\tuples(\sigma(M_i))| = 1\}$, whenever $\Sigma^+(\mathcal{AQ})\ne \emptyset$, we have:

\smallskip\centerline{
$glb \geq |\bigcup_{\sigma \in \Sigma^+_1(\mathcal{AQ})}\tuples(\sigma(M_i))|$.}

\smallskip\noindent
Denoting by $count\_min$ the cardinality shown above, we notice that, as for $count\_max$, Algorithm~\ref{algo:proc-aggr} and Algorithm~\ref{algo:cons-answer-no-groupby} can be modified so as to compute $count\_min$.

\smallskip
Another important remark regarding {\tt distinct} is that if for every $\sigma$ in $m\_Chase(T)$, we have $|\tuples(\sigma(M_i))|=1$ (i.e., the dependency $\mathbb{K} \to M_i$ is satisfied), then $\Sigma^+_1(\mathcal{AQ})=\Sigma^+(\mathcal{AQ})$, and: 

\smallskip\begin{tabular}{ll}
$\bullet$ $count\_max = lub =  |\bigcup_{\sigma \in \Sigma(\mathcal{AQ})}\tuples(\sigma(M_i))|$&\qquad$(1)$\\
$\bullet$ $count\_min = glb = |\bigcup_{\sigma \in \Sigma^+(\mathcal{AQ})}\tuples(\sigma(M_i))|$&\qquad$(2)$
\end{tabular}

\smallskip\noindent
The above remark applies to the query $\mathcal{AQ}$ of Example~\ref{ex:cons-answer-no-groupby} if $aggr=count({\tt distinct}\,M_1)$, because $\tuples(\sigma_i(M_1))$ are singletons for every $i=1,2,3$. More precisely,  $|\bigcup_{\sigma \in \Sigma(\mathcal{AQ})}\tuples(\sigma(M_i))|=|\Sigma(\mathcal{AQ})|=3$ (because the values involved are pairwise distinct), and $|\Sigma^+(\mathcal{AQ})|=1$, showing that Algorithm~\ref{algo:cons-answer-no-groupby} would return the exact value of $\consans(\mathcal{AQ})$, that is the interval $[1,3]$.
\begin{example}\label{ex:distinct}
\rm{
In the context of Example~\ref{ex:star}, let

\smallskip\centerline{$\mathcal{AQ}:$ {\tt select} $count({\tt distinct}\,M_1)$ {\tt from}~$T$}

\smallskip\noindent
be an analytic query involving a {\tt distinct} clause, but no selection condition. In this case, we have $\Sigma(\mathcal{AQ})=\Sigma^+(\mathcal{AQ})$ and it can be seen from Figure~\ref{fig:m-chase} that the only m-tuples $\sigma$  in $m\_Chase(T)$ such that $\mathbb{K} \cup sch(\mathcal{AQ}) \subseteq sch(\sigma)$ are the first three m-tuples in $m\_Chase(T)$ that we denote by $\sigma_1$, $\sigma_2$ and $\sigma_3$. Thus we have $\Sigma(\mathcal{AQ})=\Sigma^+(\mathcal{AQ})=\{\sigma_1, \sigma_2, \sigma_3\}$, along with $\tuples(\sigma_1(M_1))=\tuples(\sigma_3(M_1))=\{m_1\}$ and $\tuples(\sigma_2(M_1))=\{m'_1, m''_1\}$.

\smallskip
Referring to the four repairs of $m\_Chase(T)$ shown in Figure~\ref{fig:repairs}, we have $\ans(\mathcal{AQ}^{[R_i]})=2$ for $i=1, \ldots , 4$, implying that $\consans(\mathcal{AQ})=2$. On the other hand, applying the approximations shown above, we have $|\Sigma(\mathcal{AQ})|=|\tuples(\sigma_1(M_1)) \cup \tuples(\sigma_2(M_1)) \cup \tuples(\sigma_3(M_1))|=3$ and thus, $count\_max=3$. Moreover, $\Sigma^+_1(\mathcal{AQ})=\{\sigma_1, \sigma_3\}$ and $|\tuples(\sigma_1(M_1)) \cup \tuples(\sigma_3(M_1))|=1$. Hence, we obtain $count\_min=1$, producing the interval $[1,3]$, which approximates, but is not equal to $\consans(\mathcal{AQ})$.

\smallskip
Changing $m'_1$ to $m_1$  in $\sigma_2$ yields that $\ans(\mathcal{AQ}^{[R_1]})=\ans(\mathcal{AQ}^{[R_2]})=1$ and $\ans(\mathcal{AQ}^{[R_3]})=\ans(\mathcal{AQ}^{[R_4]})=2$. Thus $\consans(\mathcal{AQ})=[1,2]$. On the other hand,  applying the approximations as above, we have $|\Sigma(\mathcal{AQ})|=3$ and $|\tuples(\sigma_1(M_1)) \cup \tuples(\sigma_2(M_1)) \cup \tuples(\sigma_3(M_1))|=2$ and so, $count\_max=2$. Moreover, $\Sigma^+_1(\mathcal{AQ})=\{\sigma_1, \sigma_3\}$ and $|\tuples(\sigma_1(M_1))\cup \tuples(\sigma_3(M_1))|=1$. Hence, we obtain again $count\_min=1$, producing the interval $[1,2]$, which  is now equal to $\consans(\mathcal{AQ})$.
}
\hfill$\Box$
\end{example}
\section{Related Work}\label{sec:rel-work}
The problem of consistent query answering in inconsistent databases has motivated considerable research efforts since its introduction in \cite{ArenasBC99}. These efforts were first focused on conjunctive and self-join free queries involving no aggregate operators. The case of queries with aggregate operators and possibly with {\tt group-by} clauses has been introduced in \cite{ArenasBCHRS03} and then further investigated in  \cite{FuxmanFM05}. Focusing on relations whose constraints are unique key-constraints, these approaches rely on earlier work (see \cite{Wijsen19}) characterizing classes of queries that can be rewritten in polynomial time, into SQL queries whose evaluation yields the consistent answer as defined in \cite{ArenasBCHRS03}. In parallel, the issue of consistent conjunctive query answering to queries involving or not aggregate operators has been addressed in  \cite{DixitK19,DixitK22}, based on a radically different technique, mainly that of reducing the problem to the well-known SAT problem.

\smallskip
By relying on $Chase$ semantics, our approach does not strictly follow the same semantics as those cited above. Indeed, as argued in \cite{JIIS,LauS25} and illustrated in Example~\ref{ex:repair}, in our approach, database consistency is not checked against the database itself but rather, against its `chased' version.  It follows that, as explained in Section~\ref{subsec:def-rep}, the notion of repair in our approach does not coincide with the usual one in the literature, where a repair is a maximal and consistent subset of the tuples in the database. It follows that, although our definitions of consistent answer to a query, whether standard or analytic, follow the usual definitions (i.e., equal to the intersection of the answers to the query in every repair for standard queries, and equal to the interval containing all aggregate values found in the repairs for analytic queries), our semantics to consistent query answering does not coincide with those in traditional approaches. Nevertheless, since in the present work we concentrate on databases operating over a star schema, our approach is a particular case of databases whose relations have a unique primary key as in \cite{DixitK19,DixitK22,FuxmanFM05,KhalfiouiW23,KhalfiouiW24}. Moreover, it has been argued (see  Proposition~\ref{prop:standard-repair} in Section~\ref{subsec:rep-star}) that, in this case, our notion of repair can be seen as the usual notion of repair applied to the set of all true tuples (which is super-set of the tuples actually stored in the database). We also notice that in \cite{DixitK19,DixitK22}, and contrary to our approach, constraints other than key dependencies are  considered since these approaches also consider {\em denial constraints}, which are known to generalize functional dependencies. 

\smallskip
Regarding queries, we notice that contrary to  \cite{ArenasBC99,FuxmanFM05,Wijsen19}, restricted disjunctions in selection conditions, called independent conditions, are allowed in our approach (see Definition~\ref{def:indpdt}). On the other hand, restricted disjunctions are also allowed in \cite{DixitK19,DixitK22}, where unions of conjunctive queries may occur in the {\tt from} clause. From a very generic point of view, comparing these two restricted cases of disjunctions amounts to compare relational expressions such as $\pi_X(\sigma_{\Gamma_1 \vee \Gamma_2}(r_1\bowtie r_2))$ (in our approach) and  $\pi_X(\sigma_{\Gamma_1}(r_1)) \cup \pi_X(\sigma_{\Gamma_2}(r_2))$ (in \cite{DixitK19,DixitK22}) where $\pi$, $\sigma$ and $\bowtie$ are respectively the relational operators for projection, selection and join. Since in general, these expressions do not produce the same results (for instance if $r_2=\emptyset$, the former expression returns $\emptyset$ and the latter one may return a non empty set of tuples),  the two restrictions are not equivalent.

\smallskip
Another important specificity of queries in our approach, whether standard queries or analytic, is that their expression involves a single table (in fact the m-table $m\_Chase(T)$), which  can be seen as an extension of the lossless join of all tables in the data warehouse.  As a consequence, using the terminology in \cite{FuxmanFM05}, standard queries are tree-queries, whose join graph is the tree having the fact table $F$ as its root and all dimension tables $D_i$ as its leaves. Hence, as stated by Theorem~1 in~\cite{FuxmanFM05}, all standard queries in our approach, are SQL-rewritable, meaning that each query $Q$ can be rewritten into an SQL query $Q'$ whose answer is the consistent answer  to $Q$. The case of analytic queries is however less clear. Indeed, although a similar result has been stated in~\cite{FuxmanFM05}, it is argued in \cite{KhalfiouiW24} that the SQL rewriting for analytic queries involving the aggregate $sum$ is flawed, because this aggregate operator is not monotonic. In fact, a consequence of the separation Theorem 1.1 in  \cite{KhalfiouiW24} is that for  any analytic query in our approach whose aggregate operator is monotonic and associative, it is possible to generate two SQL queries, each of them computing a bound of the consistent answer.

The proof of this flaw, that can be found in \cite{DixitK22} or in \cite{KhalfiouiW24}, relies on a database schema expressed in our approach as follows: the database has three relations $D_1$, $D_2$ and $F$, respectively, defined over attributes sets $K_1A_1$, $K_2A_2$ and $K_1K_2M_1$, over which the two functional dependencies $K_1\to A_1$ and $K_2 \to A_2$ are assumed. That is, $K_1$, respectively $K_2$, is the key of $D_1$, respectively $D_2$, and the key of $F$ is $K_1K_2M_1$ (i.e., $F$ has no non-key attributes). Hence, this schema can be seen as a star schema, whose fact table has no embedded functional dependency.  In this context, it is shown  in \cite{DixitK22} and in \cite{KhalfiouiW24} that computing the consistent answer to the analytic query $\mathcal{AQ}:$ {\tt select}  $sum(M_1)$ {\tt from} $T$ {\tt where} $A_1=a_1$ {\tt and} $A_2=a_2$ (where $a_1$ and $a_2$ are constants respectively in $dom(A_1)$ and $dom(A_2)$) is NP-hard.

Although the above case does not follow the context of our work, since the functional dependencies $\mathbb{K} \to M_i$ are missing, we argue that all results stated earlier also hold in this case. Indeed, if $FD$ contains no functional dependency of the form $\mathbb{K} \to M_i$ then for every $\sigma$ in $m\_Chase(T)$ and every $i=1, \ldots, p$, $\tuples(\sigma(M_i))$ is a singleton. Immediate consequences are that, in this context:
\begin{enumerate}
\item 
$FD$ is normalized and acyclic, and so, the results stated in \cite{LauS25} hold.
\item 
Proposition~\ref{prop:star-m-chase}(2) is not true: m-tuples in $m\_Chase(T)$ may be distinct and have the same $\mathbb{K}$-value.
\item 
{\sf Step P2} in process {\sf (P)} is irrelevant and so, should be ignored. Moreover, the second item of {\sf Step P3} should be removed because for every $\sigma$ such that $\mathbb{K}M_i \subseteq sch(\sigma)$ for $i=1, \ldots ,p$, $\tuples(\sigma(M_i))$ is reduced to one tuple, and this tuple is consistent. However, it should be emphasized that Proposition~\ref{prop:all-repairs} still holds in this setting because adapting the proof amounts to discard all arguments related to inconsistencies of tuples involving  $\mathbb{K}$-values associated with measure values.  
\item 
The second item about the characterization of tuples in $\inc_{\min}(\mathcal{T})$ that follows Proposition~\ref{prop:truth-value} should be ignored because $|\tuples(\sigma(M_i)|>1$ cannot happen. On the other hand, it turns out that no tuple in $\inc_{\min}(\mathcal{T})$ can involve a measure value, because it is easy to see that if $t=t'm$ is in $\inc(\mathcal{T})$ and $m$ is in $adom(M_i)$, then $t'$ is in $\inc(\mathcal{T})$.
\item 
Despite the first paragraph in the proof of Proposition~\ref{prop:indpdt} (mentioning that this proof relies on the fact that $m\_Chase(T)$ cannot contain distinct m-tuples with the same $\mathbb{K}$-value), the proposition still holds. Indeed, as Proposition~\ref{prop:all-repairs} still applies,  for any of the two items in Proposition~\ref{prop:indpdt}, the modified process {\sf (P)} generates a repair which satisfies the conditions in the proposition.
\item
Dealing with aggregate operator $count({\tt distinct}\,.)$ in this context yields the {\em exact} consistent answer, instead of an approximation. This is so because in this case all $\sigma(M_i)$ contain one value, implying that the equalities (1) and (2) in the previous sub-section hold.
\end{enumerate}
An important consequence is that Algorithm~\ref{algo:proc-aggr}, and thus Algorithm~\ref{algo:cons-answer-no-groupby} and Algorithm~\ref{algo:cons-answer-groupby}, work in this context, showing that consistent answers to analytic queries can be computed in quasi-linear time, even when no functional dependency of the form $\mathbb{K} \to M_i$ is considered.  As a consequence, based on the results in \cite{DixitK22} or in \cite{KhalfiouiW24}, determining whether analytic queries in a star schema according to our approach is SQL-rewritable, is an open issue to be investigated.

On the other hand, this also shows that our approach extends approaches from the literature, in the sense that we allow missing values in the input tables and we can handle some disjunctive selection conditions (namely independent selection conditions involving no key attributes), at the cost of approximate consistent answers when the aggregate operator is $sum$, operating on positive or negative numbers, or when the aggregate operator is of the form {\tt $count$(distinct .)}.

\smallskip
To sum up the above remarks, we recall that our approach works according to different semantics than in existing works dealing with consistent answering. Although this prevents from accurate comparisons, we emphasize the following two main important features of our work: 
\begin{enumerate}
\item
Contrary to any other approach, ours supports the presence of missing values in the data warehouse,  at the cost of approximate consistent answers when the aggregate operator is $sum$, operating on positive or negative numbers or when the aggregate operator is of the form {\tt $count$(distinct .)}.
\item
Our complexity results outperform those from the literature, since we have shown that consistent answers are computed in quasi-linear time, i.e., in $\mathcal{O}(W\cdot\log(W))$, where $W$ is the size of the data warehouse.
\end{enumerate}
\section{Concluding Remarks}\label{sec:conclusion}
Data analytics requires the collection and integration of data coming from different sources and stored in a database operating under a specific type of schema called a `star schema'. Such a database is referred to in the literature as a {\em data warehouse} \cite{inmon96}. As a consequence, more often than not, such integrated data contain inconsistencies and missing values. In this paper, we have seen an approach to efficiently compute consistent answers to analytic queries in data warehouses. Extending earlier work concerning the consistent query answering of standard, non-analytic queries in multi-table databases, we presented in this paper an approach to computing exact consistent answers of analytic queries over star schemas under the assumption that the selection condition is independent. Our main contribution is to propose a framework for star schemas in which missing values are allowed, and in which computing the consistent answer to a usual projection-selection-join query, or to an analytic query, is  quasi-linear in the size of the data warehouse, i.e. in $\mathcal{O}(W \cdot \log(W))$ where $W$ is the size of the data warehouse. 

\smallskip
Our current work follows two main lines of research. First, the implementation of our approach is an important issue under consideration. We notice in this respect that, as $m\_Chase(T)$ has to be `re-computed' after each update, the question of incrementally implementing the changes in $m\_Chase(T)$ is crucial regarding efficiency. This issue is currently investigated based on Algorithm~\ref{algo:chase-star} and on the statements in Proposition~\ref{prop:star-m-chase}.

A second line of research is investigating the introduction of hierarchies over dimensions. Indeed, hierarchies play a central role in data warehouse querying, because they allow for data analyses at different `levels' (for instance at the level of states instead of cities in a location dimension). From a practical point of view, introducing hierarchies means introducing new functional dependencies among dimensional attributes. So, if these hierarchies have inconsistencies and/or missing values, the results on the computation of consistent answers presented here have to be revisited. Moreover, the case of queries involving the operators {\tt roll-up} or {\tt drill-down} (that allow for `navigating' within hierarchies) has to be investigated in this context.

\bigskip\noindent
{\bf Declarations}\\
The two authors contributed to the study, conception and design. Both read and approved the manuscript.
\\
No funds, grants, or other support was received for conducting this study.

\bigskip\noindent
{\bf Acknowledments}\\
The authors thank the reviewers for their thorough reading of our manuscript. Their comments and suggestions helped us improve the contents and the presentation of the modified version.



\appendix
\section{Proof of Proposition~\ref{prop:star-m-chase}}\label{append:prop-star-m-chase}
\noindent{\bf Proposition~\ref{prop:star-m-chase}}~~{\em
Let $T$ be a star-table over universe $U$. The following hold:
\begin{enumerate}
\item 
For every $\sigma$ in $m\_Chase(T)$ and every $i=1, \ldots ,n$, if $K_i \in sch(\sigma)$ then $|\tuples(\sigma(K_i))|=1$. Consequently,  if $\mathbb{K} \subseteq sch(\sigma)$ then $|\tuples(\sigma(\mathbb{K}))|=1$.
\item 
For every tuple $k$ over $\mathbb{K}$ in $\mathcal{T}$, there exists at most one $\sigma$ in $m\_Chase(T)$ such that $\mathbb{K} \subseteq sch(\sigma)$ and $\tuples(\sigma(\mathbb{K})) = (k)$.
\item
Moreover, $m\_Chase(T)$ contains the following two kinds of m-tuples:
\begin{enumerate}
\item $\sigma$ for which there exists $i_0 \in \{1, \ldots ,n\}$ such that:
\begin{itemize}
\item $sch(\sigma) \subseteq sch(D_{i_0})$, $\tuples(\sigma(K_{i_0}))= (k_{i_0})$ and for every $t \in F$, $t.K_{i_0} \ne k_{i_0}$,
\item for every $A \in sch^*(D_{i_0})$, $\sigma(A) = \{a~|~(\exists q \in D_{i_0})(q.K_{i_0} = k_{i_0} \wedge q.A = a)\}$.
\end{itemize}
\item $\sigma$ such that $\mathbb{K} \subseteq sch(\sigma)$ and $\tuples(\sigma(\mathbb{K}))=(k)$, and
\begin{itemize}
\item for every $M_j \in \mathbb{M}$, $\sigma(M_j)=\{m_j~|~(\exists t \in F)(t.\mathbb{K}=\sigma(\mathbb{K}) \wedge t.M_j = m_j)\}$,
\item for every $i =1, \ldots , n$, for every $A \in sch^*(D_i)$,\\
\rightline{$\sigma(A) = \{a~|~(\exists t \in D_i)(t.K_i = k.K_i \wedge t.A = a)\}$.\qquad\qquad\qquad\quad}
\end{itemize}
\end{enumerate}
\end{enumerate}
}

\smallskip\noindent
{\sc Proof.} {\bf 1.} The result comes from the fact that,  to generate multi-valued components over an attribute $A$, the algorithm $m\_Chase$ considers a functional dependency whose right hand-side is $A$. As in the case of a star-table, no dependency in $FD$ has its right hand-side in $\mathbb{K}$, it impossible to generate conflicts on key attributes. The proof of this item is therefore complete.

\smallskip\noindent
{\bf 2.} Let $\sigma_1$ and $\sigma_2$ be in $m\_Chase(T)$ such that $\mathbb{K} \subseteq sch(\sigma_i)$ for $i=1,2$ and $\sigma_1(\mathbb{K})=\sigma_2(\mathbb{K})$. Since $\sigma_1 \ne \sigma_2$ (the algorithm eliminates duplicates) there exists $A$ not in $\mathbb{K}$ such that $\sigma_1(A) \ne \sigma_2(A)$. Since $A$ is not in $\mathbb{K}$, either there is $i_0$ in $\{1, \ldots ,n\}$ such that $A \in sch^*(D_{i_0})$ or $A$ is in $\mathbb{M}$. Hence, $FD$ contains a dependency $X \to A$ where $X=K_{i_0}$ in the former case, or $X=\mathbb{K}$ in the latter case. Since, as stated above, $|\sigma_1(X)|=|\sigma_2(X)| = 1$, we have $\sigma_1(X)=\sigma_2(X)$. Thus, applying the {\sf m\_Chase rule} to $m\_Chase(T)$ changes $\sigma_1(A)$ and $\sigma_2(A)$ into $\sigma_1(A)\cup \sigma_2(A)$. As by definition of $m\_Chase(T)$, $m\_Chase(m\_Chase(T))=m\_Chase(T)$, we obtain a contradiction which completes the proof of this item.

\smallskip\noindent
{\bf 3.} To prove the results, we let $\left(\Sigma_l\right)_{l \geq 0}$ be the sequence of m-tables successively generated when running Algorithm~\ref{algo:chase}, and we let $\lambda$ be the least integer such that $\Sigma_\lambda = \Sigma_{\lambda+1}=m\_Chase(T)$. Since $\Sigma_0$ is obtained from $T$ by `converting' tuples in the data warehouse into m-tuples, according to the restrictions on missing values earlier stated, every m-tuple $\sigma$ in $\Sigma_0$ satisfies the following:
\begin{itemize}
\item[(A)] If $\mathbb{K} \not\subseteq sch(\sigma)$ then $\sigma$ has been obtained from a tuple in a dimension table. Thus, there exists $i_0 \in \{1, \ldots , n\}$ such that $sch(\sigma) \subseteq sch(D_{i_0})$, $\tuples(\sigma(K_{i_0}))= (k_{i_0})$ and or every $A \in sch^*(D_{i_0})$, $\sigma(A) \subseteq \{a~|~(\exists q \in D_{i_0})(q.K_{i_0} = k_{i_0} \wedge q.A = a)\}$.
\item[(B)] Otherwise, $\sigma$ has been obtained from a tuple in the fact table, and thus, for every $M_j \in \mathbb{M}$, $\sigma(M_j)\subseteq \{m_j~|~(\exists t \in F)(t.\mathbb{K}=\sigma(\mathbb{K}) \wedge t.M_j = m_j)\}$. Moreover, for every $i =1, \ldots , n$, for every $A \in sch^*(D_i)$, we have $\sigma(A)=\emptyset$ (because $A \not\in sch(F)$), implying that $\sigma(A) \subseteq \{a~|~(\exists t \in D_i)(t.K_i = k.K_i \wedge t.A = a)\}$.
\end{itemize}
We show by induction that, for $l \geq 0$, the m-tuples in $\Sigma_l$ satisfy one of the two items (A) or (B) above. Indeed, assuming that all m-tuples in $\Sigma_l$ satisfy (A) and (B), let $\sigma$ and $\sigma'$ be in $\Sigma_l$ and $Y \to B$ the dependency in $FD$ used by the {\sf m\_Chase rule} when generating $\Sigma_{l+1}$. Thus, $YB \subseteq sch(\sigma)$ and  $YB \subseteq sch(\sigma')$, and applying the {\sf m\_Chase rule} to $\sigma$ and $\sigma'$ modifies $\sigma(B)$ and $\sigma'(B)$ to produce $\sigma_1$ and $\sigma'_1$ such that $\sigma_1(B)=\sigma'_1(B)=\sigma(B) \cup\sigma'(B)$. Since $B \not\in \mathbb{K}$, it follows that $sch(\sigma) \cap \mathbb{K}=sch(\sigma_1) \cap \mathbb{K}$ and $sch(\sigma') \cap \mathbb{K}=sch(\sigma'_1) \cap \mathbb{K}$, thus that $\sigma_1$ falls in case (A), respectively in case (B), if and only if $\sigma$ falls in case (A), respectively in case (B), and the same happens for $\sigma'$ and $\sigma'_1$. Moreover, since by our induction hypothesis, $\sigma$ and $\sigma'$ satisfy the inclusions stated in items (A) and (B), it is easy to see that such is also the case for $\sigma_1$ and $\sigma'_1$. Therefore, it turns out that the m-tuples in $\Sigma_{l+1}$ satisfy the inclusions stated in items (A) and (B) above; in particular, such is the case for $m\_Chase(T)$.

We now prove that, when considering $m\_Chase(T)$, the inclusions involving $\sigma(A)$ in item (A) or (B) are in fact equalities. In case of item (A), let $\sigma \in m\_Chase(T)$ and $i_0 \in \{1, \ldots , n\}$ such that $K_{i_0} \in sch (\sigma)$, $\mathbb{K} \not\subseteq sch(\sigma)$. If there exists  $A \in sch^*(D_{i_0})$ and $q \in D_{i_0}$ such that $A \in sch(q)$, $\sigma(K_{i_0})=q.K_{i_0}$ and $q.A \not\in \sigma(A)$, then by construction of $m\_Chase(T)$ there must exist $\sigma_q \in m\_Chase(T)$ such that $q \in \tuples(\sigma_q(sch(q)))$. It follows that $\sigma(A) \ne \sigma_q(A)$ whereas $\sigma(K_{i_0}) = \sigma_q(K_{i_0})$, thus that applying the {\sf m\_Chase rule} to $\sigma$, $\sigma_q$ and $K_{i_0} \to A$ in $FD$, changes $m\_Chase(T)$. In other words this implies that $m\_Chase(m\_Chase(T)) \ne m\_Chase(T)$, which is a contradiction. Since similar arguments hold in the case where $\sigma$ satisfies item (B), this part of the proof is complete.

The last step in this proof consists in showing that for every $\sigma$ in $m\_Chase(T)$ of kind (a) in the proposition and such that $\sigma(K_{i_0})=k_{i_0}$, $k_{i_0}$ does not occur in $F$. Indeed, if such is the case, $m\_Chase(T)$ contains $\sigma'$ such that $\mathbb{K} \subseteq sch(\sigma')$ and $\sigma'(K_{i_0})=\sigma(K_{i_0})=k_{i_0}$. Since  $m\_Chase(T)$ is closed under the {\sf m\_Chase rule}, every attribute $A$ in $sch(\sigma) \setminus K_{i_0}$ belongs to $D_{i_0}^*$ and so, as $K_{i_0} \to A$ is in $FD$, $A$ is in $sch(\sigma')$ and $\sigma(A)=\sigma'(A)$. Hence, $\sigma \sqsubseteq \sigma'$, and so, applying the {\sf m\_Chase rule} to $\sigma$ and $\sigma'$ changes $m\_Chase(T)$ by removing $\sigma$, a contradiction to the fact that $m\_Chase(T)$ is closed under the {\sf m\_Chase rule}. The proof is therefore complete.\hfill$\Box$ 

\section{Proof of Lemma~\ref{lemma:repair1}}\label{append:lemma-repair1}
\noindent{\bf Lemma~\ref{lemma:repair1}}~~{\em Let $T$ be a star-table over universe $U$, $S$ a subset  of $\true(\mathcal{T})$ and $t$ a tuple in $\cons(\mathcal{T})$. If $S \models FD$ then $S \cup \{t\} \models FD$.
}

\smallskip\noindent
{\sc Proof.} Since $S \models FD$, every $\sigma$ in the m-table $m\_Chase(S)$ is such that, for every $A$ in $sch(\sigma)$, $\sigma(A)=(a)$. Thus $m\_Chase(S)$ can be seen as a table that we denote by $S^*$. Moreover, denoting by $S_t$ the table $S \cup \{t\}$, we have $m\_Chase(S_t) = m\_Chase(S^* \cup \{t\})$. Let $S^*_t= S^* \cup \{t\}$, and let us consider the computation of $m\_Chase(S^* _t)$.

To this end, given $q_1$ and $q_2$ in $S^*_t$ and $X \to A$ in $FD$ such that $q_1.X=q_2.X$, the only possible cases are as follows:
\\
$-$ If $q_1$ and $q_2$ are in $S^*_t$, then either $q_1$ and $q_2$ are not defined over $A$ or,  they are both defined such that $q_1.A=q_2.A$. In this case $m\_Chase$ does not change  $S^*_t$.
\\
$-$  If $q_1 \in S^*_t$, $q_2=t$, and $q_1$ and $q_2$ are not defined over $A$, then $m\_Chase$ does not change  $S^*_t$.
\\
$-$ If $q_1 \in S^*_t$, $q_2=t$, and $q_1$ and $q_2$ are defined over $A$. Since $t \in \cons(\mathcal{T})$, it is not possible that $q_1.A \ne t.A$. Thus in this case again, $m\_Chase$ does not change  $S^*_t$.
\\
$-$ If $q_1 \in S^*_t$,  $q_2=t$, $q_1$ is defined over $A$ and $q_2$ is not defined over $A$. Then $m\_Chase$ changes $t$ into $ta$ where $a=q_1.A$.
\\
$-$ If $q_1 \in S^*_t$,  $q_2=t$, $q_1$ is not defined over $A$ and $q_2$ is defined over $A$. Then $m\_Chase$ changes $q_1$ into $qa$ where $a=t.A$.

\smallskip
Based on the previous cases, we denote by $\Sigma$ the table obtained from $S^*_t$ by the following transformations:
\\
$(1)$ For every $q$ in $S^*$ and every $X \to A$ in $FD$ such that $q.X=t.X$, $q$ is not defined over $A$ and $t.A=a$, in $\Sigma$, $q.A$ is set to $a$.
\\
$(2)$ For every $X \to A$ in $FD$ such that $q.X=t.X$, $q.A=a$ and $t$ is not defined over $A$, in $\Sigma$, $t.A$ is set to $a$.

\smallskip
Since in a star schema, for every attribute $A$ in $U$ there is {\em at most} one functional dependency whose right hand-side is $A$, the construction of $\Sigma$ cannot generate conflicts, thus entailing that $\Sigma$ contains no conflicts. We now show that $m\_Chase(\Sigma)=\Sigma$.

Indeed, let $q_1$ and $q_2$ be in $\Sigma$ and $X \to A$ be in $FD$. The only possibility for $m\_Chase$ to change $\Sigma$ is that $q_1.X=q_2.X$, $q_1.A=a$ and $q_2$ is not defined over $A$. As this does not happen in $S^*$ (because $m\_Chase(S^*)=S^*$), and as this does not happen in $\Sigma$ (by definition of $\Sigma$), we obtain that $m\_Chase(\Sigma)=\Sigma$. It follows that $m\_Chase(S^*_t) =\Sigma$, and since we have seen that no conflicts occur in $\Sigma$, $S^*_t \models FD$ holds. The proof is therefore  complete.
\hfill$\Box$
\section{Proof of Proposition~\ref{new-prop-repairs}}\label{append:new-prop-repairs}
\noindent
{\bf Proposition~\ref{new-prop-repairs}}~~{\em
Let $T$ be a star-table over universe $U$. For a given $X \to A$ in $FD$ and a given tuple $x$ over $X$, let $\{xa_1, \ldots , xa_q\}$ be the set of all tuples over $XA$ that belong to $\true(\mathcal{T})$. Then, for every $R$ in $\rep(T)$, there exists $q_0$ in $\{1, \ldots ,q\}$ such that $xa_{q_0}$ is in $\true(\mathcal{R})$.
}

\smallskip\noindent
{\sc Proof.} It is easy to show that the proposition holds for $q=1$. Indeed, in this case $xa_1$ is in $\cons(\mathcal{T})$, and thus in $\true(\mathcal{R})$ for every $R$ in $\rep(T)$. We now assume that $q>1$, which implies that for every $\sigma$ in $m\_Chase(T)$ such that $XA \subseteq sch(\sigma)$ and $x$ is in $\tuples(\sigma(X))$, $\tuples(\sigma(XA))=(xa_1, \ldots , xa_q)$ with $q>1$. Moreover, given the form of functional dependencies in $FD$, either $(a)$ $X= \mathbb{K}$ and $A$ is in $\mathbb{M}$, or $(b)$ there exists $i_0$ in $\{1, \ldots ,n\}$ such that $X=K_{i_0}$ and $A$ is in $sch^*(D_{i_0})$.

Let $R$ be in $\rep(T)$ such that $\true(\mathcal{R})$ contains no $xa$ from $\{xa_1, \ldots , xa_q\}$, and let $q_0$ in $\{1, \ldots ,q\}$. Denoting by $R'$ the table $R \cup \{xa_{q_0}\}$, we show that in either case $(a)$ or $(b)$ above, $R' \models FD$. To this end, we first notice that, since for every $\mathbb{K}$-value $k$ and for any $K_i$-value $k_i$, $k$ and $k_i$ are in $\cons(\mathcal{T})$, the tuple $x$ is in $\true(\mathcal{R})$. Hence, $\true(\mathcal{R})$ contains tuples whose $X$-value is $x$ and for every such tuple, the $A$-value is missing. The cases $(a)$ and $(b)$ are then handled as follows when computing $m\_Chase(R')$: every $t$ in $\true(\mathcal{R}')$ such that $t.X=x$ is considered and its $A$-value is set to $a_{q_0}$. Denoting by $\Sigma$ the resulting m-table, let $X' \to A'$ be a dependency  in $FD$, and let $\sigma_1$ and $\sigma_2$ be two m-tuples in $\Sigma$ such that $\sigma_1(X') \cap \sigma_2(X')\ne \emptyset$ where $X'$ is either $\mathbb{K}$ or $K_j$. Since $R$ is in $\rep(T)$, $R$ satisfies $FD$ and thus, all components of all m-tuples in $m\_Chase(T)$ are singletons. Since this also holds in $\Sigma$, it follows that $\sigma_1(X') \cap \sigma_2(X')\ne \emptyset$ implies that $\sigma_1(X') =\sigma_2(X')$. Then, the following holds:

\smallskip\noindent
$-$ If $X'=X$ then by definition of $\Sigma$, $\sigma_1(A) = \sigma_2(A) =(a_{q_0})$. Thus $\Sigma$ is not changed.
\\
$-$ If $X'= K_j$, $X=K_i$ and $i \ne j$, $\sigma_1$ and $\sigma_2$ are both in $R'$ and their $A'$-components have not been changed when building $\Sigma$ (because $K_j \ne K_i$ implies $A' \ne A$).  Since $m\_Chase(\true(\mathcal{R}'))= \true(\mathcal{R}')$ and $R' \models FD$,  either $A'$ is in $sch(\sigma_1) \cap sch(\sigma_2)$ and $\sigma_1(A')=\sigma_2(A')$, or $A'$ is neither in  $sch(\sigma_1)$ nor in $sch(\sigma_2)$. In either case, $\Sigma$ is not changed.
\\
$-$ If $X'=K_j$ and $X=\mathbb{K}$, since as above we have $A' \ne A$, $m\_Chase$ does not change $\Sigma$.
\\
$-$ If $X'=\mathbb{K}$ and $X = K_i$, then in $\true(\mathcal{R}')$, we have $\sigma_1(X) = \sigma_2(X) =x$, and  $A$ is neither in $sch(\sigma_1)$ nor in $sch(\sigma_2)$. By definition of $\Sigma$, $\sigma_1(A)$ and $\sigma_2(A)$ are set to $(a_{q_0})$, and then here again, $m\_Chase$ does not change $\Sigma$.

\smallskip
We thus obtain that $\Sigma=m\_Chase(\mathcal{R}')$ and that $R' \models FD$, because $R \models FD$ and no conflicts occur in $\Sigma$. Moreover, as $\true(\mathcal{R}) \subseteq \true(\mathcal{R}')$, we have $\cons(\mathcal{T}) \subseteq \true(\mathcal{R}')$. It follows from Definition~\ref{def:repair} that $R$ can not be in $\rep(T)$ since the inclusion $\true(\mathcal{R}) \subset \true(\mathcal{R}')$ is strict (because $xa_{q_0}$ is in $\true(\mathcal{R}')$ but not in $\true(\mathcal{R})$). As this is a contradiction with our hypothesis that $R$ is in $\rep(T)$, the proof is complete.\hfill$\Box$ 

\section{Proof of Proposition~\ref{prop:all-repairs}}\label{append:all-repairs}
\noindent
{\bf Proposition~\ref{prop:all-repairs}}~~{\em Let $T$ be a star-table over $U$. $R$ is a repair of $T$ if and only if there is a $\varphi$ as defined above such that $\true(\mathcal{R})=\true(\mathcal{R}_\varphi)$.
}

\smallskip\noindent
{\sc Proof.} We first notice that $\true(\mathcal{R}_\varphi) \subseteq \true(\mathcal{T})$ holds because all tuples in $R_\varphi$ are in $\true(\mathcal{T})$. To show that $R_\varphi \models FD$, we denote by $T_\varphi$ the table $\{t_\varphi(\sigma)~|~\sigma \in m\_Chase(T)\}$, and we recall that $R_\varphi = T_\varphi \cup \cons(\mathcal{T})$.  We first prove that $m\_Chase(T_\varphi)=T_\varphi$.

Indeed, let $X \to A$ in $FD$, $t_\varphi(\sigma)$ and $t_\varphi(\sigma')$ in $T_\varphi$ such that $t_\varphi(\sigma).X = t_\varphi(\sigma').X$. If $\sigma$ and $\sigma'$ are defined over $A$, then, by construction of $T_\varphi$, we have $t_\varphi(\sigma).A = t_\varphi(\sigma').A$, in which case $m\_Chase$ does not change $T_\varphi$. The case whereby $\sigma$ is defined over $A$ and $\sigma'$ is not defined over $A$ is not possible in $m\_Chase(T)$, and thus it is not possible that  $t_\varphi(\sigma)$ is defined over $A$ while $t_\varphi(\sigma')$ is not. Therefore, we have  $m\_Chase(T_\varphi)=T_\varphi$, and this implies that $T_\varphi \models FD$, because no conflicts can occur in $T_\varphi$.

\smallskip
Given an m-table $\Sigma$ over universe $U$, we denote by $\tau(\Sigma)$ the set of all tuples occurring in $\Sigma$. More formally: $\tau(\Sigma)=\{q \in \mathcal{T}~|~(\exists \sigma \in \Sigma)(\exists t \in \tuples(\sigma))(q \sqsubseteq t)\}$.

Recalling from \cite{LauS25} that, when $FD$ is acyclic, we also have $m\_Chase(\cons(\mathcal{T}))=\cons(\mathcal{T})$ and $\cons(\mathcal{T})\models FD$, we prove by induction that, at each step $k$ of the computation of $m\_Chase(R_\varphi)$ the obtained m-table $\Sigma^k$ is such that $(1)$ $\Sigma^k$ contains no conflict and $(2)$ $\tau(\Sigma^k)=\tau(R_\varphi)$.

\smallskip\noindent
$\bullet$ For $k=0$, i.e., $\Sigma^0=R_\varphi$, $(2)$ is obvious. As for $(1)$, assume that $R_\varphi$ contains $t$ and $t'$ such that for $X \to A$ in $FD$ we have $t.X=t'.X$ and $t.A \ne t'.A$. In this case, as $t$ and $t'$ cannot be both in $T_\varphi$ or both in $\cons(\mathcal{T})$, we consider that $t \in T_\varphi$ and that $t' \in \cons(\mathcal{T})$. As $t$ is in $\true(\mathcal{T})$, this implies that $t'$ is in $\inc(\mathcal{T})$, which is a contradiction. We therefore obtain that $t.A=t'A$, and thus, $R_\varphi$ contains no conflicts.

\smallskip\noindent
$\bullet$ Assuming now that the result holds for some $k >0$, we prove that such is also the case for $k+1$. As $\Sigma^k$ contains no conflicts, its rows can be seen as tuples. So consider $t$ and $t'$ in $\Sigma^k$ and $X \to A$ in $FD$ such that $t.X=t'.X$ and $t$ and $t'$ are defined over $A$. As $\Sigma^k$ contains no conflict, we have $t.A=t'.A$ and so,  $\Sigma^{k+1}=\Sigma^k$. Thus $(1)$ and $(2)$ are trivially satisfied.

Consider now $t$ and $t'$ in $\Sigma^k$ and $X \to A$ in $FD$ such that $t.X=t'.X$ and $t.A =a$ and $A \not\in sch(t')$. In this case, the tuple $t'$ is set to $t'_A$ such that $sch(t'_A)=sch(t) \cup \{A\}$, $t'_A.A=a$ and $t'_A.sch(t)=t$. Thus contrary to the previous case, $\Sigma^{k+1}=\Sigma^k$ does not hold. We however notice that this transformation generates not conflicts, and thus that $(1)$ is satisfied by $\Sigma^{k+1}$. We now argue that $t'_A$ is a tuple in $\tau(R_\varphi)$, which combined with the inductive hypothesis that $\tau(\Sigma^k)=\tau(R_\varphi)$, implies that $\tau(\Sigma^{k+1})=\tau(R_\varphi)$. As $t$ and $t'$ are in $\Sigma^k$, they both belong to $\tau(R_\varphi)$, and so there are $q$ and $q'$ in $R_\varphi$ such that $t \sqsubseteq q$ and $t' \sqsubseteq q'$. To show that $t'_A$ is in $\tau(R_\varphi)$, we successively consider the following cases:
\\
$-$ If $q$ and $q'$ are both in $T_\varphi$ (respectively both in $\cons(\mathcal{T})$): as these sets are closed under the {\sf m\_Chase rule}, $t$ and $t'$ are both in $T_\varphi$ (respectively both in $\cons(\mathcal{T})$), and so $t'_A$ is in $T_\varphi$ (respectively in $\cons(\mathcal{T})$), because $t'_A$ is obtained through the {\sf m\_Chase rule}.
\\
$-$ If $q$ is in $\cons(\mathcal{T})$ and $q'$ is in $T_\varphi \cap \inc(\mathcal{T})$: in this case $xa$ is in $\cons(\mathcal{T})$ (because $xa \sqsubseteq q$). If $\sigma'$ is the m-tuple in $m\_Chase(T)$ such that $q'=t_\varphi(\sigma')$, $XA \subseteq sch(\sigma')$ and $xa$ is in $\tuples(\sigma'(XA))$. Hence $t'_A \sqsubseteq t_\varphi(\sigma')$, and thus $t'_A$ is in $\tuples(R_\varphi)$.
\\
$-$ If $q$ is in $T_\varphi \cap \inc(\mathcal{T})$ and $q'$ in $\cons(\mathcal{T})$: in this case, assume first that $xa$ is in $\cons(\mathcal{T})$. Then, $t'_A$ is in $\cons(\mathcal{T})$, because for every $Y \to B$ in $FD$ other than $X \to A$, if $YB \subseteq sch(t'_A)$ then $YB \subseteq sch(t')$.

Assuming now that $xa$ is in $\inc(\mathcal{T})$, we have $a=\varphi(x)$ and so, for every $Y \to B$ such that $YB \subseteq sch(t'_A)$, $t'_A.B = \varphi(t'_A.Y)$ (because as $t'$ is in $\cons(\mathcal{T})$,  for every $Y \to B$ such that $YB \subseteq sch(t')$, $t'.B = \varphi(t'.Y)$). Hence, there exists $\sigma'$ in $m\_Chase(T)$ such that $t'_A \sqsubseteq t_\varphi(\sigma')$. Therefore $t'_A$ is in $\tau(R_\varphi)$.

\smallskip
As a consequence, $\Sigma^{k+1}$ satisfies $(1)$ and $(2)$, and thus, so does $m\_Chase(R_\varphi)$, meaning that $\true(R_\varphi)=\tau(R_\varphi)$ and $R_\varphi \models FD$. By Definition~\ref{def:repair}, in order to show that $R_\varphi$ is in $\rep(T)$, we have to show that $\true(\mathcal{R}_\varphi)$ is maximal.

To this end, we let $S$ be such that $\true(\mathcal{R}_\varphi) \subseteq \true(\mathcal{S}) \subseteq \true(\mathcal{T})$ and $S \models FD$, and we let $q$ be a tuple in $\true(\mathcal{S}) \setminus \true(\mathcal{R}_\varphi)$. As $q$ is in  $ \true(\mathcal{T})$, there exist $\sigma$ in $m\_Chase(T)$ and $t$ in $\tuples(\sigma)$ such that $q \sqsubseteq t$. Since $\cons(\mathcal{T}) \subseteq \true(\mathcal{R}_\varphi)$, it follows that $q$ is in $\inc(\mathcal{T})$, implying that there exists $X \to A$ in $FD$ such that  $XA \subseteq sch(q)$ and $\true(\mathcal{T})$ contains $xa'$ such that $x = q.X$ and $a' \ne q.A$. By construction of $R_\varphi$, $\true(\mathcal{R}_\varphi)$ contains a tuple $t'$ from $\tuples(\sigma)$, and so, we have $t.X =q.X=t'.X$ (because for every $X \to A \in FD$, $|\tuples(\sigma(X))|=1$) and $t.A \ne t'.A$. As $t.A =q.A$ and $t'$ is in $\true(\mathcal{S})$, this implies that $S \not\models FD$, which is a contradiction. This part of the proof is therefore complete.

\smallskip
Conversely, let $R$ be in $\rep(T)$. By Proposition~\ref{new-prop-repairs}, for every $X \to A$ in $FD$, for a given $X$-value $x$ in $\true(\mathcal{T})$, $\true(\mathcal{R})$ contains one tuple among all tuples of the form $xa_i$ from $\true(\mathcal{T})$ where $a_i$ is in $adom(A)$. According to the steps of Process {\sf (P)}, these tuples  define a repair,  denoted by $R_\varphi$, and we show that $\true(\mathcal{R})=\true(\mathcal{R}_\varphi)$. Since $R$ and $R_\varphi$ are in $\rep(T)$, $\cons(\mathcal{T})$ is a subset of $\true(\mathcal{R})$ and of $\true(\mathcal{R}_\varphi)$. Now, let $t$ be in $\inc(\mathcal{T})$. Then, for every $X \to A$ in $FD$ such that $XA \subseteq sch(t)$, $t.XA$ is in $\true(\mathcal{R})$ if and only if $t.XA$ is in $\true(\mathcal{R}_\varphi)$ because $t.X$ is in $\cons(\mathcal{T})$, associated with the same $A$-value in $R$ and $R_\varphi$.  It follows that when $t$ is in $\inc(\mathcal{T})$, $t$ is in $\true(\mathcal{R})$ if and only if $t$ is in $\true(\mathcal{R}_\varphi)$. Hence, we obtain that $\true(\mathcal{R})=\true(\mathcal{R}_\varphi)$, which completes the proof.\hfill$\Box$

\section{Proof of Proposition~\ref{prop:cons-ans-ana}}\label{append:cons-ans-ana}
\noindent
{\bf Proposition~\ref{prop:cons-ans-ana}}~~{\em Let $T$ be a star-table over universe $U$, and let $Q: {\tt select}$~$X$~{\tt from~$T$~where~}$\Gamma$ be a query such that $\Gamma$ is an independent selection condition. Then $\consans(Q)$ is the set of all tuples $x$ over $X$ for which there exists $\sigma$ in $m\_Chase(T)$ such that 

\begin{enumerate}
\item $sch(Q) \subseteq sch(\sigma)$, $x \in \sigma(X)$ and $\sigma(sch(\Gamma)) \cap Sat(\Gamma) \ne \emptyset$,
\item for every $Y \to B$ in $FD$ such that $YB \subseteq sch(Q)$:
\begin{enumerate}
\item if $B \in X$, then $|\tuples(\sigma(B))|=1$,
\item if $B \in sch(\Gamma)$, then $\tuples(\sigma(B)) \subseteq Sat(\Gamma(B))$.
\end{enumerate}
\end{enumerate}
}
\smallskip\noindent
{\sc Proof.} Assume first that $\sigma$ in $m\_Chase(T)$ satisfies the items of the proposition. To show that $x=\sigma(X)$ is in $\consans(Q)$, we consider a repair $R$ in $\rep(T)$ and we show that $\true(\mathcal{R})$ contains a tuple $x\gamma$ where $\gamma$ is in $Sat(\Gamma)$.

By Proposition~\ref{prop:all-repairs}, we can assume that $R$ has been obtained by applying process {\sf (P)}, which implies that $\true(\mathcal{R})$  contains a tuple $t_\sigma$ in $\tuples(\sigma)$. Moreover, by Proposition~\ref{prop:truth-value}, item 1 and item 2(b) imply that $|\sigma(X)|=1$, thus $t_\sigma.X=x$ and $x$ is in $\cons(\mathcal{T})$. By Theorem~\ref{theo:repair}, it follows that $x$ is in $\true(\mathcal{R})$.  Moreover, since $\Gamma$ is independent, item~1 implies that for every $B$ in $sch(\Gamma)$, there exits $b$ in $\sigma(B)$ such that $b$ is in $Sat(\Gamma(B))$. Depending on whether $B$ occurs as a right hand-side of a dependency in $FD$ or not, we have the following:\\
(1) If $B$ is the right hand-side of no dependency in $FD$ (which happens if $B$ is a key attribute in $\mathbb{K}$), then we have $|\sigma(B)|=1$ and thus $\sigma(B)=(b)$, implying that $t_\sigma.B$ is in $Sat(\Gamma(B))$.
\\
(2) Otherwise,  if $B$ is the right hand-side of  a dependency in $FD$, by item 2(b), $t_\sigma.B$ is obviously  in $Sat(\Gamma(B))$.

It follows that in any case, $t_\sigma.B$ is in $Sat(\Gamma(B))$, thus that $t_\sigma.sch(\Gamma)$ is in $Sat(\Gamma)$ since $\Gamma$ is independent. Hence, for $\gamma = t_\sigma.sch(\Gamma)$, $x \gamma$ is in $\true(\mathcal{R})$, showing that  $x$ is in $\consans(Q)$.

\smallskip
Conversely, let $x$ be in $\consans(Q)$. By Proposition~\ref{prop:approx}, $x$ belongs to $\wconsans(Q)$, meaning that there exists $t$ in $\true(\mathcal{T})$ such that $sch(Q) \subseteq sch(t)$, $t.X=x$, $x \in \cons(\mathcal{T})$ and $t.sch(\Gamma) \in Sat(\Gamma)$. As for every $t$ in $\true(\mathcal{T})$, there exists $\sigma$ in $m\_Chase(T)$ such that $t$ is in $\tuples(\sigma(sch(t)))$, there must exist $\sigma$ in $m\_Chase(T)$ such that $sch(Q) \subseteq sch(\sigma)$, $x \in \sigma(X)$, $\sigma(sch(\Gamma)) \cap  Sat(\Gamma) \ne \emptyset$ and for every $Y \to B$ in $FD$ such that $YB \subseteq X$, $|\tuples(\sigma(B))|=1$ (because $x$ is in $\cons(\mathcal{T})$). In other words, there exists $\sigma$ in $m\_Chase(T)$ satisfying  tem 1 and item 2(a) of the proposition.

Denoting by $\Sigma(Q)$ the set of all these m-tuples, let $\sigma$ be in $\Sigma(Q)$ for which item~2(b) is not satisfied. Then, let $Y \to B$ in $FD$ such that  $B \in sch(\Gamma)$ and $b$ in $\tuples(\sigma(B)) \setminus Sat(\Gamma(B))$. By item~1, $\tuples(\sigma(sch(Q)))$ contains $t$ such that $t.X=x$ and $t.B=b$. As $t$ is in $\true(\mathcal{T})$, by Proposition~\ref{prop:repairs}, there exists a repair $R$ such that $t$ is in $\true(\mathcal{R})$. Then, for every $x \gamma$ such that $\gamma$ is in $Sat(\Gamma)$, we have $t.Y=x.Y$ and $t.B \ne \gamma.B$ (since $\gamma.B \in Sat(\Gamma(B))$ and $t.B \not\in Sat(\Gamma(B))$). Consequently, $t$ and $x\gamma$ can not both belong to $\true(\mathcal{R})$ (because $R \models FD$). Therefore, if $\Sigma(Q)$ contains one m-tuple not satisfying item~2(b), then $x$ cannot belong to $\consans(Q)$. We moreover notice that if one $\sigma$ in $\Sigma(Q)$ does not satisfy item~2(b) then for any other $\sigma'$ in $\Sigma(Q)$, we have $\sigma'(Y)=\sigma(Y)$ and thus $\sigma'(B)=\sigma(B)$. It thus follows that $\sigma'$ does not satisfy item~2(b) either. As a consequence, we obtain that if $x$ is in $\consans(Q)$ then $m\_Chase(T)$ contains an m-tuple satisfying all three items in the proposition. The proof is therefore complete.\hfill$\Box$

\section{Proof of Proposition~\ref{prop:indpdt}}\label{append:prop-indpt}
\noindent
{\bf Proposition~\ref{prop:indpdt}}~~{\em Let $T$ be a star-table over universe $U$ and $\mathcal{AQ}:~${\tt select [$X$]},$\,aggr(M_i)$~{\tt from $T$ where} $\Gamma$ {\tt [group by~$X$]} an analytic query where $\Gamma$ is independent, i.e., $\Gamma=\Gamma(A_1) \wedge\ldots \wedge  \Gamma(A_k)$ where $sch(\Gamma)=A_1 \ldots A_k$. If $\Sigma(\mathcal{AQ}) \setminus \Sigma^+(\mathcal{AQ}) \ne \emptyset$, then there exist $R_1$ and $R_2$ in $\rep(T)$ such that:
\begin{enumerate}
\item  $(\forall \sigma \in \Sigma(\mathcal{AQ}) \setminus \Sigma^+(\mathcal{AQ}))(\exists t \in \tuples(\sigma) \cap \true(\mathcal{R}_1))(t.sch(\Gamma)\not\in Sat(\Gamma))$,
\item  $(\forall \sigma \in \Sigma(\mathcal{AQ}) \setminus \Sigma^+(\mathcal{AQ}))(\exists t \in \tuples(\sigma) \cap \true(\mathcal{R}_2))(t.sch(\Gamma)\in Sat(\Gamma))$.
\end{enumerate}
}

\smallskip\noindent
{\sc Proof.} The proof relies on the fact that for every repair $R$ of $T$ and every $\sigma$ in $\Sigma(\mathcal{AQ})$, $\true(\mathcal{R})$ contains exactly one tuple $t$ such that  $t \in \tuples(\sigma)$ and $t.K\in \sigma(\mathbb{K})$ (the existence is a consequence of Proposition~\ref{prop:repairs}, unicity follows from Proposition~\ref{prop:star-m-chase}(1) because $|\sigma(\mathbb{K})|=1$, $\mathbb{K}$ is a key of $R$ and $R \models FD$). As a consequence of Proposition~\ref{prop:all-repairs}, the existence of repairs $R_1$ and $R_2$ is explicitly shown using the process {\sf (P)}.

\smallskip\noindent
{\bf 1.} Regarding $R_1$, we build $\varphi$ as follows: For every $\sigma$ in $\Sigma(\mathcal{AQ})$:\\
{\sf Step P1.} For every $A_p^k$ in $sch^*(D_p) \cap sch(\sigma)$ such that $\sigma(K_p)=k_p$ and $\varphi_p^k$ has not been assigned:
\\
$(a)$ If $\sigma \in \Sigma^+(\mathcal{AQ})$ then choose an $A_p^k$-value in $\sigma(K_pA_p^k)$ and let $\varphi_p^k(k_p)=a_p^k$.
\\
$(b)$ If $\sigma \in \Sigma(\mathcal{AQ}) \setminus \Sigma^+(\mathcal{AQ})$ and  $\tuples(\sigma(A_p^k)) \not\subseteq Sat(\Gamma(A_k^p))$, choose an $A_p^k$-value $a^k_p$ in $\sigma(K_pA_p^k) \setminus Sat(\Gamma(A_k^p))$ and let $\varphi_p^k(k_p)=a_p^k$. If  $\tuples(\sigma(A_p^k)) \subseteq Sat(\Gamma(A_k^p))$, choose any $A_p^k$-value $a^k_p$ in $\sigma(K_pA_p^k)$ and let $\varphi_p^k(k_p)=a_p^k$.

\smallskip\noindent
{\sf Step P2.} If $\mathbb{K} \subseteq sch(\sigma)$, for every $M_i$ in $\mathbb{M} \cap sch(\sigma)$, if $\varphi_i(k)$ (where $k=\sigma(\mathbb{K})$) has not been assigned, then\\
$(a)$ If $M_i \not\in sch(\Gamma)$ or if $M_i \in sch(\Gamma)$ and $\sigma \in \Sigma^+(\mathcal{AQ})$, choose an $M_i$-value $m$ in $\sigma(M_i)$ and let $\varphi_i(k)=m$.\\
$(b)$ Otherwise (i.e., $M_i \in sch(\Gamma)$ and $\sigma \in \Sigma(\mathcal{AQ})\setminus \Sigma^+(\mathcal{AQ})$), choose an $M_i$-value $m$ in $\sigma(M_i) \setminus Sat(\Gamma(M_i))$ for $\varphi_i(k)$.

\smallskip\noindent
Once all m-tuples in $\Sigma(\mathcal{AQ})$ have been considered, the two steps above are completed by considering all non-processed $K_i$- or $\mathbb{K}$-values as done in the generic description of process {\sf (P)}. Consequently, the corresponding set $\varphi$ in process {\sf (P)} is properly defined. 

\smallskip\noindent
{\sf Step P3.} Build up the tuples $t_\varphi(\sigma)$ for every $\sigma$ in $m\_Chase(T)$. We notice that, according to the definitions of $\Sigma(\mathcal{AQ})$  and $\Sigma^+(\mathcal{AQ})$, the following holds:
\\
$-$ if $\sigma \in  \Sigma^+(\mathcal{AQ})$ then $t_\varphi(\sigma).sch(\Gamma) \in Sat(\Gamma)$,
\\
$-$  if $\sigma \in  \Sigma(\mathcal{AQ}) \setminus \Sigma^+(\mathcal{AQ})$ then $t_\varphi(\sigma).sch(\Gamma) \not\in Sat(\Gamma)$.

\smallskip\noindent
By Proposition~\ref{prop:all-repairs}, we therefore obtain a repair $R_1$ satisfying item~1 in the proposition.

\smallskip\noindent
{\bf 2.} Regarding the item~2, the repair $R_2$ is obtained as above, by changing $(b)$ in {\sf Step P1} and {\sf Step P2} as follows: instead of choosing a value {\em not} in the corresponding set $Sat(\Gamma(A))$, choose a value that belongs to $Sat(\Gamma(A))$. We notice that, according to the definitions of $\Sigma(\mathcal{AQ})$  and $\Sigma^+(\mathcal{AQ})$, such a choice is always possible.

\smallskip
Then, as above, these two steps are completed by considering all non-processed $K_i$- or $\mathbb{K}$-values, and {\sf Step P3} is considered. In this case, by Proposition~\ref{prop:all-repairs}, we obtain a repair $R_2$ such that for every $\sigma$ in $\Sigma(\mathcal{AQ})$, $t_\varphi(\sigma).sch(\Gamma)$ is in $Sat(\Gamma)$. Hence $R_2$ satisfies item~2 in the proposition, and the proof is complete.
\hfill$\Box$

\section{Proof of Proposition~\ref{prop:compute-cons-ans-no-group-by}}\label{append:compute-cons-ans-no-group-by}
\noindent
{\bf Proposition~\ref{prop:compute-cons-ans-no-group-by}}~~{\em Let $T$ be a star-table over universe $U$ and $\mathcal{AQ} :~${\tt select} $aggr(M_i)$~{\tt from $T$ where} $\Gamma$ an  analytic query with no {\tt group-by} clause.
\begin{itemize}
\item 
If $aggr$ is $min$, $max$ or $count$, then $\consans(\mathcal{AQ})=[min\_ans,max\_ans]$ where $min\_ans$ and $max\_ans$ are returned by  Algorithm~\ref{algo:cons-answer-no-groupby}.
\item 
If $aggr=sum$ and if $\consans(\mathcal{AQ})=[glb, lub]$, then  $min\_ans$ and $max\_ans$ as returned by  Algorithm~\ref{algo:cons-answer-no-groupby} satisfy that $min\_ans \leq glb$ and $max\_ans \geq lub$.
\\
Moreover, if {\em for every $m \in adom(M_i)$, $m \geq 0$}, then $\consans(\mathcal{AQ})=[min\_ans,max\_ans]$.
\item 
For every aggregate operator and every selection condition $\Gamma$,  $ans^*$ as returned by  Algorithm~\ref{algo:cons-answer-no-groupby} is equal to $\consans^*(\mathcal{AQ})$.
\end{itemize}
}

\smallskip\noindent
{\sc Proof.} We separately consider two cases, depending on whether the test line~\ref{line:final-test} in Algorithm~\ref{algo:cons-answer-no-groupby} succeeds or not. If the test fails, i.e., if $change\_min\_max$ as returned by the call of {\sf Compute\_aggregate} has value {\tt false}, this means that $m\_Chase(T)$ contains no $\sigma$ such that $\mathbb{K} \cup sch(Q) \subseteq sch(\sigma)$, $|\tuples(\sigma(X))|=1$ and $\tuples(\sigma(sch(\Gamma)))  \subseteq Sat(\Gamma)$. By Corollary~\ref{coro:cons-ans-ana}, this holds if and only if the consistent answer to $Q: ~${\tt select} $\mathbb{K},X$~{\tt from $T$ where} $\Gamma$ is empty. Thus, the test fails if and only if there exists a repair $R$ of $T$ for which $\ans(Q^{[R]})=\emptyset$. In this case $\consans(\mathcal{AQ})$ is {\tt NULL}, which is expressed by the fact that the values of $min\_ans$, $max\_ans$ and $ans^*$ returned by the call of {\sf Compute\_aggregate} are respectively {\tt -dummy}, {\tt +dummy} and {\tt -dummy}. Hence, in this case, Algorithm~\ref{algo:cons-answer-no-groupby} provides the correct answer.

\smallskip
Suppose now that the test line~\ref{line:final-test} succeeds, i.e., that the value of $change\_min\_max$ returned by the call of {\sf Compute\_aggregate} is {\tt true}. The statement line~\ref{line:proc-check-change-1} of Algorithm~\ref{algo:proc-aggr} shows that $m\_Chase(T)$ contains at least one $\sigma$ such that $\mathbb{K} \cup sch(Q) \subseteq sch(\sigma)$, $|\tuples(\sigma(X))|=1$ and $\tuples(\sigma(sch(\Gamma)))  \subseteq Sat(\Gamma)$. By Corollary~\ref{coro:cons-ans-ana}, this holds if and only if the consistent answer to $Q:~${\tt select} $\mathbb{K},X$~{\tt from $T$ where} $\Gamma$ is not empty. Thus, the test succeeds if and only if for every repair $R$ of $T$, $\ans(Q^{[R]}) \ne \emptyset$, in which case, $min\_ans$, $max\_ans$ and $ans^*$ are proper values, either values from $adom(M_i)$ if $aggr \ne count$, or positive integers if $aggr=count$. In this case, it can be seen that the m-tuples $\sigma$ in $m\_Chase(T)$ that contribute to the construction of the interval are such that $\mathbb{K} \cup X \cup sch(\Gamma) \subseteq sch(\sigma)$ and $\tuples(\sigma(sch(\Gamma))) \cap Sat(\Gamma) \ne \emptyset$. This is so, because of our assumption on missing values in the fact table $F$ (every tuple with an $M_i$-value must be defined over all key attributes); and because any $\sigma$ not defined over some attributes in $X \cup sch(\Gamma)$ or such that $\tuples(\sigma(sch(\Gamma))) \cap Sat(\Gamma) = \emptyset$ cannot contribute in the consistent answer to $\mathcal{AQ}$. This explains why, when the test line~\ref{line:test-schema} of Algorithm~\ref{algo:proc-aggr} fails, no action is taken.

On the other hand, the m-tuples $\sigma$  such that $\mathbb{K} \cup X \cup sch(\Gamma) \subseteq sch(\sigma)$ and $\tuples(\sigma(sch(\Gamma))) \cap Sat(\Gamma) \ne \emptyset$ are precisely those in $\Sigma(\mathcal{AQ})$, and recalling that $\Sigma^+(\mathcal{AQ})$ is the set of m-tuples in $\Sigma(\mathcal{AQ})$ such that $\tuples(\sigma(sch(\Gamma))) \subseteq  Sat(\Gamma)$, we have the following for every $\sigma$ in $\Sigma(\mathcal{AQ})$:
\\
$(1)$ By Proposition~\ref{prop:repairs}, for every $t\in \tuples(\sigma)$, there exists a repair $R$ such that $t \in \true(\mathcal{R})$. Moreover, $t$ is the unique tuple over $sch(\sigma)$ in $\true(\mathcal{R})$ having $t.\mathbb{K}$ as a $\mathbb{K}$-value. Notice that since $t.sch(\Gamma)$ may not be in $Sat(\Gamma)$, it is possible that $t.M_i$ does not contribute in the computation of the aggregate in $R$.
\\
$(2)$ If $\sigma$ is in $\Sigma^+(\mathcal{AQ})$, by Corollary~\ref{coro:cons-ans-ana}, every repair $R$ is such that $\true(\mathcal{R})$ contains a tuple $t$ in $\tuples(\sigma)$. In this case, since $t.sch(\Gamma)$ is in $Sat(\Gamma)$, $t.M_i$ does contribute in the computation of the aggregate in $R$.

\smallskip
Before showing that the values returned by Algorithm~\ref{algo:cons-answer-no-groupby} are as stated in Definition~\ref{def:anal-query}, we mention that  the aggregate operators $min$, $max$ and $sum$ are not defined when their argument is empty, which we write as $aggr(\emptyset)={\tt NULL}$. Otherwise, if $v_1$, $v_2$ and $v_3$ are values to which $aggr$ applies, then

{\em Commutativity:} $aggr(\{v_1, v_2\})=aggr(\{v_2,v_1\})$.

{\em Associativity:} $aggr(\{v_1, aggr(\{v_2,v_3\})\})=aggr(\{aggr(\{v_1,v_2\},v_3\})=aggr(\{v_1,v_2,v_3\})$.

{\em Monotonicity:} If $aggr \ne count$ and $v_2 \leq v_3$ then $aggr(\{v_1,v_2\}) \leq aggr(\{v_1,v_3\})$.

\smallskip\noindent
The first two properties show that aggregate values do not depend on the order the elementary values are considered and how they are grouped during the computation. Moreover, the third property shows that, if $aggr \ne count$, the higher the values, the higher the aggregate values. In our context, recalling that elementary values are values in $\sigma(M_i)$ for $\sigma$ in $\Sigma(\mathcal{AQ})$, this last property shows that, when the aggregate is different than $count$, for a fixed $\sigma$, the least, respectively the highest, aggregate value is obtained by considering the least, respectively the highest, possible $M_i$-value. These values, respectively denoted by $min_\sigma$ and $max_\sigma$ are computed lines~\ref{min-max-sigma-start}-\ref{min-max-sigma-end}  of Algorithm~\ref{algo:proc-aggr}.

Since the second property shows that aggregates can be seen as operating over a {\em set} of values we recall the following standard properties that will be used in this proof. Let $S_1$ and $S_2$ be sets of values to which $aggr$ applies, and such that $S_1 \subseteq S_2$, then: $\min(S_1) \geq \min(S_2)$, $\max(S_1) \leq \max(S_2)$, $count(S_1) \geq count(S_2)$, and {\em if all values} in $S_1$ or $S_2$ are positive, $sum(S_1) \leq sum(S_2)$. We notice that regarding the last property, if values in $S_1$ or $S_2$ can be positive or negative, no generic comparison can be stated.

We now explain, for each aggregate, how the bounds of the interval of the consistent answer can be obtained, and see why Algorithm~\ref{algo:cons-answer-no-groupby} returns (or not) these values. First, as for every $\sigma \in  \Sigma^+(\mathcal{AQ})$, every repair contains {\em exactly} one tuple in $\tuples(\sigma)$, the $M_i$-values in $\sigma(M_i)$ if $M_i \not\in sch(\Gamma)$, or in $\sigma(M_i) \cap Sat(sch(\Gamma))$ otherwise, contribute in the computation of the aggregate. Moreover, since by monotonicity, $min_\sigma$ contributes in the computation of $min\_ans$ and $max_\sigma$ contributes in
 the computation of $max\_ans$, we have the following, given that $\Sigma^+(\mathcal{AQ}) \subseteq \Sigma(\mathcal{AQ})$:
 \begin{itemize}
 \item
For $aggr=min$, $min\_ans = \min\{min_\sigma~|~\sigma \in \Sigma(\mathcal{AQ})\}$ and $max\_ans = \min\{max_\sigma~|~\sigma \in \Sigma^+(\mathcal{AQ})\}$. These values are computed respectively in lines~\ref{Xmin} and \ref{Ymin-max} of Algorithm~\ref{algo:proc-aggr}.
 \item
For $aggr=max$, $min\_ans = \max\{min_\sigma~|~\sigma \in \Sigma^+(\mathcal{AQ})\}$ and $max\_ans = \max\{max_\sigma~|~\sigma \in \Sigma(\mathcal{AQ})\}$. These values are computed respectively in lines~\ref{Ymin-max} and \ref{Zmax} of Algorithm~\ref{algo:proc-aggr}.
 \item
For $aggr=count$, $min\_ans$, respectively  $max\_ans$ is the cardinality of $\Sigma^+(\mathcal{AQ})$, respectively of $\Sigma(\mathcal{AQ})$. These values are computed respectively in lines~\ref{Ymin-max} and \ref{Zcount} of Algorithm~\ref{algo:proc-aggr}.
 \item
For $aggr=sum$, $min\_ans$, respectively $max\_ans$, is the minimal, respectively maximal, sum that can be obtained by adding one $M_i$-value in every $\sigma(M_i)$ for $\sigma \in \Sigma^+(\mathcal{AQ})$ and then possibly one $M_i$-value in every $\sigma(M_i)$ for $\sigma \in \Sigma^+(\mathcal{AQ})$, respectively of $\Sigma(\mathcal{AQ})$. These values are computed  in lines~\ref{Ymin-max} and \ref{Xsum}, respectively in lines~\ref{Ymin-max} and \ref{Zsum} of Algorithm~\ref{algo:proc-aggr}. Notice that if all $M_i$-values are positive, knowing that adding $0$ is neutral, then the test line~\ref{test-minsum} always fails and thus, $proc\_min$ is the sum of all minimal $M_i$-values in $\sigma(M_i)$ for $\sigma \in \Sigma^+(\mathcal{AQ})$. Similarly, in this case, the test line~\ref{test-maxsum} always succeeds, except for $0$, and thus, $proc\_max$ is the sum of all maximal $M_i$-values in $\sigma(M_i)$ for $\sigma \in \Sigma(\mathcal{AQ})$.
\end{itemize}
To complete the proof that the returned values $min\_ans$ and $max\_ans$ are as stated in the proposition, we have to show that $\rep(T)$ contains repairs $R_{min}$ and $R_{max}$ such that $\ans(\mathcal{AQ}^{[R_{min}]})=min\_ans$ and $\ans(\mathcal{AQ}^{[R_{max}]})=max\_ans$. These results are consequences of Proposition~\ref{prop:indpdt}  by considering $R_{min}=R_2$ and $R_{max}=R_1$ if $aggr = min$, and $R_{min}=R_1$ and $R_{max}=R_2$ if $aggr$ is $max$ or $count$ or if $M_i$-values are positive and $aggr=sum$. We also mention that if $aggr=sum$ with no restriction on the $M_i$-values, $min\_ans$, respectively $max\_ans$, is the least, respectively the largest, sum that can be obtained using all relevant sets $\sigma(M_i)$. We thus have $min\_ans \leq glb$ and $max\_ans \geq lub$, which concludes this part of the proof.

\smallskip
The last item in the proposition regarding $\consans^*(\mathcal{AQ})$ and $ans^*$  is an immediate consequence of Corollary~\ref{coro:cons-ans-ana}. The proof of the proposition  is thus complete.
\hfill$\Box$

\end{document}